\documentclass[final]{siamltex}

\usepackage{epsfig}
\usepackage{amsmath}
\usepackage{amsfonts}

\title{Numerical modeling of elastic waves\\ across imperfect contacts.}

\author{Bruno Lombard\thanks{Laboratoire de M\'ecanique et d'Acoustique, 31 chemin Joseph Aiguier, 13402 Marseille, France({\tt lombard@lma.cnrs-mrs.fr}).}
        \and Jo\"el Piraux\thanks{Laboratoire de M\'ecanique et d'Acoustique, 31 chemin Joseph Aiguier, 13402 Marseille, France({\tt piraux@lma.cnrs-mrs.fr}).}}

\begin{document}

\maketitle

\begin{abstract}
A numerical method is described for studying how elastic waves interact with imperfect contacts such as fractures or glue layers existing between elastic solids. These contacts have been classicaly modeled by interfaces, using a simple rheological model consisting of a combination of normal and tangential linear springs and masses. The jump conditions satisfied by the elastic fields along the interfaces are called the "spring-mass conditions". By tuning the stiffness and mass values, it is possible to model various degrees of contact, from perfect bonding to stress-free surfaces. The conservation laws satisfied outside the interfaces are integrated using classical finite-difference schemes. The key problem arising here is how to discretize the spring-mass conditions, and how to insert them into a finite-difference scheme: this was the aim of the present paper. For this purpose, we adapted an interface method previously developed for use with perfect contacts [J. Comput. Phys. 195 (2004) 90-116]. This numerical method also describes closely the geometry of arbitrarily-shaped interfaces on a uniform Cartesian grid, at negligible extra computational cost. Comparisons with original analytical solutions show the efficiency of this approach.
\end{abstract}



\pagestyle{myheadings}
\thispagestyle{plain}
\markboth{B. LOMBARD AND J. PIRAUX}{ELASTIC WAVES AND IMPERFECT CONTACTS}

\section{Introduction}

Here it is proposed to study the propagation of mechanical waves in an elastic medium divided into several subdomains. The wavelengths are assumed to be much larger than the thickness of the contact zones between subdomains, or {\it interphases} \cite{ROKHLIN1}. Each interphase is replaced by a zero-thickness {\it interface}, where elastic fields satisfy jump conditions. In elastodynamics, the contacts between elastic media are usually assumed to be perfect \cite{ACHENBACH}. They can therefore be modeled by perfect jump conditions, such as perfectly bonded, perfectly lubricated, or stress-free conditions. For example, perfectly bonded conditions will mean that both elastic displacements and normal elastic stresses are continuous across the interface at each time step.

In practice, contacts are often imperfect because of the presence of microcracks or interstitial media in the interphase. Take, for example, fractures in the earth, which are filled with air or liquid, and where jumps occur in the elastic displacements and elastic stresses. The simplest imperfect conditions are the {\it spring-mass conditions} (which are sometimes called "linear slip displacements"): these conditions are realistic in the case of incident waves with very small amplitudes \cite{PYRAK90}. The spring-mass conditions have been extensively studied, both theoretically and experimentally \cite{ROKHLIN1,SCHOENBERG80, TATTERSAL73}. This approach has been applied in various disciplines, such as nondestructive evaluation of materials \cite{BAIK1,ROUSSEAU03} and geophysics \cite{PYRAK90}. 

However, very few studies have dealt so far with the numerical simulation of wave propagation across imperfect contacts described by spring-mass conditions. To our knowledge, only three approaches have been proposed for this purpose. First, Gu {\it et al.} developed a boundary integral method which can be applied to arbitrarily-shaped interfaces \cite{GU2}; but this method requires knowing the Green's functions on both sides of the interface, which complicates the study of realistic heterogeneous media. Secondly, a finite-element method has been proposed by Haney and Sneider \cite{HANEY03}. The jump conditions are incorporated automatically here into the numerical scheme, via the variational formulation, and attempts have been made to perform numerical analysis. The main drawback of this approach is that it requires adapting the mesh to the interface, which increases the computational effort. Thirdly, other authors have approached this problem by implicitly accounting for the boundary conditions by using an equivalent medium, and then deriving new finite-difference formulas \cite{BAREN01,COATES95}. This approach can be applied to arbitrarily-shaped interfaces on a uniform Cartesian grid, but the accuracy is low, and this method involves explicitly changing the numerical scheme near the interface. Note that the inertial effects were not investigated in any of these three cases, although they may be important factors \cite{ROKHLIN1}.

The aim of this paper is to describe a procedure for incorporating the spring-mass conditions into existing finite-difference schemes, on a regular Cartesian grid. The geometry of arbitrarily-shaped interfaces is properly taken into account, reducing the unwanted diffraction classically induced by the Cartesian grid. Lastly, the extra computational cost is low. For this purpose, we adapted the {\it explicit simplified interface method} (ESIM) previously developed for dealing with perfect contacts in 1D \cite{BIBLE1} and 2D \cite{BIBLE2}. A study has also dealt with imperfect contacts in 1D \cite{ALIMENTAIRE1}. In the present study, this approach is extended to 2-D configurations. The focus here is on the description of imperfect contacts; to avoid additional complications, we take media with simple constitutive laws (elastic and isotropic media), but the procedure should be suitable for dealing with more realistic media. The inertial effects are taken into account. 

This paper is organized as follows. In section 2, the problem is stated in terms of the configuration, the conservation laws and the spring-mass conditions. In section 3, a numerical strategy is described for integrating the conservation laws on the whole computational domain: the same scheme is used throughout the domain, but near an interface, {\it modified values} of the solution are used, which implicitly account for the spring-mass conditions. Section 4 is the core part of this paper: it describes in detail how to compute the modified values. Numerical experiments are described in section 5, and comparisons with original analytical solutions show the efficiency of the method. Note that although no rigorous mathematical proof of the validity of the algorithms was obtained, the results of the numerical experiments performed were extremely satisfactory.

\section{Problem statement}

\subsection{Configuration}

\begin{figure}[htbp]
\begin{center}
\includegraphics[scale=0.9]{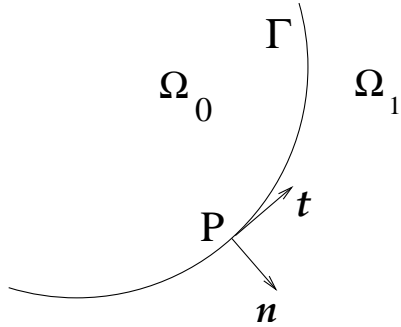}
\caption{Interface $\Gamma$ between two elastic media $\Omega_0$ et $\Omega_1$.}
\label{Patate1}
\end{center}
\end{figure}

Let us consider two isotropic elastic media $\Omega_0$ and $\Omega_1$ separated by a stationary interface $\Gamma$ (figure \ref{Patate1}). We study a two-dimensional configuration with plane strains, and adopt Cartesian coordinates $x$ and $y$ pointing rightward and upward, respectively. The interface is described by a parametric description $(x(\tau),y(\tau))$. The unit tangential vector $\boldsymbol{t}$ and the unit normal vector $\boldsymbol{n}$ are 
\begin{equation}
\boldsymbol{t}=
\frac{\textstyle 1}{\textstyle \sqrt{x^{'2}+y^{'2}}}\,
\left(
\begin{array}{c}
x^{'}\\
\\
y^{'}
\end{array}
\right),\qquad
\boldsymbol{n}=
\frac{\textstyle 1}{\textstyle \sqrt{x^{'2}+y^{'2}}}\,
\left(
\begin{array}{c}
-y^{'}\\
\\
x^{'}
\end{array}
\right),
\label{NT}
\end{equation}
where $x^{'}=\frac{d\,x}{d\,\tau}$ and  $y^{'}=\frac{d\,y}{d\,\tau}$. $\Gamma$ is assumed to be sufficiently smooth to ensure that $x(\tau)$, $y(\tau)$ and their successive spatial derivatives are continuous all along $\Gamma$, up to a given order of derivation. 

The physical parameters are the density $\rho$, the elastic speed of the compressional P-waves $c_p$, and the elastic speed of the shear SV-waves $c_s$. For the sake of simplification, these parameters are taken to be piecewise constant; however, they may be discontinuous across $\Gamma$
\begin{equation}
(\rho,\,c_p,\,c_s)=
\left\{
\begin{array}{l}
(\rho_0,\,c_{p0},\,c_{s0}) \quad \mbox{ if } \quad (x,\,y) \in \Omega_0, \\
\\
(\rho_1,\,c_{p1},\,c_{s1}) \quad \mbox{ if } \quad (x,\,y) \in \Omega_1. 
\end{array}
\right.
\label{Piecewise}
\end{equation}
The elastic fields are the two components of the elastic velocity $\boldsymbol{v}(v_1,v_2)$ and the three independent components of the elastic stress tensor $\boldsymbol{\sigma}(\sigma_{11}, \sigma_{12}, \sigma_{22})$. The projections, normal and tangential to the interface, of the elastic displacement $\boldsymbol{u}(u_1,u_2)$, those of the velocity  $\boldsymbol{v}$, and those of the normal stress $\boldsymbol{\sigma.n}$ are denoted by 
\begin{equation}
\left|
\begin{array}{l}
\displaystyle
u_N= \boldsymbol{u.n} ,\\
\\
\displaystyle
u_T=\boldsymbol{u.t}, 
\end{array}
\right. \qquad
\left|
\begin{array}{l}
\displaystyle
v_N=\boldsymbol{v.n} ,\\
\\
\displaystyle
v_T=\boldsymbol{v.t}, 
\end{array}
\right. \qquad
\left|
\begin{array}{l}
\displaystyle
\sigma_N=(\boldsymbol{\sigma.n}).\boldsymbol{n},\\
\\ 
\displaystyle
\sigma_T=(\boldsymbol{\sigma.n}).\boldsymbol{t}.
\end{array}
\right.
\label{U_NT}
\end{equation}
Let $P$ be a point on $\Gamma$ (figure \ref{Patate1}), and $t$ be the time. Given a function $f(x,\,y,\,t)$, the limit values of $f$ at $P$ on both sides of $\Gamma$ are written
\begin{equation}
f_l(P,\,t)=\lim_{M\rightarrow P,M\in \Omega_l}f(M,\,t),
\label{Flim}
\end{equation}
where $l=0,1$. The jump of $f$ across $\Gamma$, from $\Omega_0$ to $\Omega_1$, is denoted by
\begin{equation}
[f(P,\,t)]=f_1(P,\,t)-f_0(P,\,t).
\end{equation}

\subsection{Conservation laws}

To study the propagation of small perturbations in $\Omega_i$ ($i=0,1$), we use a velocity-stress formulation of elastodynamic equations. Setting
\begin{equation}
\boldsymbol{U}=\,^T(v_1,v_2, \sigma_{11}, \sigma_{12}, \sigma_{22}),
\label{U_sol}
\end{equation}
the linearization of mechanics equations gives a first-order linear hyperbolic system in each subdomain
\begin{equation}
\frac{\textstyle \partial}{\textstyle \partial\,t}\,\boldsymbol{U}+\boldsymbol{A}_l\,\frac{\textstyle \partial}{\textstyle \partial\,x}\,\boldsymbol{U}+\boldsymbol{B}_l\,\frac{\textstyle \partial}{\textstyle \partial\,y}\,\boldsymbol{U}=\boldsymbol{0}, \qquad l=0,1,
\label{LC0}
\end{equation}
which is satisfied outside $\Gamma$. The $5 \times 5$ piecewise-constant matrices $\boldsymbol{A}_l$ and $\boldsymbol{B}_l$ ($l=0,1$) are \cite{ACHENBACH}
\begin{equation*}
\boldsymbol{A}_l=-\left(
\begin{array}{ccccc}
0      &     0   &  \hspace{0.6cm}\displaystyle\frac{\textstyle 1}{\textstyle \rho} \hspace{0.5cm} &  \hspace{0.5cm} 0 \hspace{0.5cm}     &  \hspace{0.5cm}  0 \hspace{0.5cm}  \\
&&&&\\
0      &     0    &    0    & \displaystyle\frac{\textstyle 1}{\textstyle \rho}      &    0\\
&&&&\\
\rho\,c_p^2   &    0    &    0    &    0    &    0\\
&&&&\\
0      &    \rho\,c_s^2   &    0    &    0    &    0\\
&&&&\\
\rho\left(c_p^2-2\,c_s^2\right)  &   0    &    0    &    0    &    0
\end{array}
\right),
\end{equation*}
\\
\begin{equation}
\boldsymbol{B}_l=-\left(
\begin{array}{ccccc}
0      &     0    &   \hspace{0.5cm}  0   \hspace{0.5cm}  & \hspace{0.5cm} \displaystyle\frac{\textstyle 1}{\textstyle \rho}  \hspace{0.5cm}  &  \hspace{0.5cm}  0 \hspace{0.5cm}  \\
&&&&\\
0      &     0    &    0      &    0    &    \displaystyle\frac{\textstyle 1}{\textstyle \rho}\\
&&&&\\
0      &    \rho\left(c_p^2-2\,c_s^2\right)    &     0    &    0    &    0\\
&&&&\\
\rho\,c_s^2   &     0    &    0      &    0    &    0\\
&&&&\\
0      &   \rho\,c_p^2    &    0    &    0    &    0
\end{array}
\right).
\label{MATAB_sol}
\end{equation}

\subsection{Spring-mass conditions}\label{SEC_sM}

An incident wave at the interface generates four other waves: a reflected P-wave, a reflected SV-wave, a transmitted P-wave, and a transmitted SV-wave. To pose the problem suitably, it is necessary to define four independent jump conditions satisfied by the fields along $\Gamma$ (figure \ref{Patate1}). {\it Perfectly bonded conditions} are generally used for this purpose, namely
\begin{equation}
\begin{array}{l}
\displaystyle
\left[u_N(P,\,t)\right]=0,\qquad
\left[\sigma_N(P,\,t)\right]=0,\\
\\
\displaystyle
\left[u_T(P,\,t)\right]=0,\qquad \,
\left[\sigma_T(P,\,t)\right]=0,
\end{array}
\label{Perfect_JC}
\end{equation}
corresponding to a perfectly bonded contact between the two solids in question.

\begin{figure}[htbp]
\begin{center}
\includegraphics[scale=0.6]{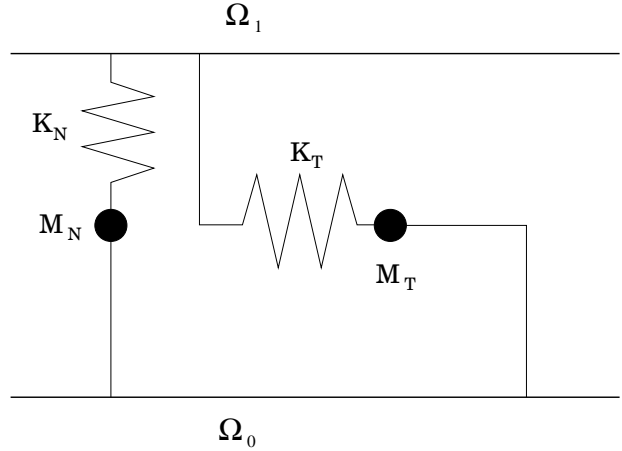}
\caption{Spring-mass rheological model of the contact.}
\label{Masse_ressort}
\end{center}
\end{figure}

To describe an imperfect contact, one can generalize (\ref{Perfect_JC}) into {\it spring-mass conditions}. With the notations defined in (\ref{Flim}), the spring-mass conditions are 
\begin{equation}
\boxed{
\begin{array}{l}
\displaystyle
\left[u_N(P,\,t)\right]=\frac{\textstyle 1}{\textstyle K_N}\,\sigma_{N0}(P,\,t),\qquad
\left[\sigma_N(P,\,t)\right]=M_N\,\frac{\textstyle \partial^2}{\textstyle \partial\,t^2}\,u_{N0}(P,\,t),\\
\\
\displaystyle
\left[u_T(P,\,t)\right]=\frac{\textstyle 1}{\textstyle K_T}\,\sigma_{T0}(P,\,t),\qquad \,\,
\left[\sigma_T(P,\,t)\right]=M_T\,\frac{\textstyle \partial^2}{\textstyle \partial\,t^2}\,u_{T0}(P,\,t),
\end{array}}
\label{SM_JC}
\end{equation}
where $K_N>0$, $K_T >0$, $M_N\geq 0$, $M_T \geq 0$, are called the normal stiffness, the tangential stiffness, the normal mass, and the tangential mass of the interface, respectively. The conditions (\ref{SM_JC}) are called "spring-mass conditions" because of analogies with the equations governing the dynamics of a spring-mass system (figure \ref{Masse_ressort}). One basic underlying assumption made here is that the elastic stresses do not affect the nature of the contact, and hence, that $K_N$, $K_T$, $M_N$, and $M_T$ do not depend on $t$ or on the fields. They can vary with space, and hence with the parameter $\tau$. 

The spring-mass conditions provide an easy way of describing a wide range of contacts between solids, from perfect contact to disconnected media. For $K_N \rightarrow+\infty$, $K_T \rightarrow+\infty$, $M_N=0$, and $M_T=0$, we obtain the perfectly bonded conditions (\ref{Perfect_JC}). For $K_N \rightarrow+\infty$, $K_T \rightarrow 0$, $M_N=0$, and $M_T=0$, we obtain $\sigma_T(P,\,t)\rightarrow0$, which amounts to a perfect slip with no friction. Lastly, for $K_N \rightarrow 0$, $K_T \rightarrow 0$, $M_N=0$, and $M_T=0$, we obtain $\sigma_N(P,\,t)\rightarrow0$ and $\sigma_T(P,\,t)\rightarrow0$, hence $\boldsymbol{\sigma.n}(P,\,t)\rightarrow\boldsymbol{0}$: the media $\Omega_0$ and $\Omega_1$ tend to have stress-free boundaries, which means that no waves are transmitted from one medium to the other. 

The spring-mass conditions entail an important property: the plane waves reflected and transmitted by a plane interface with conditions (\ref{SM_JC}) are frequency-dependent \cite{TATTERSAL73}. These waves therefore show a distorted profile that is quite different from the profile of the incident wave, even below the critical angle. In addition, even if the incident wave is spatially bounded in the direction of propagation, the reflected and transmitted waves are not spatially bounded: a "coda" follows each of these waves. Phenomena of this kind, which do not occur with perfect conditions, are observed experimentally (see e.g. \cite{PYRAK90}). 

By choosing appropriate values of $K_N$, $K_T$, $M_N$, and $M_T$, it is possible to model realistic configurations. The spring-mass conditions (\ref{SM_JC}) can also be obtained quite rigorously in some cases; the values of $K_N$, $K_T$, $M_N$, and $M_T$ will therefore depend on the physical and geometrical properties of the interphase. Take a plane elastic layer sandwiched between two homogeneous isotropic half-spaces. If the thickness of the intermediate layer is much smaller than the wavelength, the spring-mass conditions can be deduced from an asymptotic analysis of the wave propagation behavior inside the layer \cite{ROKHLIN1}. 

The spring-mass conditions model (\ref{SM_JC}) has some limitations. First, if $K_N<+\infty$ and $M_N\neq 0$, or if $K_T<+\infty$ and $M_T\neq 0$, these conditions are asymmetrical. In Appendix A, we establish that the influence of this asymmetry is either null (in the case of reflected waves) or negligible (in that of transmitted waves) in the one-dimensional context. We have checked numerically that this is also the case in the 2-D context. Setting up symmetrical conditions would considerably complicate the jump conditions, in return for absolutely negligible effects.

The second drawback follows from the first equation given by (\ref{SM_JC}): there is no reason why a negative jump of $u_N$ should not occur, whith an absolute value greater than the real thickness of the interphase. Since a penetration of both faces on the interphase is not physically realistic, the conditions (\ref{SM_JC}) are valid only in the case of very small perturbations. With larger perturbations, finer modeling procedures are required, based on nonlinear contact laws \cite{ZHAO01}. Jump conditions of this kind, which are currently under study, require more complex numerical methods. 
 
\section{Time-marching}

\subsection{Numerical scheme}\label{SEC_TM}

To integrate the hyperbolic system (\ref{LC0}), we introduce a uniform lattice of grid points: $(x_i,y_j,t_n)=(i\,\Delta\,x,j\,\Delta\,y,n\,\Delta\,t)$, where $\Delta\,x=\Delta\,y$ are the spatial mesh sizes, and $\Delta\,t$ is the time step. The approximation $\boldsymbol{U}_{i,j}^n$ of $\boldsymbol{U}(x_i,y_j,t_n)$ is computed using explicit two-step, spatially-centred finite-difference schemes. The time-stepping of these schemes is written symbolically
\begin{equation}
\boldsymbol{U}_{i,j}^{n+1}=\boldsymbol{H}_{i,j}\left(\boldsymbol{U}_{i+{\tilde i},\,j+{\tilde j}}^n,\,({\tilde i}, \,{\tilde j})\in\Sigma\right).
\label{TM}
\end{equation}
$\boldsymbol{H}_{i,j}$ is a discrete operator, and $\Sigma$ is the stencil of the scheme. The subscripts in $\boldsymbol{H}_{i,j}$ refer to the physical parameters at $(x_i,\,y_j)$. See \cite{LEV90} for a review of the huge body of literature on numerical methods for conservation laws.

The interface $\Gamma$ is immersed in the regular meshing, so that one can distinguish between two sets of grid points: the {\it regular points}, where the stencil of the scheme involves a single medium, e.g. $\Omega_0$ or $\Omega_1$, and the {\it irregular points}, where the stencil of the scheme crosses $\Gamma$. The distribution of irregular points along $\Gamma$ obviously depends on the geometry of $\Gamma$ and on the stencil of the scheme. At regular points, the scheme (\ref{TM}) is applied classically, as in homogeneous media. At irregular points, however, the scheme (\ref{TM}) is modified to take the spring-mass conditions (\ref{SM_JC}) into account. This modification is carried out using an interface method, and the main aim of the present article is to describe this procedure, which will be presented in detail in subsequent sections.

In the numerical experiments described in section 5, we use a second-order scheme: the Wave Propagation Algorithm (WPALG), originally developed by LeVeque in the field of computational fluid dynamics \cite{WPALG2}. Its stencil is $({\tilde i}=-2\,...2,\,{\tilde j}=-2\,...2)$ and $({\tilde i},\,{\tilde j})\neq(\pm 2,\pm 2)$. WPALG is a useful tool for dealing with linear elastic wave propagation, for at least three reasons. First, it involves the use of nonlinear flux limiters that prevent numerical dispersion. Secondly, this scheme reduces the numerical anisotropy introduced by the Cartesian grid. Thirdly, WPALG is stable in 2D up to CFL=1. For the convergence measurements performed in section 5 (test 1), we also used the standard second-order Lax-Wendroff scheme.

Note that other schemes can be used: in particular, we have successfully combined the interface method with staggered schemes, such as \cite{SAENGER00}. It should therefore be possible to adapt most solvers for use with the interface method described in the forthcoming discussion.

\subsection{Interface method}

\begin{table}[htbp]
\begin{center}
\begin{tabular}{|l||ll|}
\hline
                  &                    &                                                    \\
                  &  Value             &  Number of                                         \\
                  &                    &                                                    \\ 
\hline   
\hline             
                  &                    &                                                    \\
$n_v$             & $5(k+1)(k+2)/2$    &  components of $\boldsymbol{U}_l^k$                \\
                  &                    &                                                    \\
$n_c$             & $2(k+1)(k+2)$      &  jump conditions                                   \\
                  &                    &                                                    \\
$n_m$             & $k(k-1)/2$         &  compatibility conditions                          \\
                  &                    &                                                    \\
$n_q$             & variable           &  grid points to estimate $\boldsymbol{U}_l^k$      \\
                  &                    &                                                    \\
\hline
\end{tabular}
\vspace{0.6cm}
\caption{$n$ with indices used throughout the text ($l=0,\,1$).}
\label{Tab_Notations}
\end{center}
\end{table}

From now on, we will focus on the time-stepping procedure near the interface. To take into account the spring-mass conditions satisfied along $\Gamma$, the scheme (\ref{TM}) is also applied at irregular points, but some of the numerical values used for the time-stepping procedure are changed. The use of so-called {\it modified values} deduced from the jump conditions is the key feature of the interface method developed in our previous studies: the "explicit simplified interface method" (ESIM) \cite{ALIMENTAIRE1, BIBLE2, BIBLE1}. At time $t_n$, the general method used to compute modified values is as follows. On each side of $\Gamma$, one defines a smooth extension $\boldsymbol{U}^*(x,\,y,\,t_n)$ of the exact solution on the other side. The extension $\boldsymbol{U}^*$ is built satisfying the same jump conditions as the exact solution $\boldsymbol{U}$. At any irregular point, the modified value is a numerical estimate of $\boldsymbol{U}^*$ at this point. 

\begin{figure}[htbp]
\begin{center}
\includegraphics[scale=0.8]{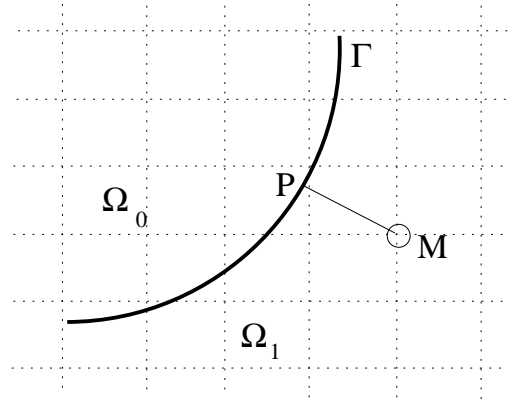}
\caption{Zoom on an irregular point $M$; orthogonal projection $P$ of $M$ onto $\Gamma$.}
\label{Patate_ESIM}
\end{center}
\end{figure}

Let us introduce some notations. Take an irregular point $M$ with coordinates $(x_I,\,y_J)$, belonging to $\Omega_1$ (the following discussion can easily be adapted to the case where $M(x_I,\,y_J)\in\Omega_0$). Let $P(x_P,\,y_P)$ be a point on $\Gamma$ near $M$, for example the closest orthogonal projection of $M$ onto $\Gamma$ (figure \ref{Patate_ESIM}). The vector containing the limit values of the exact solution $\boldsymbol{U}(x,y,t_n)$ and those of its spatial derivatives at $P$ up to the $k$-th order is denoted by 
\begin{equation}
\boldsymbol{U}_l^k=\lim_{M \rightarrow P,M \in \Omega_l}
\,^T\left(
^T\boldsymbol{U},...,
\frac{\textstyle \partial^\alpha}{\textstyle \partial\,x^{\alpha-\beta}\,\partial\,y^\beta}\,^T\boldsymbol{U},...,
\frac{\textstyle \partial^k}{\textstyle \partial\,y^k}\,^T\boldsymbol{U}
\right),
\label{UK}
\end{equation}
where $l$ = 0 or 1, $\alpha=0,...,k$ and $\beta=0,...,\alpha$. This vector has $n_v=5(k+1)(k+2)/2$ components. Throughout the text, many $n$ with indices have been used; to avoid any confusion, they are summed up in table \ref{Tab_Notations}. To obtain concise expressions for the $k$-th order Taylor expansions at $P$ of quantities at $(x_i,\,y_j)$, we define the $5 \times n_v$ matrix
\begin{equation}
\boldsymbol{\Pi}_{i,j}^k=\left(\boldsymbol{I}_5,...,
\frac{\textstyle 1}{\textstyle \beta \,!\,(\alpha-\beta)\,!}\,(x_i-x_P)^{\alpha-\beta}(y_j-y_P)^\beta\boldsymbol{I}_5,...,
\frac{\textstyle (y_j-y_P)^k}{\textstyle k\,!}\,\boldsymbol{I}_5
\right),
\label{PI}
\end{equation}
where $\boldsymbol{I}_5$ is the $5 \times 5$ identity matrix, $\alpha=0,...,k$ and $\beta=0,...,\alpha$. The modified value $\boldsymbol{U}^*_{I,J}$ is then defined as a numerical estimate of the smooth extension 
\begin{equation}
\boldsymbol{U}^*(x_I,y_J,t_n)=\boldsymbol{\Pi}_{I,J}^k \,\boldsymbol{U}_0^k.
\label{VM0}
\end{equation}
Note that $\boldsymbol{U}_0^k$ is the limit value of the solution and its spatial derivatives on the other side of $\Gamma$ with respect to $M$. 

The modified values at all the irregular points surrounding $\Gamma$ are computed in a similar way at $t_n$. The time-stepping at an irregular point $(i,j)$ is written symbolically
\begin{equation}
\boldsymbol{U}_{i,j}^{n+1}=\boldsymbol{H}_{i,j}\left(\hat{\boldsymbol{U}}_{i+{\tilde i},\,j+{\tilde j}}^n,\,({\tilde i}, \,{\tilde j})\in\Sigma\right),
\label{TMESIM}
\end{equation}
where
\begin{equation}
\hat{\boldsymbol{U}}_{i+{\tilde i},\,j+{\tilde j}}=
\left\{
\begin{array}{l}
\boldsymbol{U}_{i+{\tilde i},\,j+{\tilde j}}^n \mbox{  if } (x_{i+{\tilde i}},\,y_{j+{\tilde j}}) \in \Omega_1,\\
\\
\boldsymbol{U}_{i+{\tilde i},\,j+{\tilde j}}^* \mbox{  if } (x_{i+{\tilde i}},\,y_{j+{\tilde j}}) \in \Omega_0.
\end{array}
\right.
\end{equation} 
The time-stepping procedure is completed over the whole computational domain by applying (\ref{TM}) at the regular points. It only remains now to calculate $\boldsymbol{U}^*$'s, since $\boldsymbol{U}_0^k$ in (\ref{VM0}) is unknown. 

\section{Calculating modified values}

\subsection{Differentiation of the spring-mass conditions}\label{SEC_DER}

In the first step towards calculating the modified value (\ref{VM0}), we look for the jump conditions satisfied by $\boldsymbol{U}_l^k$ (\ref{UK}), for any $k$. These conditions are deduced from the spring-mass conditions (\ref{SM_JC}) satisfied by $u_{N,T}$ and $\sigma_{N,T}$. Before describing the procedure, let us introduce a new notation. The vector containing the limit values of the $(k+1)$-th spatial derivatives of $\boldsymbol{U}(x,y,t)$ at $P$ is denoted by
\begin{equation}
\overline{\boldsymbol{U}}_l^{k+1}=\lim_{M \rightarrow P,M \in \Omega_l}
\,^T\left(
\frac{\textstyle \partial^{k+1}}{\textstyle \partial\,x^{k+1}}\,^T\boldsymbol{U},
...,
\frac{\textstyle \partial^{k+1}}{\textstyle \partial\,x^{k+1-\alpha}\,\partial\,y^\alpha}\,^T\boldsymbol{U},
...,
\frac{\textstyle \partial^{k+1}}{\textstyle \partial\,y^{k+1}}\,^T\boldsymbol{U}
\right),
\label{UKbar}
\end{equation}
where $l$ = 0 or 1, $\alpha=0,...,k+1$. This vector has $5(k+2)$ components. Once again, the point $P$ considered and the instant $t$ are omitted.

For $k=0$, the two equations in (\ref{SM_JC}) that deal with the jump in the elastic displacement are differentiated in terms of $t$. The geometry of $\Gamma$ and stiffness and mass values do not depend on $t$; since $\boldsymbol{v}=\frac{\partial\boldsymbol{u}}{\partial\,t}$, we obtain 
\begin{equation}
\begin{array}{ll}
\displaystyle
\left[v_N(P,\,t)\right]=\frac{\textstyle 1}{\textstyle K_N}\,\frac{\textstyle \partial}{\textstyle \partial\,t}\,\sigma_{N0}(P,\,t),\qquad
\left[\sigma_N(P,\,t)\right]=M_N\,\frac{\textstyle \partial}{\textstyle \partial\,t}\,v_{N0}(P,\,t),\\
\\
\displaystyle
\left[v_T(P,\,t)\right]=\frac{\textstyle 1}{\textstyle K_T}\,\frac{\textstyle \partial}{\textstyle \partial\,t}\,\sigma_{T0}(P,\,t),\qquad\,\
\left[\sigma_T(P,\,t)\right]=M_T\,\frac{\textstyle \partial}{\textstyle \partial\,t}\,v_{T0}(P,\,t).
\end{array}
\label{SM_JC0}
\end{equation}
The time derivatives in (\ref{SM_JC0}) are replaced by spatial derivatives thanks to the conservation laws (\ref{LC0}). We sum up the relations thus obtained in matrix terms
\begin{equation}
\boldsymbol{C}_1^0\,\boldsymbol{U}_1^0=\boldsymbol{C}_0^0\,\boldsymbol{U}_0^0+\boldsymbol{E}_0^0\,\overline{\boldsymbol{U}}_0^1.
\label{SM0}
\end{equation} 
$\boldsymbol{C}_l^0$ ($l=0,1$) are $4 \times 5$ matrices; $\boldsymbol{U}_l^0$ are the vectors (\ref{UK}) for $k=0$ (i.e. the limit values of (\ref{U_sol})); $\boldsymbol{E}_0^0$ is a $4 \times 10$ matrix; lastly, $\overline{\boldsymbol{U}}_0^1$ is the vector (\ref{UKbar}) for $l=0$ and $k=0$. Matrices $\boldsymbol{C}_l^0$ describe the perfectly bonded conditions (\ref{Perfect_JC}). Matrix $\boldsymbol{E}_0^0$ describes the correction induced by the springs and masses in (\ref{SM_JC}). Both matrices $\boldsymbol{C}_l^0$ and $\boldsymbol{E}_0^0$ depend on $\tau$, but they are independent of $t$; they are dealt with in greater detail in Appendix B.

To compute the conditions satisfied up to $k=1$, we differentiate (\ref{SM0}) in terms of $t$ and $\tau$. First, the differentiation of (\ref{SM0}) in terms of $t$ yields 
\begin{equation}
\boldsymbol{C}_1^0\, \frac{\textstyle \partial}{\textstyle \partial\,t} \,\boldsymbol{U}_1^0=\boldsymbol{C}_0^0\,\frac{\textstyle \partial}{\textstyle \partial\,t}\,\boldsymbol{U}_0^0+\boldsymbol{E}_0^0\, \frac{\textstyle \partial}{\textstyle \partial\,t}\,\overline{\boldsymbol{U}}_0^1.
\label{SM0t}
\end{equation}
The time derivatives in (\ref{SM0t}) are replaced by spatial derivatives, using the conservation laws (\ref{LC0}); with the notations (\ref{UK}) and (\ref{UKbar}), one readily obtains ($l=0,1$)
$$
\frac{\textstyle \partial}{\textstyle \partial\,t} \,\boldsymbol{U}_l^0=
\left(
\begin{array}{ccc}
\boldsymbol{0} & -\boldsymbol{A}_l & -\boldsymbol{B}_l
\end{array}
\right)\boldsymbol{U}_l^1,\qquad
\frac{\textstyle \partial}{\textstyle \partial\,t} \,\overline{\boldsymbol{U}}_0^1=
\left(
\begin{array}{ccc}
-\boldsymbol{A}_0 & -\boldsymbol{B}_0 & \boldsymbol{0}\\
\boldsymbol{0} & -\boldsymbol{A}_0 & -\boldsymbol{B}_0 
\end{array}
\right)\overline{\boldsymbol{U}}_0^2,
$$
which are injected into (\ref{SM0t}). Secondly, the differentiation of (\ref{SM0}) in terms of $\tau$ yields
\begin{equation}
\left(\frac{\textstyle d}{\textstyle d\,\tau}\,\boldsymbol{C}_1^0\right)\,\boldsymbol{U}_1^0+\boldsymbol{C}_1^0\,\frac{\textstyle \partial}{\textstyle \partial\,\tau}\,\boldsymbol{U}_1^0=
\left(\frac{\textstyle d}{\textstyle d\,\tau}\,\boldsymbol{C}_0^0\right)\,\boldsymbol{U}_0^0+\boldsymbol{C}_0^0\,\frac{\textstyle \partial}{\textstyle \partial\,\tau}\,\boldsymbol{U}_0^0+
\left(\frac{\textstyle d}{\textstyle d\,\tau}\,\boldsymbol{E}_0^0\right)\,\overline{\boldsymbol{U}}_0^1+\boldsymbol{E}_0^0\,\frac{\textstyle \partial}{\textstyle \partial\,\tau}\,\overline{\boldsymbol{U}}_0^1.
\label{SM0tau}
\end{equation}
Since $\boldsymbol{U}_l^0$ ($l=0,1$) and $\overline{\boldsymbol{U}}_0^1$ depend on $x(\tau)$ and $y(\tau)$, the chain-rule gives
$$
\begin{array}{lll}
\displaystyle
l=0,1,\qquad \frac{\textstyle \partial}{\textstyle \partial\,\tau}\,\boldsymbol{U}_l^0&=&
\displaystyle x^{'}\frac{\textstyle \partial}{\textstyle \partial\,x}\,\boldsymbol{U}_l^0+y^{'}\frac{\textstyle \partial}{\textstyle \partial\,y}\,\boldsymbol{U}_l^0,\\
&&\\
&=&\displaystyle
\left(
\begin{array}{ccc}
\boldsymbol{0} & x^{'}\boldsymbol{I}_5 & y^{'}\boldsymbol{I}_5
\end{array}
\right)\boldsymbol{U}_l^1,
\end{array}
$$ 
and 
$$
\frac{\textstyle \partial}{\textstyle \partial \,\tau}\,\overline{\boldsymbol{U}}_0^1=
\left(
\begin{array}{ccc}
x^{'}\boldsymbol{I}_5 & y^{'}\boldsymbol{I}_5 & \boldsymbol{0}\\
\boldsymbol{0} & x^{'}\boldsymbol{I}_5 & y^{'}\boldsymbol{I}_5 
\end{array}
\right)\overline{\boldsymbol{U}}_0^2.
$$
Due to the normalization of vectors $\boldsymbol{n}$ and $\boldsymbol{t}$ in (\ref{NT}) and (\ref{U_NT}), special care must be taken with the differentiation procedure $\frac{d}{d\,\tau}\,\boldsymbol{E}_0^0$ in (\ref{SM0tau}) (see Appendix B). From the previous discussion, one builds three $12 \times 15$ matrices $\boldsymbol{C}_0^1$, $\boldsymbol{C}_1^1$ and $\boldsymbol{D}_0^1$, and one $12 \times 15$ matrix $\boldsymbol{E}_0^1$, so that    
\begin{equation}
\boldsymbol{C}_1^1\,\boldsymbol{U}_1^1=\left(\boldsymbol{C}_0^1+\boldsymbol{D}_0^1\right)\,\boldsymbol{U}_0^1+\boldsymbol{E}_0^1\,\overline{\boldsymbol{U}}_0^2.
\label{SM1}
\end{equation} 
Matrices $\boldsymbol{C}^1_l$ describe the influence of perfectly bonded conditions. Matrices $\boldsymbol{D}_0^1$ and $\boldsymbol{E}_0^1$ describe the changes introduced by the springs and masses.

By iterating a similar procedure $(k-1)$ times, one can find matrices such that 
\begin{equation}
\boxed{
\boldsymbol{C}_1^k\,\boldsymbol{U}_1^k=\left(\boldsymbol{C}_0^k+\boldsymbol{D}_0^k\right)\,\boldsymbol{U}_0^k+\boldsymbol{E}_0^k\,\overline{\boldsymbol{U}}_0^{k+1},}
\label{SMk}
\end{equation} 
where $\boldsymbol{C}_0^k$, $\boldsymbol{C}_1^k$ and $\boldsymbol{D}_0^k$ are $n_c \times n_v$ matrices, with $n_c=2(k+1)(k+2)$ and $n_v=5(k+1)(k+2)/2$; $\boldsymbol{E}_0^k$ is a $n_c \times 5(k+2)$ matrix (in practice and as shown in section \ref{SEC_U*}, the last matrix is never computed). This is a tedious procedure, however, even with low values of $k$ (e.g. $k=1$). This task can be carried out automatically by developing appropriate formal calculus tools; the simulations shown in section 5 were obtained in this way.

\subsection{Compatibility conditions}

Some components of the spatial derivatives of $\boldsymbol{U}$ are not independent. We set
\begin{equation}
\alpha_1=\frac{\textstyle \lambda+2\,\mu}{\textstyle 4\,(\lambda+\mu)}=\frac{\textstyle c_p^2}{\textstyle 4\left(c_p^2-c_s^2\right)},\qquad
\alpha_2=\frac{\textstyle -\lambda}{\textstyle 4\,(\lambda+\mu)}=\frac{\textstyle 2\,c_s^2-c_p^2}{\textstyle 4\left(c_p^2-c_s^2\right)},
\end{equation}
where $\lambda$ and $\mu$ are the Lam\'e coefficients. Then, the standard plane elasticity {\it compatibility conditions} of Saint-Venant \cite{GERMAIN,LOVE} are, in terms of stresses, 
\begin{equation}
\alpha_2\, \frac{\textstyle \partial^2 \,\sigma_{11}}{\textstyle \partial \,x^2}+\alpha_1\, \frac{\textstyle \partial^2 \,\sigma_{22}}{\textstyle \partial \,x^2}
-\frac{\textstyle \partial^2 \,\sigma_{12}}{\textstyle \partial\,x\,\partial \,y}+
\alpha_1 \,\frac{\textstyle \partial^2 \,\sigma_{11}}{\textstyle \partial \,y^2}+\alpha_2 \,\frac{\textstyle \partial^2 \,\sigma_{22}}{\textstyle \partial \,y^2}=0.
\label{CCS0}
\end{equation}
Equation (\ref{CCS0}) is differentiated $(k-2)$-times in terms of $x$ and $y$, giving the $n_m=k(k-1)/2$ relations  
\begin{equation}
\begin{array}{l}
\displaystyle
\alpha_2 \,\frac{\textstyle \partial^k \,\sigma_{11}}{\textstyle \partial \,x^{k-j}\,\partial\,y^j}+\alpha_1 \,\frac{\textstyle \partial^k \,\sigma_{22}}{\textstyle \partial \,x^{k-j}\,\partial\,y^j}
-\frac{\textstyle \partial^k \,\sigma_{12}}{\textstyle \partial\,x^{k-j-1}\,\partial \,y^{j+1}}\\
\\
\displaystyle
+\alpha_1\, \frac{\textstyle \partial^k \,\sigma_{11}}{\textstyle \partial \, x^{k-j-2}\,\partial \,y^{j+2}}+\alpha_2\, \frac{\textstyle \partial^k \,\sigma_{22}}{\textstyle \partial \, x^{k-j-2}\,\partial \,y^{j+2}}=0,\quad k\geq 2, \,j=0,...,k-2.
\label{CCSK}
\end{array}
\end{equation}
Conditions (\ref{CCSK}) are satisfied at each point in $\Omega_l$ ($l=0,1$), especially at $P$ (see figure \ref{Patate1}). One can therefore express the limit values $\boldsymbol{U}_l^k$ in terms of a vector $\boldsymbol{V}_l^k$ with $n_v-n_m$ independent components
\begin{equation}
\boxed{
\boldsymbol{U}_l^k = \boldsymbol{G}_l^k \, \boldsymbol{V}_l^k,\quad l=0,1
}
\label{UGV}
\end{equation}
where $\boldsymbol{G}_l^k$ is a $n_v \times (n_v-n_m)$ matrix deduced from (\ref{CCSK}). The relation (\ref{UGV}) is useful to reduce the number of components in (\ref{SMk}), as seen in the next subsection. An algorithm has been proposed in \cite{BIBLE2} for computing the non-null components of $\boldsymbol{G}_l^k$ but there was a mistake for $k\geq3$. The correct algorithm is ($l=0,\,1$)
\begin{equation}
\left|
\begin{array}{l}
\alpha=0,\quad \beta=0,\\[5pt]
\mbox{for } \gamma=0,...,k, \mbox{ for } \delta=0,...,\gamma\\[5pt]
\qquad \mbox{if } \delta=0 \mbox{ then for }\varepsilon=1,...,5\\[5pt]
\qquad \qquad \alpha=\alpha+1,\quad  \beta=\beta+1,\,\quad\boldsymbol{G}_l^k[\alpha,\beta]=1\\[5pt]
\qquad \mbox{if } \gamma \neq 0 \mbox{ and } \delta \neq 0 \mbox{ and } \gamma \neq \delta \mbox{ then}\\[5pt]
\qquad \qquad \mbox{if } \gamma = 2 \mbox{ then }\nu=0,\,\eta=0,\\[5pt]
\qquad \qquad \mbox{else if } \delta = 1 \mbox{ then }\nu=0,\,\eta=1,\\[5pt]
\qquad \qquad \mbox{else if } \delta = \gamma-1 \mbox{ then }\nu=1,\,\eta=0,\\[5pt]
\qquad \qquad \mbox{else } \nu=1,\,\eta=1,\\[5pt]
\qquad \qquad \alpha=\alpha+1,\quad \beta=\beta+1,\quad \qquad\boldsymbol{G}_l^k[\alpha,\beta]=1\\[5pt]
\qquad \qquad \alpha=\alpha+1,\quad \beta=\beta+1,\quad\qquad \boldsymbol{G}_l^k[\alpha,\beta]=1\\[5pt]
\qquad \qquad \alpha=\alpha+1,\quad \beta=\beta+1,\quad\qquad \boldsymbol{G}_l^k[\alpha,\beta]=1\\[5pt]
\qquad \qquad \alpha=\alpha+1,\quad \beta=\beta-5+\nu,\quad\,\, \boldsymbol{G}_l^k[\alpha,\beta]=\alpha_2\\[5pt]
\qquad \qquad \qquad \qquad \quad\quad \beta=\beta+2-\nu,\quad\, \boldsymbol{G}_l^k[\alpha,\beta]=\alpha_1\\[5pt]
\qquad \qquad \qquad \qquad \quad\quad \beta=\beta+7,\quad \qquad \boldsymbol{G}_l^k[\alpha,\beta]=\alpha_1\\[5pt]
\qquad \qquad \qquad \qquad \quad\quad \beta=\beta+2-\eta,\quad\, \boldsymbol{G}_l^k[\alpha,\beta]=\alpha_2\\[5pt]
\qquad \qquad \alpha=\alpha+1,\quad \beta=\beta-5+\eta,\quad \,\,\boldsymbol{G}_l^k[\alpha,\beta]=1\\[5pt]
\qquad \mbox{if } \gamma \neq 0 \mbox{ and } \gamma = \delta \mbox{ then for }\varepsilon=1,...,5\\[5pt]
\qquad \qquad \alpha=\alpha+1,\quad  \beta=\beta+1,\quad\,\boldsymbol{G}_l^k[\alpha,\beta]=1.
\end{array}
\right.
\end{equation}

\subsection{Solution of underdetermined systems}

It is now required to express $\boldsymbol{V}_1^k$ in terms of $\boldsymbol{V}_0^k$. For this purpose, we inject the compatibility conditions (\ref{UGV}) into the jump conditions (\ref{SMk}). Setting
\begin{equation}
\boldsymbol{S}_1^k=\boldsymbol{C}_1^k\,\boldsymbol{G}_1^k,\qquad \boldsymbol{S}_0^k=\left(\boldsymbol{C}_0^k\,+\boldsymbol{D}_0^k\right)\boldsymbol{G}_0^k,
\label{MatS}
\end{equation}
we obtain the underdetermined system
\begin{equation}
\boldsymbol{S}_1^k\, \boldsymbol{V}_1^k=\boldsymbol{S}_0^k\, \boldsymbol{V}_0^k+\boldsymbol{E}_0^k\, \overline{\boldsymbol{U}}_0^{k+1}.
\label{SVD0}
\end{equation}
To find the full range of solutions of (\ref{SVD0}), a singular value decomposition (SVD) of $\boldsymbol{S}_1^k$ \cite{NRPAS} is computed. Setting $(\boldsymbol{S}_1^k)^{-1}$ to denote the ($n_v-n_m) \times n_c$ generalized inverse of $\boldsymbol{S}_1^k$, and $\boldsymbol{R}_{s_1}^k$ to denote the $(n_v-n_m)\times(n_v-n_m-n_c)$ matrix associated with the kernel of $\boldsymbol{S}_1^k$, we obtain
\begin{equation}
\boxed{
\boldsymbol{V}_1^k= \left(\left(\boldsymbol{S}_1^k\right)^{-1}\boldsymbol{S}_0^k\,|\,\boldsymbol{R}_{s_1}^k\right)
\left(
\begin{array}{c}
\boldsymbol{V}_0^k\\
\\
\boldsymbol{\Lambda}^k
\end{array}
\right)+\left(\boldsymbol{S}_1^k\right)^{-1}\,\boldsymbol{E}_0^k\,\overline{\boldsymbol{U}}_0^{k+1},}
\label{SVD1}
\end{equation}
where $\boldsymbol{\Lambda}_k$ is a $(n_v-n_m-n_c)$ vector of Lagrange multipliers. A similar procedure can be used to express $\boldsymbol{V}_0^k$ in terms of $\boldsymbol{V}_1^k$. 

\subsection{Numerical estimates of limit fields}\label{SEC_U*}

Consider a set ${\mathcal B}$ of $n_q$ grid points surrounding $P$; for practical purposes, we define ${\mathcal B}$ as the set of grid points whose distance from $P$ is less than a given distance $q$. We successively examine each grid point in ${\mathcal B}$, writing the $k$-th order Taylor expansions of $\boldsymbol{U}(x_i,y_j,t_n)$ at $P$. If $(x_i,\,y_j)$ is included in $\Omega_0$, we deduce from (\ref{PI}) and (\ref{UGV}) that
\begin{equation}
\begin{array}{lll}
(x_i,\,y_j) \in  {\mathcal B}\cap \Omega_0, \quad \boldsymbol{U}(x_i,y_j,t_n) &=& \boldsymbol{\Pi}_{i,j}^k\,\boldsymbol{U}_0^k+ \boldsymbol{O}(\Delta\,x^{k+1})\\
&&\\
&=& \boldsymbol{\Pi}_{i,j}^k\,\boldsymbol{G}_0^k\,\boldsymbol{V}_0^k+ \boldsymbol{O}(\Delta\,x^{k+1})\\
&&\\
&=& \boldsymbol{\Pi}_{i,j}^k\,\boldsymbol{G}_0^k\,\left(\boldsymbol{1}\,|\,\boldsymbol{0}\right)
\left(
\begin{array}{c}
\boldsymbol{V}_0^k\\
\\
\boldsymbol{\Lambda}^k
\end{array}
\right)+ \boldsymbol{O}(\Delta\,x^{k+1}),
\end{array}
\label{TAY1}
\end{equation}
where $\boldsymbol{1}$ is the identity matrix $(n_v-n_m)\times(n_v-n_m)$, and $\boldsymbol{0}$ is the null matrix $(n_v-n_m)\times(n_v-n_m-n_c)$. If $(x_i,\,y_j)$ is included in $\Omega_1$, we deduce from (\ref{PI}), (\ref{UGV}), and (\ref{SVD1}) that
\begin{equation}
\begin{array}{lll}
(x_i,\,y_j) \in  {\mathcal B}\cap \Omega_1, \quad \boldsymbol{U}(x_i,y_j,t_n) &=& \boldsymbol{\Pi}_{i,j}^k\,\boldsymbol{U}_1^k + \boldsymbol{O}(\Delta\,x^{k+1})\\
&&\\
&=& \boldsymbol{\Pi}_{i,j}^k\,\boldsymbol{G}_1^k\,\boldsymbol{V}_1^k+ \boldsymbol{O}(\Delta\,x^{k+1})\\
&&\\
&=& \boldsymbol{\Pi}_{i,j}^k\,\boldsymbol{G}_1^k\,
\left(\left(\boldsymbol{S}_1^k\right)^{-1}\boldsymbol{S}_0^k\,|\,\boldsymbol{R}_{s_1}^k\right)
\left(
\begin{array}{c}
\boldsymbol{V}_0^k\\
\\
\boldsymbol{\Lambda}^k
\end{array}
\right) \\
&&+\boldsymbol{\Pi}_{i,j}^k\,\boldsymbol{G}_1^k\,\left(\boldsymbol{S}_1^k\right)^{-1}\,\boldsymbol{E}_0^k\,\overline{\boldsymbol{U}}_0^{k+1}+\boldsymbol{O}(\Delta\,x^{k+1}).
\end{array}
\label{TAY2}
\end{equation}
Relations (\ref{TAY1}) and (\ref{TAY2}) are summed up in matrix terms as follows
\begin{equation}
\left(
\boldsymbol{{\mathcal U}}^n
\right)_{\mathcal B}
=\boldsymbol{M}
\left(
\begin{array}{c}
\boldsymbol{V}_0^k\\
\\
\boldsymbol{\Lambda}^k
\end{array}
\right)
+
\boldsymbol{N}\,\overline{\boldsymbol{U}}_0^{k+1}
+
\left(
\begin{array}{c}
\boldsymbol{O}(\Delta\,x^{k+1})\\
\vdots\\
\boldsymbol{O}(\Delta\,x^{k+1})
\end{array}
\right),
\label{U=MV}
\end{equation}
where $\left(\boldsymbol{{\mathcal U}}^n\right)_{\mathcal B}$ refers to the set of exact values $\boldsymbol{U}(x_i,y_j,t_n)$ at the grid points of ${\mathcal B}$, $\boldsymbol{M}$ is a $5\,n_q \times(2\,n_v-2\,n_m-n_c)$ matrix, and $\boldsymbol{N}$ is a $5\,n_q \times 5(k+2)$ matrix. The value of $n_q$ depends on ${\mathcal B}$; this set is chosen so that (\ref{U=MV}) is overdetermined. In view of Table \ref{Tab_Notations} and Table \ref{Tab_Matrice}, this means that
$$
\begin{array}{lll}
n_q&\geq&(2\,n_v-2\,n_m-n_c)/5,\\
&&\\
&\geq&2\left(k^2+5\,k+3\right)/5.
\end{array}
$$
To ensure this inequality, 
\begin{equation}
q=(k+0.2)\,\Delta\,x
\end{equation}
can be used as a radius to obtain ${\mathcal B}$. From now on, numerical values and exact values are used indiscriminately in (\ref{U=MV}). We also remove  $\boldsymbol{N}\,\overline{\boldsymbol{U}}_0^{k+1}$ and the remainder terms. The least-squares inverse $\boldsymbol{M}^{-1}$ of $\boldsymbol{M}$ is computed using classical techniques, such as LU or SVD \cite{NRPAS}. The set of Lagrange multipliers $\boldsymbol{\Lambda}^k$ is not useful for computing modified values; the restriction $\overline{\boldsymbol{M}^{-1}}$ of $\boldsymbol{M}^{-1}$ is therefore defined by keeping only the $(n_v-n_m)$ first lines of $\boldsymbol{M}^{-1}$. This gives the least-squares numerical estimate
\begin{equation}
\boldsymbol{V}_0^k=\overline{\boldsymbol{M}^{-1}}\left(\boldsymbol{{\mathcal U}}^n\right)_{\mathcal B}.
\label{V0k}
\end{equation}

\subsection{Modified value}

We now have all the tools required to be able to compute the modified value at $M(x_I,\,y_J)$, as defined in section 3.2. The modified value at $(x_I,\,y_J)$ is deduced from (\ref{VM0}), (\ref{UGV}), and (\ref{V0k})
\begin{equation}
\boxed{
\boldsymbol{U}_{I,J}^*=\boldsymbol{\Pi}_{I,J}^k\,\boldsymbol{G}_0^k\,\overline{\boldsymbol{M}^{-1}}\left(\boldsymbol{{\mathcal U}}^n\right)_{\mathcal B}.}
\label{UIJ*}
\end{equation}
A similar procedure is applied at each irregular point surrounding $\Gamma$. All the matrices required to compute the modified values are recalled in Table \ref{Tab_Matrice}, with their sizes.

\begin{table}[htbp]
\begin{center}
\begin{tabular}{|l||ll|}
\hline
                                 &                                     &                                                   \\
Matrix                           &  Size                               &  Comment                                          \\
                                 &                                     &                                                   \\ 
\hline   
\hline             
                                 &                                     &                                                   \\
$\boldsymbol{C}_l^k$             & $n_c \times n_v$                    & jump conditions: perfect contact (\ref{SMk})      \\
                                 &                                     &                                                   \\
$\boldsymbol{D}_l^k$             & $n_c \times n_v$                    & jump conditions: imperfect contact (\ref{SMk})    \\
                                 &                                     &                                                   \\
$\boldsymbol{G}_l^k$             & $n_v \times (n_v-n_m)$              & compatibility conditions (\ref{UGV})              \\
                                 &                                     &                                                   \\
$\boldsymbol{S}_l^k$             & $n_c \times (n_v-n_m)$              & (\ref{MatS})                                      \\
                                 &                                     &                                                   \\
$\boldsymbol{R}_{s_l}^k$         & $(n_v-n_m) \times (n_v-n_c-n_m)$    & kernel of $\boldsymbol{S}_l^k$ (\ref{SVD1})       \\
                                 &                                     &                                                   \\
$\boldsymbol{\Pi}_{i,j}^k$       & $5 \times n_v$                      & Taylor expansions (\ref{PI})                      \\
                                 &                                     &                                                   \\
$\boldsymbol{1}$                 & $(n_v-n_m) \times (n_v-n_m)$        & identity (\ref{TAY1})                             \\
                                 &                                     &                                                   \\
$\boldsymbol{0}$                 & $(n_v-n_m) \times (n_v-n_c-n_m)$    & null (\ref{TAY1})                                 \\
                                 &                                     &                                                   \\
$\boldsymbol{M}$                 & $5\,n_q \times (2\,n_v-2\,n_v-n_c)$ & (\ref{U=MV})                                      \\
                                 &                                     &                                                   \\
$\overline{\boldsymbol{M}^{-1}}$ & $(n_v-n_q) \times 5 \,n_q$          & restriction of $\boldsymbol{M}^{-1}$ (\ref{UIJ*})      \\
                                 &                                     &                                                   \\
\hline
\end{tabular}
\vspace{0.6cm}
\caption{Matrices used for computing the modified values ($l=0,\,1$).}
\label{Tab_Matrice}
\end{center}
\end{table}

\subsection{Comments about the algorithm}

This algorithm can easily be adapted to existing codes without having to change the scheme. At each time step, one only needs to compute a few "modified values" (\ref{UIJ*}), which do not depend on the scheme selected. 

Since the interface is stationary and the nature of the contacts does not vary with time, most of the algorithm can be built up during a preprocessing step, consisting in determining irregular points, deriving interface conditions (\ref{SMk}), solving underdetermined systems (\ref{SVD1}), and computing the matrices involved in (\ref{UIJ*}). All these quantities are stored for future use. At each time step, we need only to carry out one matrix-vector multiplication (\ref{UIJ*}) at each irregular point, before running time-stepping procedure. The matrices involved here are small: $5\times 5\,n_q$ components, usually with $n_q\approx$ 20. In addition, the number of irregular points is much smaller than the number of grid points. The extra computational cost is therefore very low. The CPU time measurements are likely to be approximately the same as those presented in \cite{BIBLE2}.

The main aim of the interface method is to incorporate the jump conditions into finite-difference schemes. But it also gives a finer resolution of the geometries than the poor stair-step description induced by the Cartesian grid. The differentiations of spring-mass conditions involve successive derivatives of $x^{'}$ and $y^{'}$ (see section \ref{SEC_DER}). In addition, the Taylor expansions in (\ref{TAY1}), (\ref{TAY2}) and (\ref{UIJ*}) give an information about the position of $P$ inside the mesh. 

In section \ref{SEC_TM}, we have stated that the interface method can be adapted to staggered grid schemes. This adaptation requires the algebra in sections 3 and 4 to be completely rewritten (although this is quite straightforward). To give an idea of the changes involved, let us focus on the staggered scheme \cite{SAENGER00}, with a grid $(x_{i},\,y_{j},\,t_n)$ for stresses and a grid $(x_{i+\frac{1}{2}},\,y_{j+\frac{1}{2}},\,t_{n+\frac{1}{2}})$ for velocities. Two different sets of irregular points now need to be stored. The jump conditions for velocities and stresses are not correlated (see (\ref{SM0}) and Appendix B); two independant systems of jump conditions (\ref{SMk}) are therefore written. Since the jump conditions satisfied by the velocities yield as many equations as unknowns, they do not require a SVD. The compatibility conditions (\ref{CCSK}) are applied only to the jump conditions satisfied by the stresses, that constitute an underdetermined system. The modified values calculated in sections 4-4 and 4-5 are also split in two parts: at time $t_{n+\frac{1}{2}}$, the modified values of the stresses are computed (based on numerical values of $\boldsymbol{\sigma}$ at $t_n$) before being inserted into the time-marching procedure performed on the velocities. In the same way, at time $t_{n+1}$, the modified values of the velocities are computed (based on numerical values of $\boldsymbol{v}$ at $t_{n+\frac{1}{2}}$) before they are used to perform the time-marching procedure on the stresses.

\subsection{Open questions}\label{SEC_OPEN}

Four questions relating to numerical analysis remain to be solved. First, what are the effects of the term $\boldsymbol{N}\,\overline{\boldsymbol{U}}_0^{k+1}$ which was neglected in (\ref{U=MV}) ? In 1D \cite{ALIMENTAIRE1}, we established that this term is of the same order as the remainder terms. The proof in 1D was based on performing explicit calculations on MAPLE; this approach seems difficult to apply in the 2-D context, because it depends on the geometry of $\Gamma$ near $P$.

Secondly, what is the local truncation error at the irregular points obtained by combining the interface method with the scheme ? In the 1-D context \cite{ALIMENTAIRE1,BIBLE1}, we established that the following was true: if $r$ is the order of the scheme, the local truncation error at irregular points is still equal to $r$ if $k\geq r$ (which means $k\geq 2$ in the case of second-order schemes such as WPALG), when this scheme is combined with the interface method. Extending this result to the 2-D context would require answering the previous question about $\boldsymbol{N}\,\overline{\boldsymbol{U}}_0^{k+1}$, and bounding the matrix $\overline{\boldsymbol{M}^{-1}}$. In practice, the convergence measurements indicate that this property is false in the 2-D context (at least for $r=2$): we need $k=3$ to ensure second-order accuracy (see section \ref{SEC_NUM1}).

Thirdly, what happens in the extreme case of a homogeneous medium (that is, $\rho_0=\rho_1$, $c_{p0}=c_{p1}$, $c_{s0}=c_{s1}$, $K_N \rightarrow+\infty$, $K_T \rightarrow+\infty$, $M_N=0$ and $M_T=0$) ? The ideal would obviously be that $\boldsymbol{U}_{i,j}^*=\boldsymbol{U}_{i,j}^n$, in order to recover the scheme in homogeneous medium. In 1-D configurations \cite{ALIMENTAIRE1}, we have shown that it is the case, given simple conditions about the set ${\mathcal B}$ used to estimate the modified values. In a 2-D context, these conditions cannot be satisfied because estimating the modified values requires least-squares computations. However, it would be interesting to estimate the difference between $\boldsymbol{U}_{i,j}^*$ and $\boldsymbol{U}_{i,j}^n$, which might lead to an optimal choice of  ${\mathcal B}$. 

Fourthly and lastly, the stability analysis still remains to be performed. With a nonlinear scheme (such as WPALG), this seems out of reach so far, but an analysis could perhaps be carried out with linear schemes (such as the Lax-Wendroff scheme). In the absence of theoretical rules, we have numerically considered many geometrical configurations, physical parameters values, stiffness and mass values. With a wide range of parameters, no instabilities are usually observed up to the CFL limit, even after very long integration times (a few thousands of time steps). However, instabilities are observed in two cases. First, instabilities can increase when the physical parameters differ considerably between the two sides of an interface. This problem was previously mentioned in \cite{BIBLE2} in connection with perfect contacts. In practice, it is not too penalizing when dealing with realistic media: even quite large differences between the impedance values, such as those encountered in the case of  Plexiglass-aluminium interfaces, yield stable computations. Secondly, the instabilities can increase at low values of $K_{N,T}$, when the two media tend to be disconnected. This may be due to  the small denominators in (\ref{SM_JC}). In this case, increasing the order $k$ provides an efficient means of improving the stability limit (see subsection \ref{SEC_NUM3}).

\section{Numerical experiments}

\subsection{Configurations}

\begin{figure}[htbp]
\begin{center}
\begin{tabular}{cc}
$(i=0)$ & $(i=0)$\\
&\\
\includegraphics[width=5.2cm,height=5.2cm]{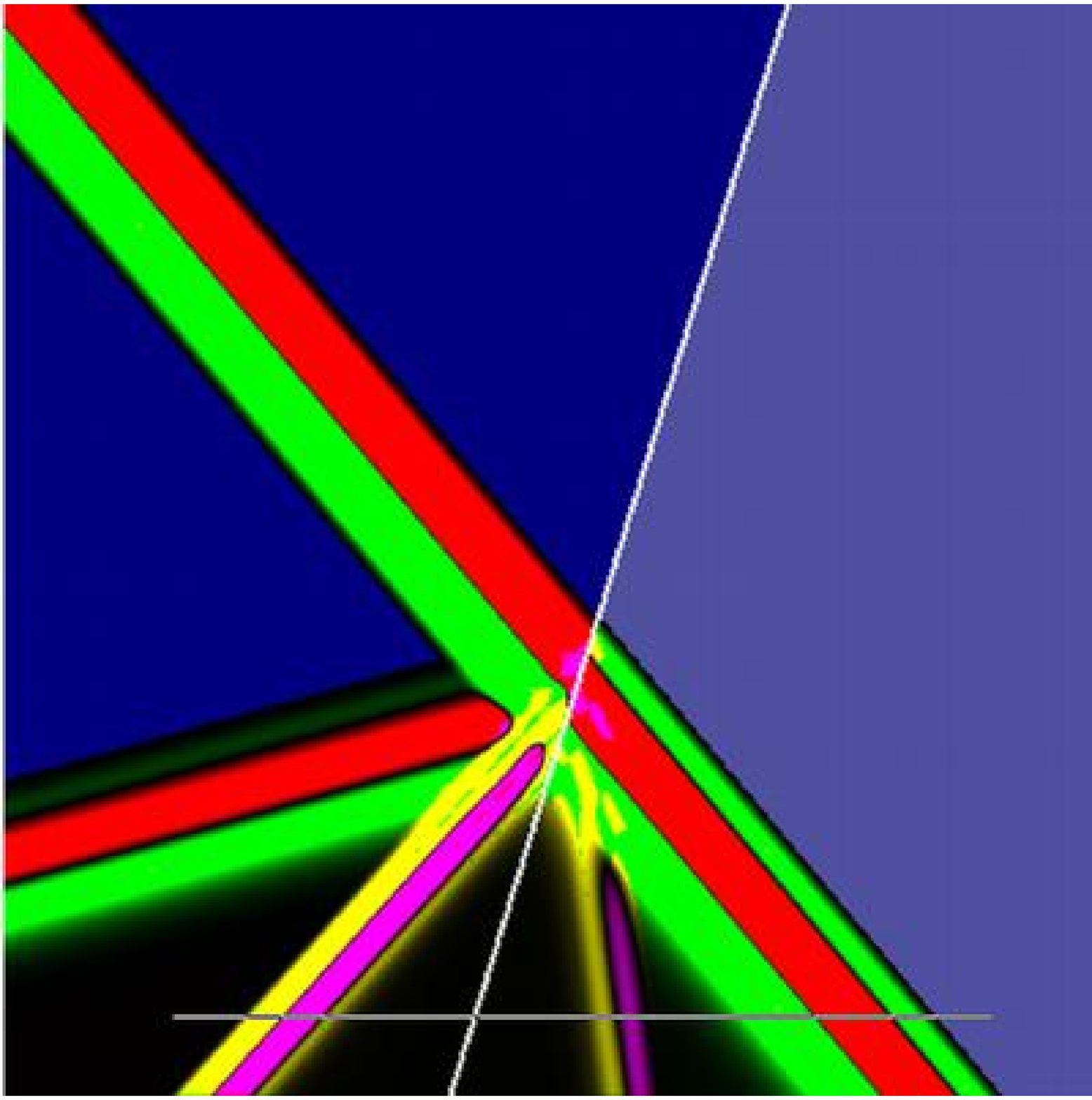}& 
\includegraphics[width=5.2cm,height=5.2cm]{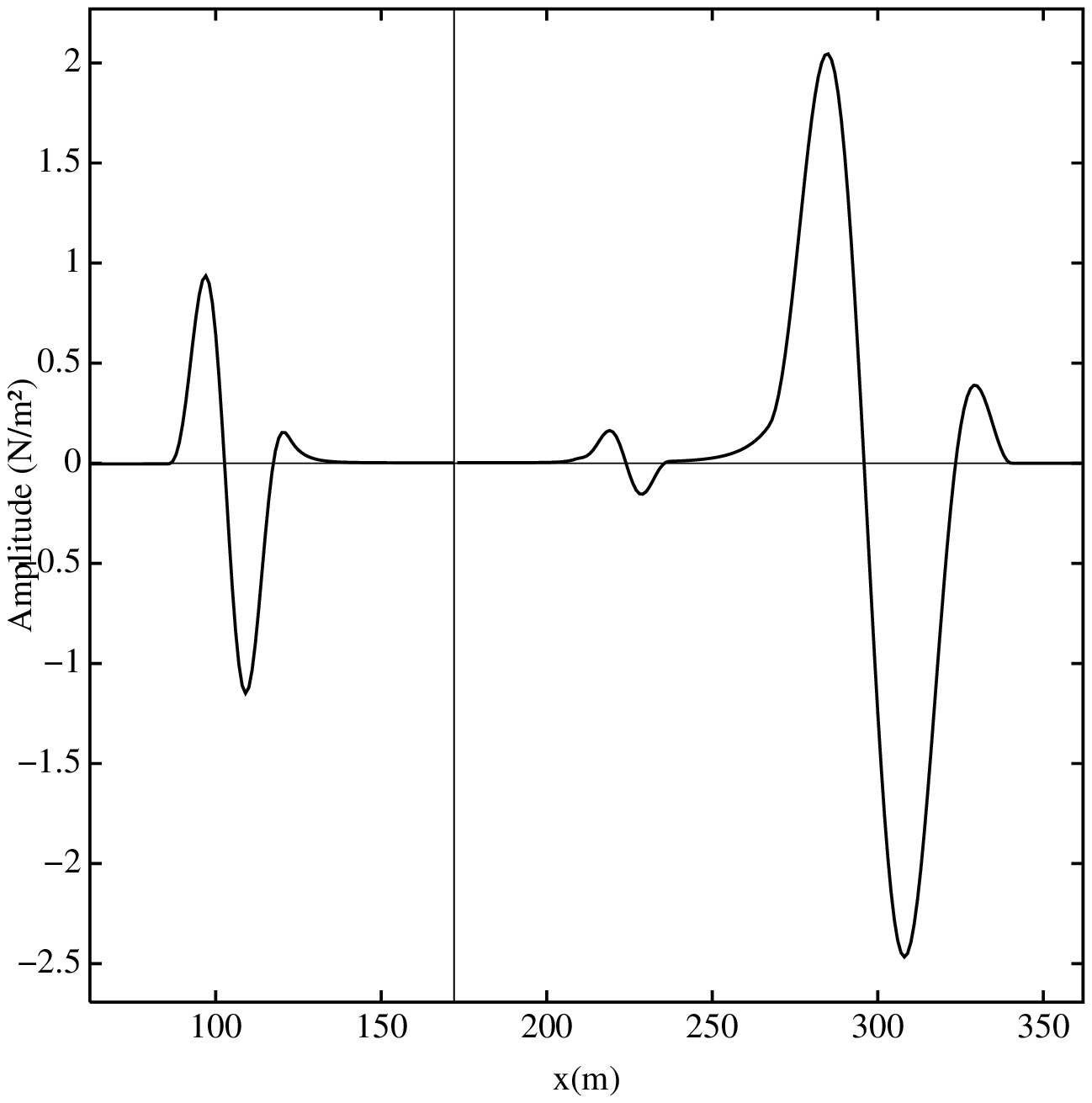}\\
&\\
$(i=1)$ & $(i=1)$\\
&\\
\includegraphics[width=5.2cm,height=5.2cm]{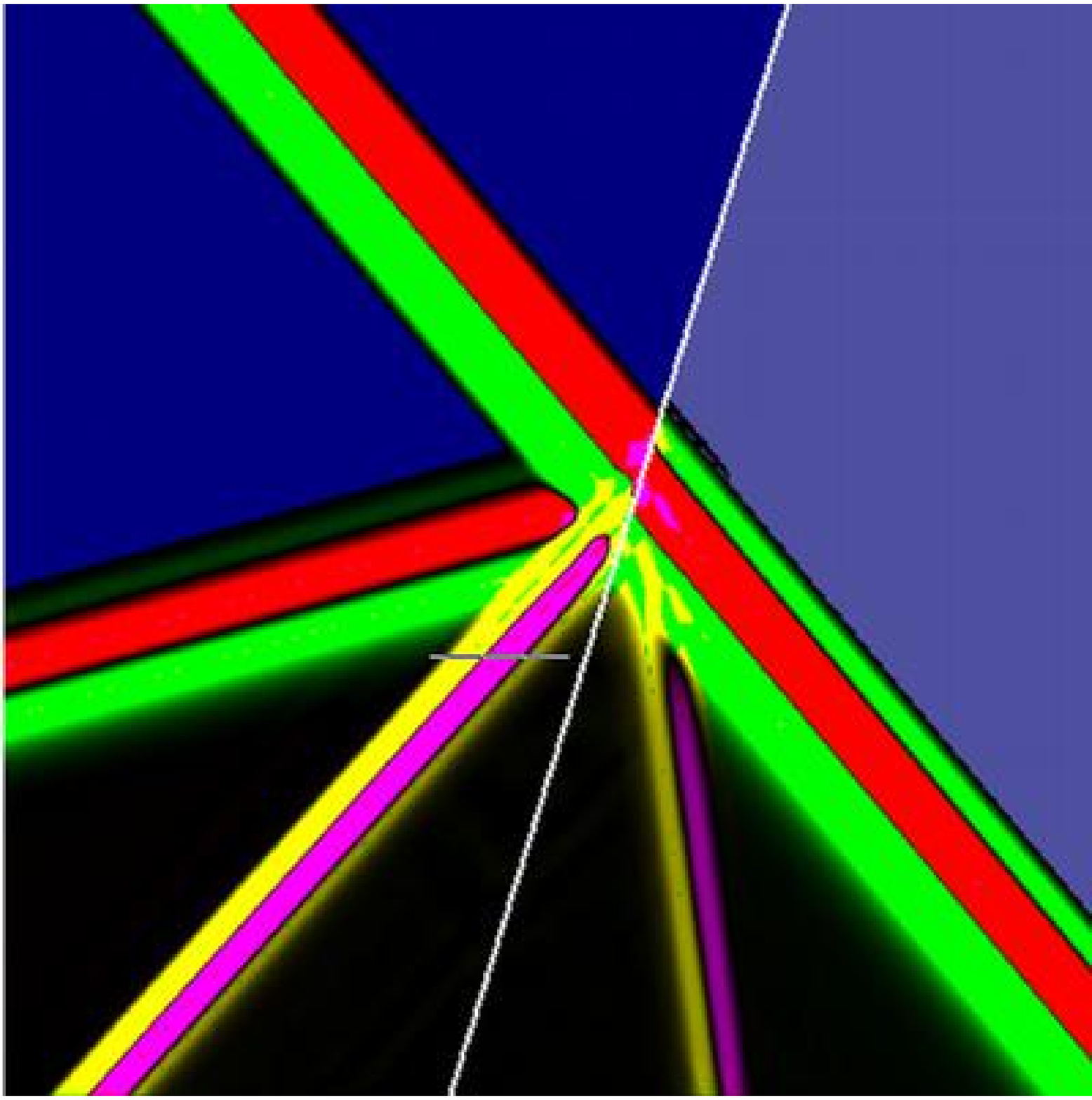}& 
\includegraphics[width=5.2cm,height=5.2cm]{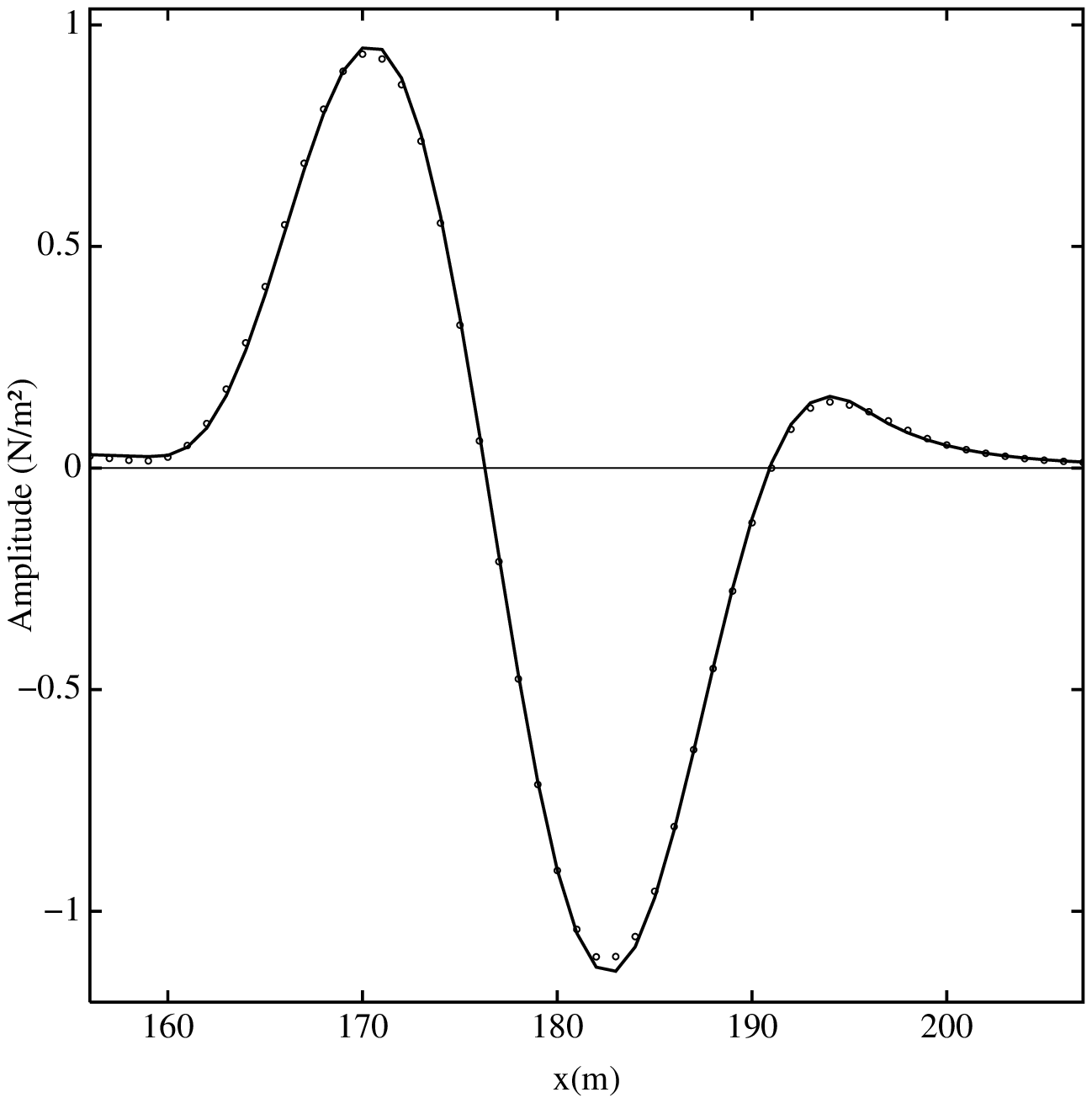}\\
&\\
$(i=2)$ & $(i=2)$\\
&\\
\includegraphics[width=5.2cm,height=5.2cm]{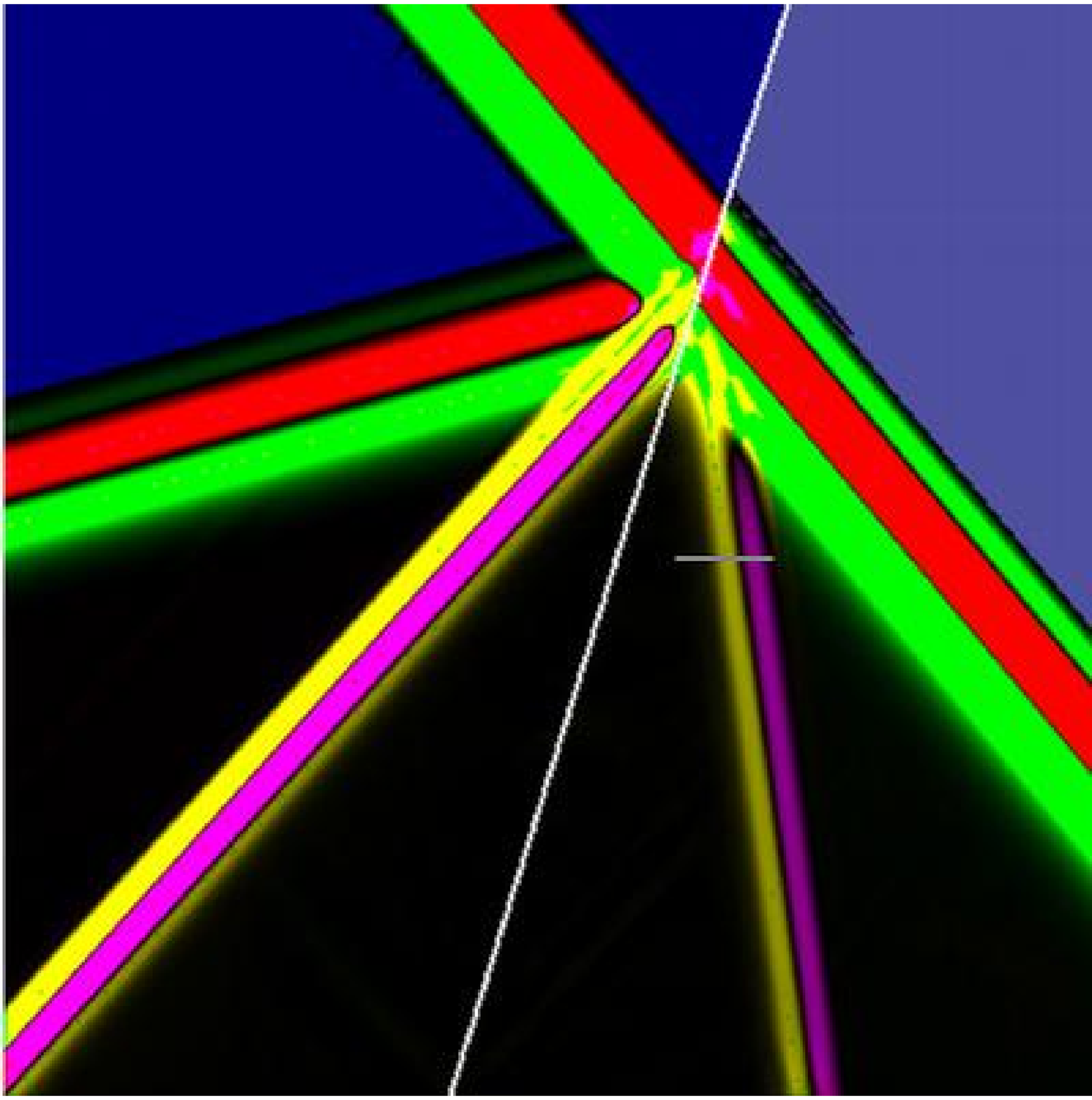}& 
\includegraphics[width=5.2cm,height=5.2cm]{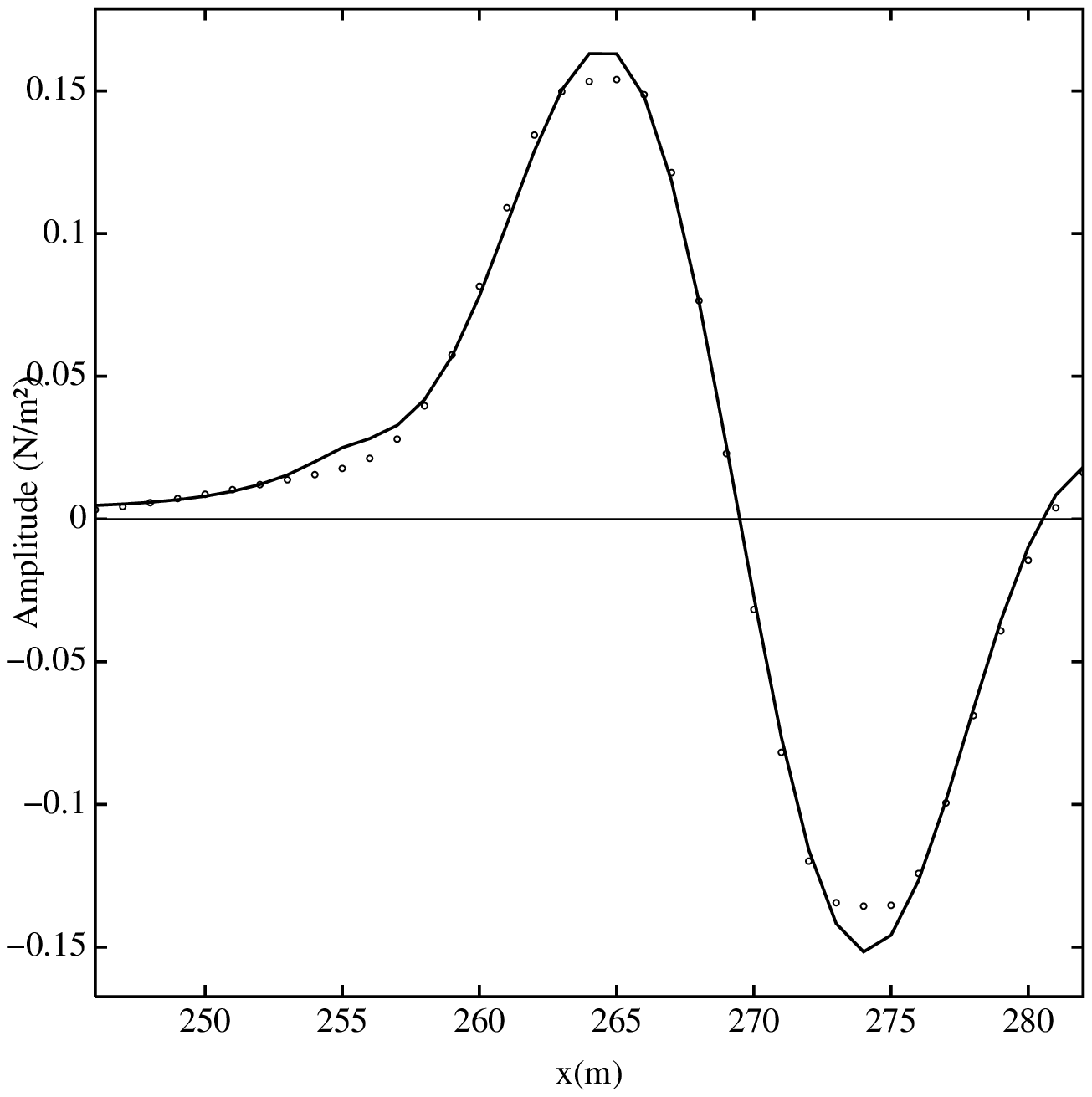}
\end{tabular}
\caption{Test 1: plane interface between identical media at $t=t_0+75\,i\,\Delta\,t$.}
\label{Test1}
\end{center}
\end{figure}

Five numerical experiments are performed here. Except for test 3, the physical parameters are the same on both sides of $\Gamma$: this enables us to underscore both the effects of the spring-mass conditions and the accuracy of the interface method, since all wave reflections and changes observed will result from the spring-mass conditions, described numerically by the interface method.

Three geometrical configurations are studied: a plane interface, a circular interface, and a non-canonical object described by cubic splines. Analytical solutions can be obtained for the first two configurations when the spring-mass conditions are constant. The analytical solution of the problem with a plane wave impinging on a plane interface with spring-mass conditions can be quite easily calculated using Fourier analysis; this procedure will therefore not be described here. We have not found articles dealing with the more intricate case of a circular interface; since this analytical solution is useful to validate the algorithm, it has been described in Appendix C. Except for the convergence measurements in Test 1, all the computations were performed with the WPALG. 

Apart from test 5, the computations are all initialized by a plane right-going P-wave
\begin{equation}
\boldsymbol{U}(x,y,t_0)=\varepsilon\,
\left(
\begin{array}{c}
\displaystyle
-\frac{\textstyle \cos \theta}{\textstyle c_{p1}}\\
\\
\displaystyle
-\frac{\textstyle \sin \theta}{\textstyle c_{p1}}\\
\\
\displaystyle
\frac{\textstyle \lambda_1+2\,\mu_1\,\cos^2 \theta}{\textstyle c_{p1}^2}\\
\\
\displaystyle
\frac{\textstyle 2\,\mu_1\,\sin \theta\,\cos \theta}{\textstyle c_{p1}^2}\\
\\
\displaystyle
\frac{\textstyle \lambda_1+2\,\mu_1\,\sin^2 \theta}{\textstyle c_{p1}^2}
\end{array}
\right)\,f\left(t_0-\frac{\textstyle x \cos \theta+y \sin \theta}{\textstyle c_{p1}}\right),
\label{Init}
\end{equation}
where $\varepsilon$ is an amplitude factor, $\mu_1=\rho_1\,c_{s1}^2$ and $\lambda_1=\rho_1(c_{p1}^2-2\,c_{s1}^2)$ are the Lam\'e coefficients, $\theta$ is the angle between the direction of propagation and the $x$-axis, and $t_0$ is the initial instant. Function $f$ is a $C^2$ spatially-bounded sinusoid
\begin{equation}
f(\xi)=
\left\{
\begin{array}{l}
\displaystyle
\sin(\omega_c\,\xi)-\frac{\textstyle 1}{\textstyle 2}\,\sin(2\,\omega_c\,\xi)\quad \mbox{ if  }\, 0<\xi<\frac{\textstyle 1}{\textstyle f_c},\\
\\
0 \,\mbox{ else}, 
\end{array}
\right.
\label{JKPS}
\end{equation}
where $f_c$ is the central frequency, and $\omega_c=2\,\pi\,f_c$.

We will conclude this section by making some comments on the figures. $\sigma_{11}$ is shown with a green-red palette for P-waves, and a magenta-yellow palette for SV-waves (in the electronic version of this article, the plates are color). The distinction between these waves depends on numerical estimates of div $\boldsymbol{v}$ and {\bf curl} $\boldsymbol{v}$. Most of the snapshots shown are accompanied by a slice; the position of a slice is denoted by a horizontal line on the corresponding snapshot. On each slice, the exact solution and the numerical solution are indicated by a solid line and by points, respectively.

\begin{table}[htbp]
\begin{center}
\begin{tabular}{cccccc}
\hline
                  &      &                     &                      &                &                 \\
Scheme            &$N_x$ & L$_{\infty}$ error  &  L$_{\infty}$ order  &   L$_1$ error  &  L$_1$ order    \\
                  &      &                     &                      &                &                 \\
\hline
\hline
                  &      &                     &                      &                &                 \\
Lax-Wendroff      & 50   &  6.265e-1           &      -               &  4.676e3       &   -             \\
                  & 100  &  1.760e-1           &     {\bf 1.832}      &  1.336e3       &   {\bf 1.807}  \\
+                 & 200  &  4.755e-2           &     {\bf 1.888}      &  3.718e2       &   {\bf 1.845}   \\
                  & 400  &  1.228e-2           &     {\bf 1.953}      &  9.437e1       &   {\bf 1.978}   \\
ESIM              & 800  &  3.164e-3           &     {\bf 1.956}      &  2.311e1       &   {\bf 2.030}   \\
                  & 1600 &  7.916e-4           &     {\bf 1.999}      &  5.667e0       &   {\bf 2.028}   \\
                  &      &                     &                      &                &                 \\
\hline
                  &      &                     &                      &                &                 \\
WPALG             & 50   &  4.570e-1           &      -               &  6.259e2       &    -            \\
                  & 100  &  1.472e-1           &     {\bf 1.634}      &  1.504e2       &   {\bf 2.058}   \\
+                 & 200  &  4.260e-2           &     {\bf 1.789}      &  3.954e1       &   {\bf 1.926}   \\
                  & 400  &  1.398e-2           &     {\bf 1.607}      &  1.071e1       &   {\bf 1.884}   \\
ESIM              & 800  &  4.464e-3           &     {\bf 1.647}      &  2.833e0       &   {\bf 1.919}   \\
                  & 1600 &  1.503e-3           &     {\bf 1.570}      &  7.369e-1      &   {\bf 1.943}   \\
                  &      &                     &                      &                &                 \\
\hline
\end{tabular}
\vspace{0.6cm}
\caption{Convergence measurements in Test 1.}
\label{TAB_pLANE}
\end{center}
\end{table}

\subsection{Test 1: plane interface between identical media}\label{SEC_NUM1}

Here we take the case of a $L_x \times L_y=400 \times 400$ m$^2$ domain, with a plane inclined interface $\Gamma$. The points ($x=$ 168 m, $y=17$ m) and ($x=241$ m, $y=253$ m) belong to $\Gamma$, and hence the angle between $\Gamma$ and the $x$-axis is roughly equal to 72.8 degrees. The physical parameters are identical on both sides of $\Gamma$ 
$$
\left\{
\begin{array}{l}
\rho_0=\rho_1=1200\mbox{ kg/m}^3,\\
\\
c_{p0}=c_{p1}=2800\mbox{ m/s},\\
\\
c_{s0}=c_{s1}=1400\mbox{ m/s},
\end{array}
\right.
$$
which corresponds to realistic values in the case of Plexiglass. The spring-mass conditions are constant along $\Gamma$; the stiffness and mass values are
$$
\left\{
\begin{array}{l}
K_N=10^{9}\,\mbox{ kg/s}^{2}, \quad K_T=10^{7}\,\mbox{ kg/s}^{2},\\
\\
M_N=2000\,\mbox{ kg/m}^{2},\quad M_T=1000\,\mbox{ kg/m}^{2}.
\end{array}
\right.
$$
The incident P-wave (\ref{Init}) is defined by: $\theta$ = 40 degrees, $f_c=49.9$ Hz, $t_0=0.1$ s, and $\varepsilon=2\,10^{-3}$. The numerical experiments are performed with $N_x \times N_y=400 \times 400$ grid points, which amounts to 56 and 28 grid points per central wavelength for P-waves and SV-waves, respectively. We take CFL=0.9, and $k=2$ for the extrapolations involved in the interface method. To initialize the computation, one also needs to compute reflected and transmitted P- and SV-waves, via discrete Fourier transforms (DFT). To obtain the required level of accuracy, we perform the DFTs on 32768 points, with a sampling frequency of 0.018125 Hz. At each time step, the exact solution is imposed at the boundaries of the computational domain.

Figure \ref{Test1} shows the solution at time steps $t_i=t_0+75\,i\,\Delta\,t$ ($i=0,1,2$). The slices are performed from
\begin{itemize}
\item $i=0$: ($x=70$ m, $y=26$ m) to ($x=360$ m, $y=26$ m),
\item $i=1$: ($x=156$ m, $y=162$ m) to ($x=207$ m, $y=162$ m),
\item $i=2$: ($x=246$ m, $y=198$ m) to ($x=282$ m, $y=198$ m).
\end{itemize}
Looking from the left to the right on the first slice ($i=0$), one can successively observe the reflected SV-wave, the transmitted SV-wave and the transmitted P-wave. As mentioned in section \ref{SEC_sM}, these waves do not have the same sinusoidal profile as the incident P-wave. A "coda" occurs after the reflected and transmitted P and SV-waves: as in \cite{ALIMENTAIRE1}, these fields are not spatially bounded in the direction of the propagation.

The second slice ($i=1$) crosses the reflected SV-wave. The agreement between exact and numerical values is good; one only observes a small amount of numerical diffusion induced by WPALG. This is also so in the case of the last slice ($i=2$) across the transmitted SV-wave. 

In table 5-1, we show convergence measurements performed with the Lax-Wendroff scheme and with WPALG, by taking $k=3$. This table was obtained by taking successively refined meshes with a constant CFL and measuring the differences between the analytical and numerical values. The orders of accuracy (2 for Lax-Wendroff, 1.5 in norm L$_\infty$ and 2 in norm L$_1$ for WPALG) are the same as in homogeneous medium. Similar results can be obtained by increasing $k$. However, for $k=2$, the convergence orders fall below 1.4. (we recall that $k=2$ sufficies to ensure second-order accuracy for perfect contacts in 2-D contexts \cite{BIBLE2}, and for imperfect contacts in 1-D contexts \cite{ALIMENTAIRE1}). 

\begin{figure}[htbp]
\begin{center}
\begin{tabular}{cc}
$(i=0)$ & $(i=0)$\\
&\\
\includegraphics[width=5.2cm,height=5.2cm]{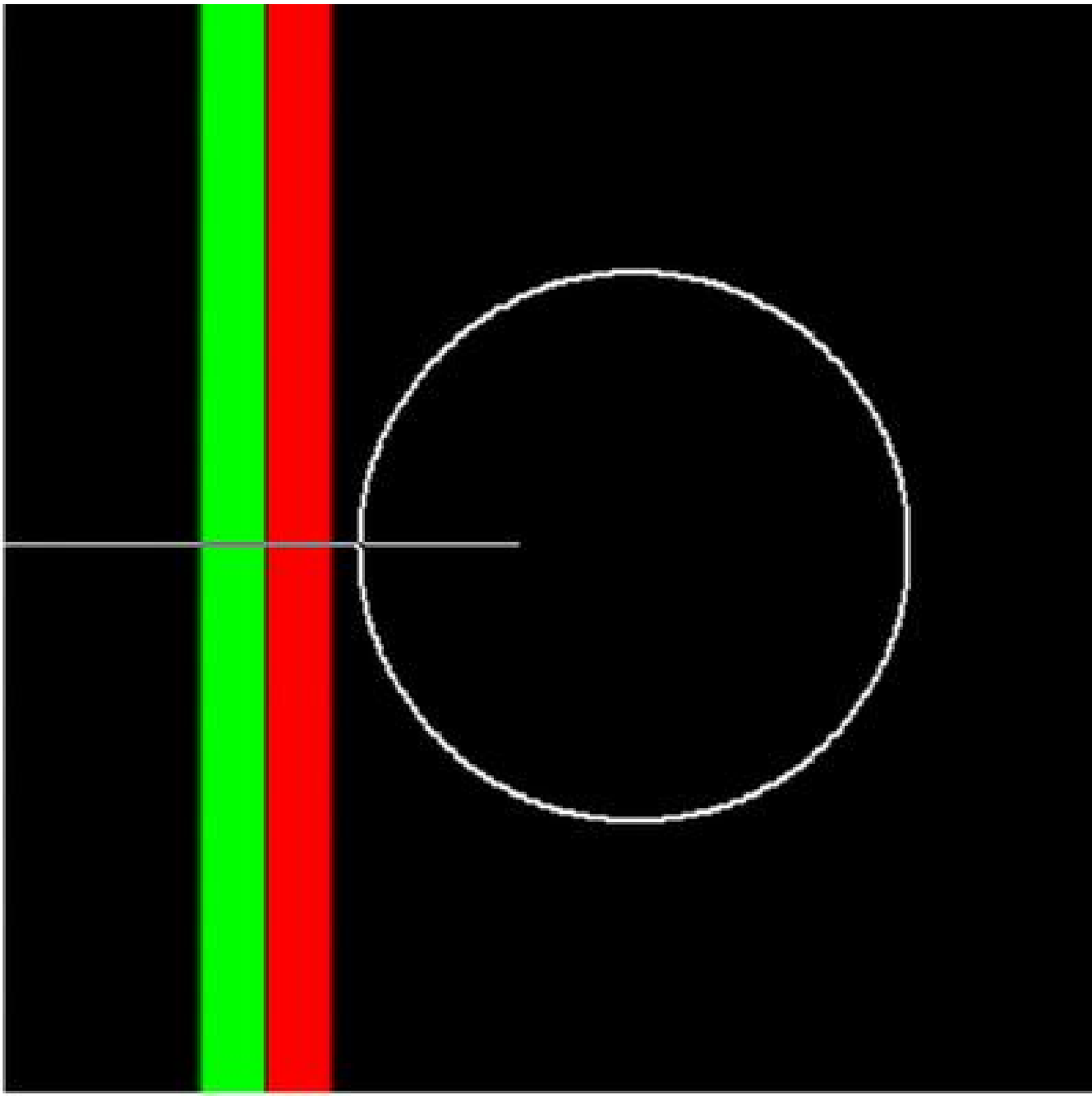}& 
\includegraphics[width=5.2cm,height=5.2cm]{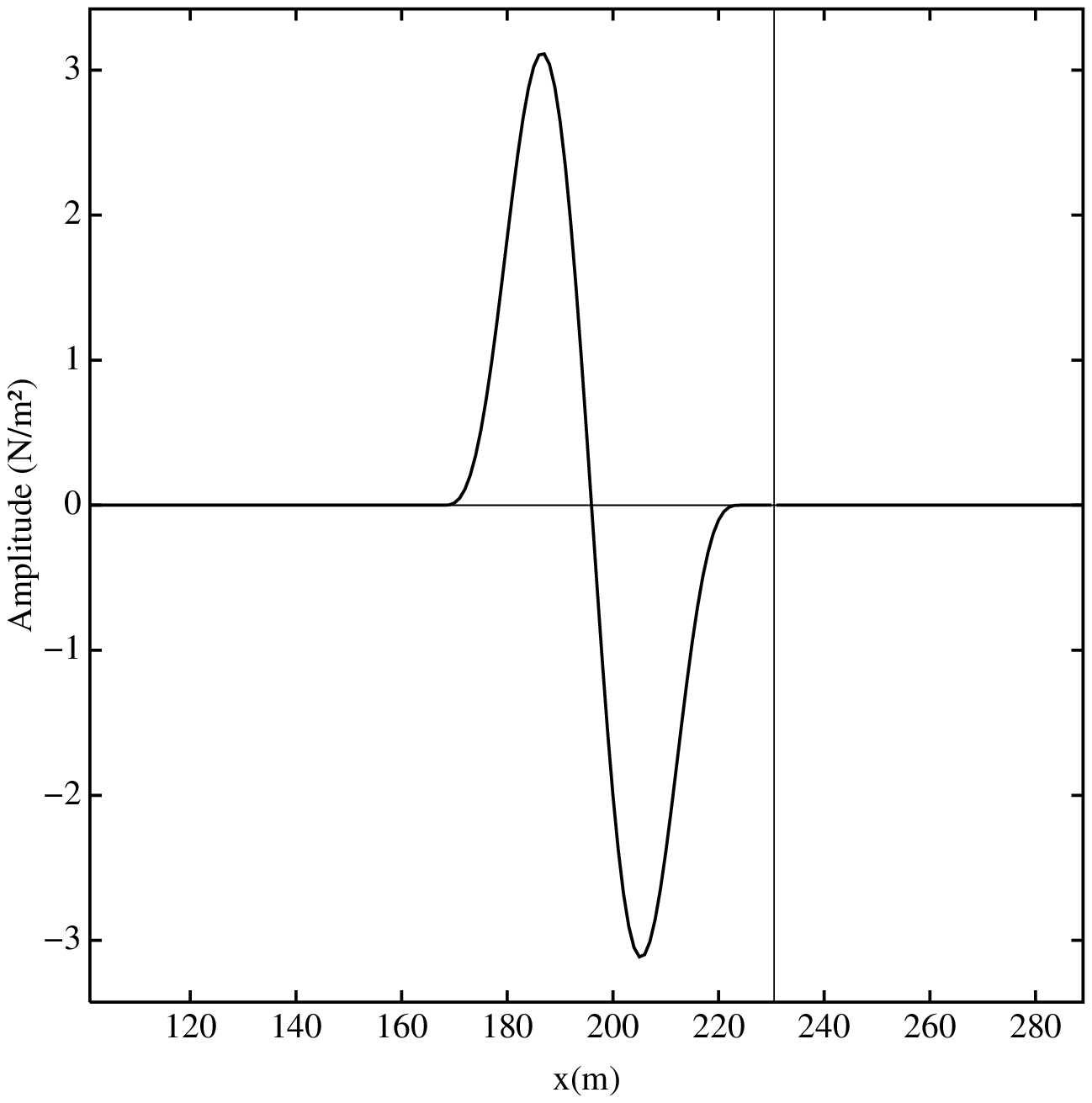}\\
&\\
$(i=1)$ & $(i=1)$\\
&\\
\includegraphics[width=5.2cm,height=5.2cm]{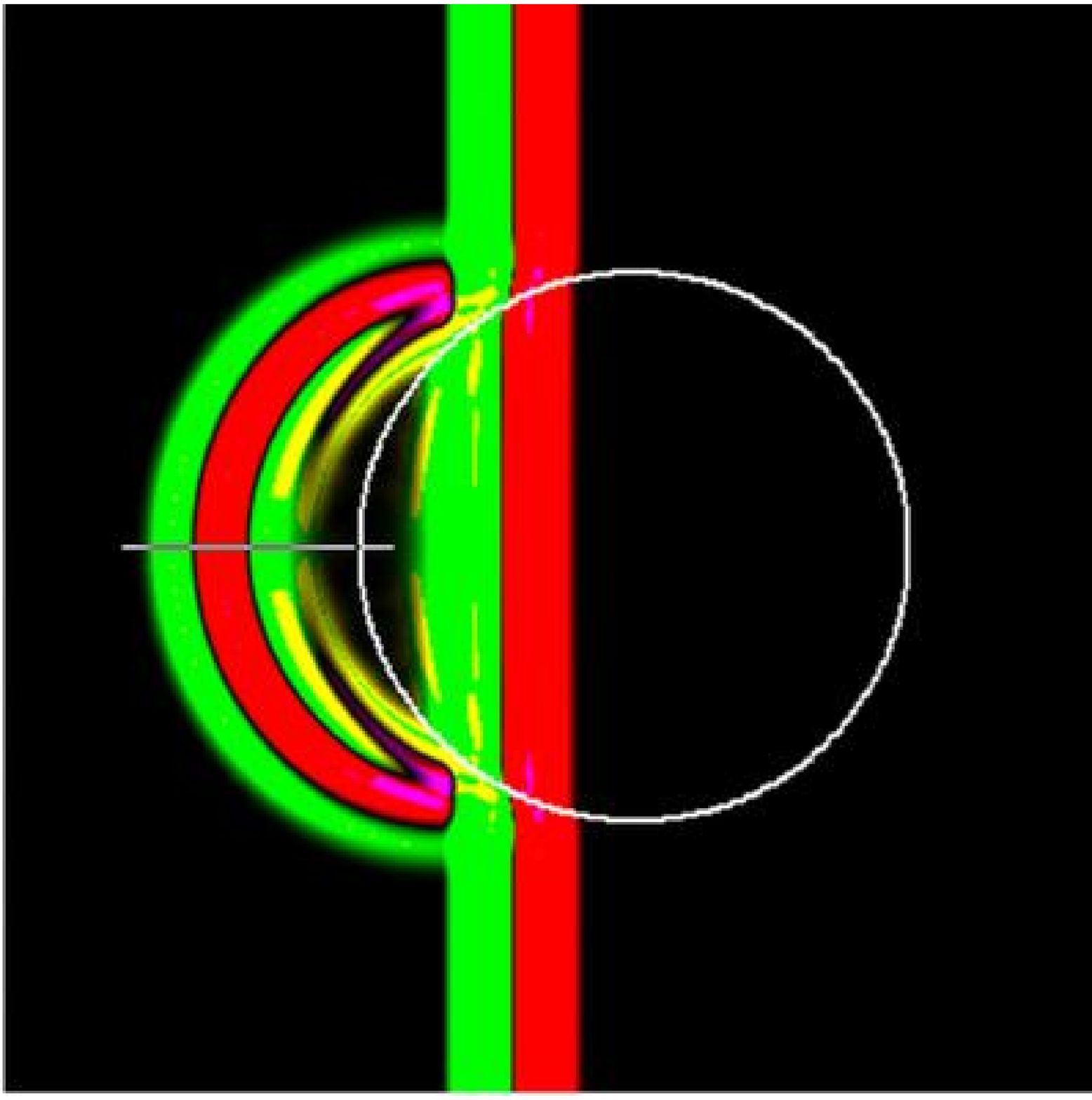}& 
\includegraphics[width=5.2cm,height=5.2cm]{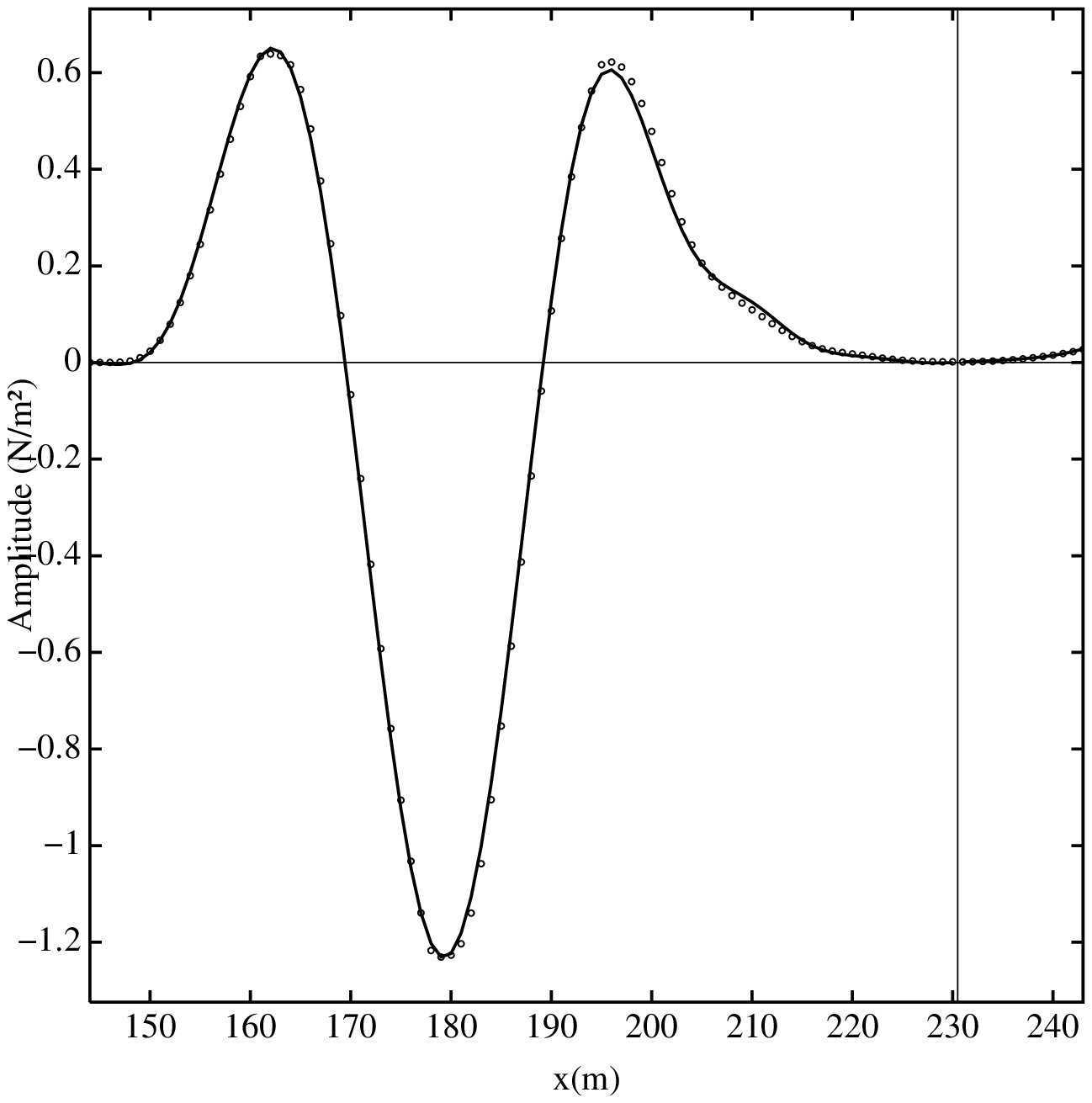}\\
&\\
$(i=2)$ & $(i=2)$\\
&\\
\includegraphics[width=5.2cm,height=5.2cm]{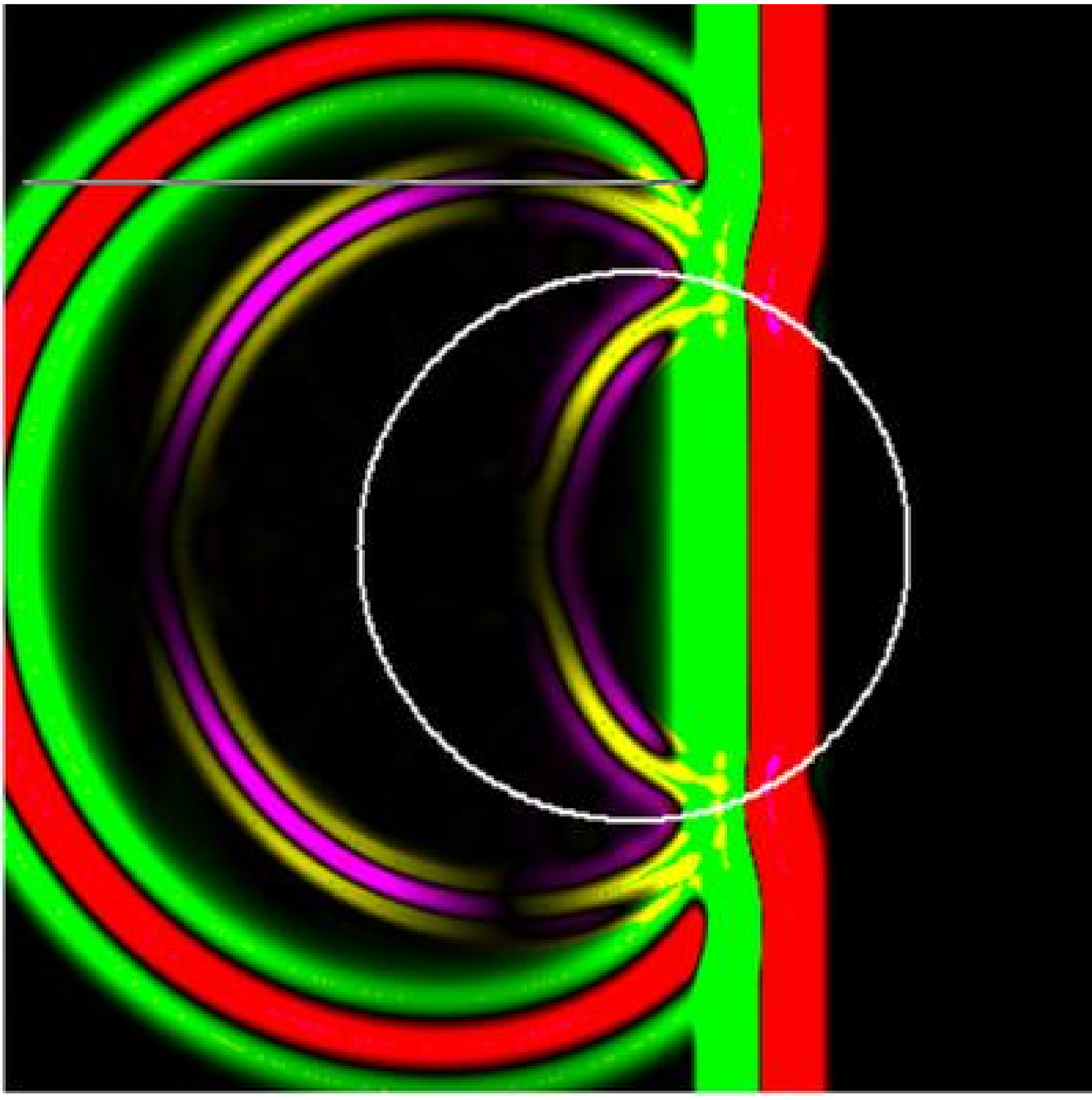}& 
\includegraphics[width=5.2cm,height=5.2cm]{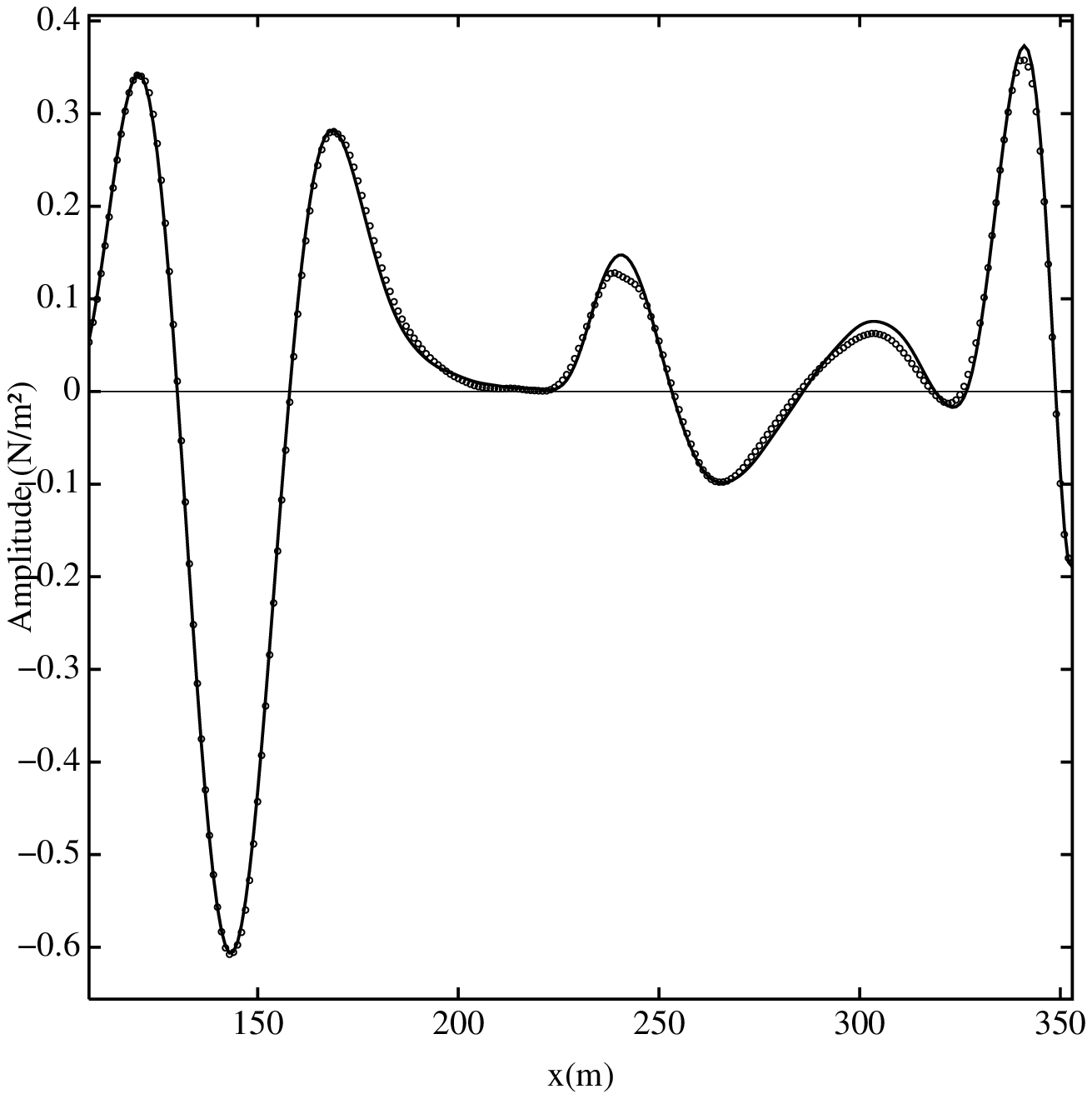}
\end{tabular}
\caption{Test 2-a: cylindar between identical media, $t=t_0+100\,i\,\Delta\,t$.}
\label{Test2_A}
\end{center}
\end{figure}

\begin{figure}[htbp]
\begin{center}
\begin{tabular}{cc}
$(i=3)$ & $(i=3)$\\
&\\
\includegraphics[width=5.2cm,height=5.2cm]{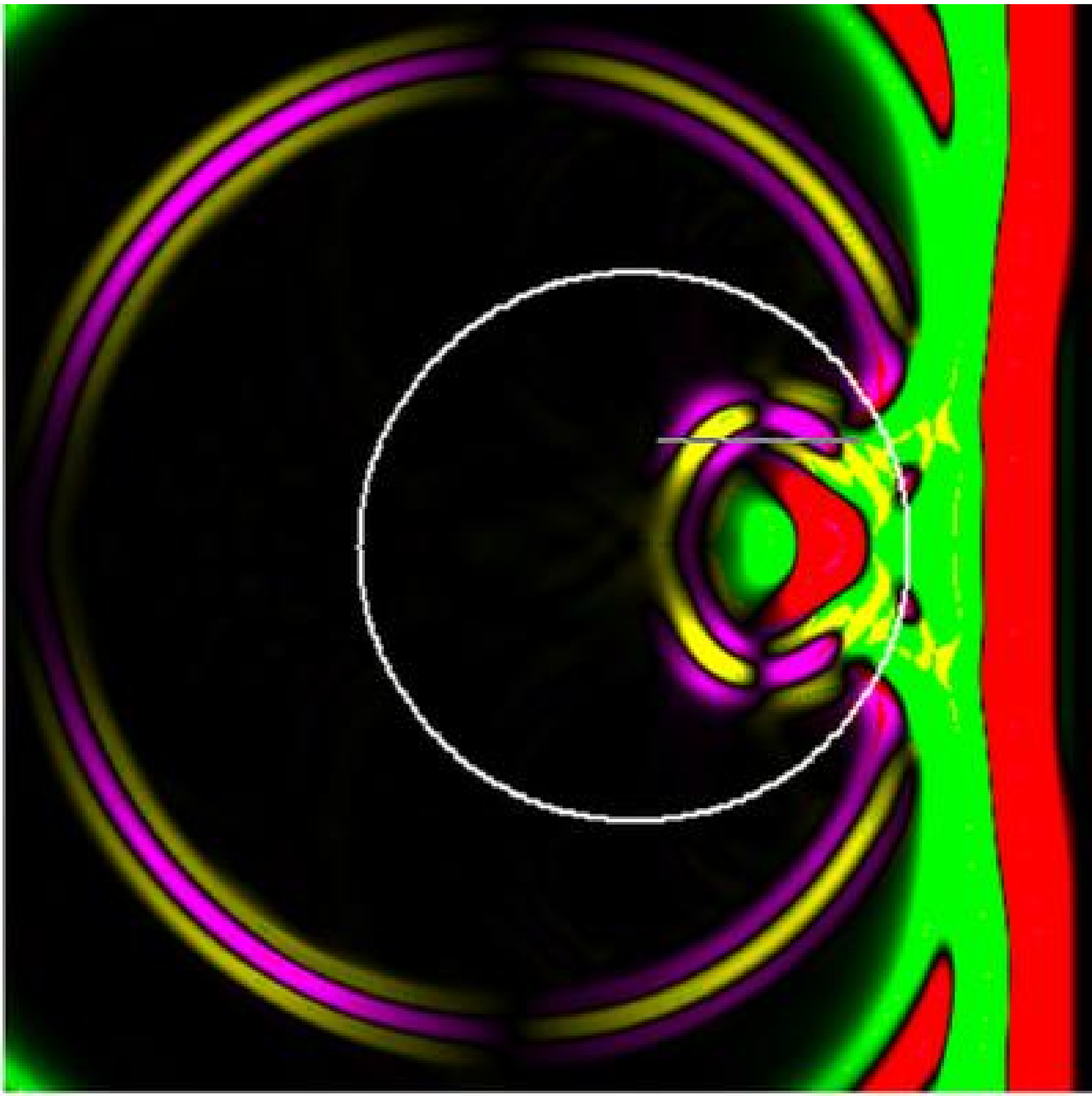}& 
\includegraphics[width=5.2cm,height=5.2cm]{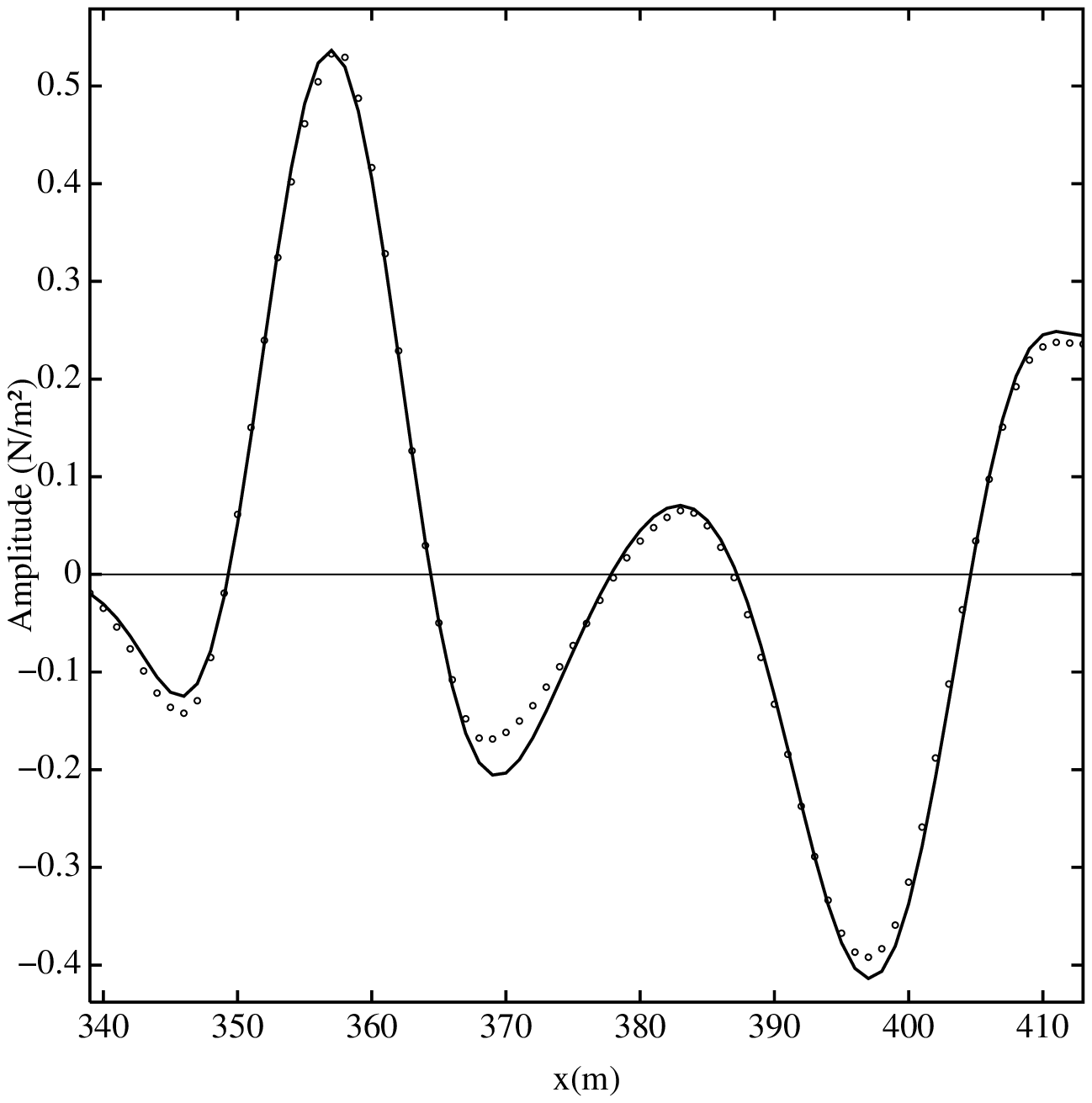}\\
&\\
$(i=4)$ & $(i=4)$\\
&\\
\includegraphics[width=5.2cm,height=5.2cm]{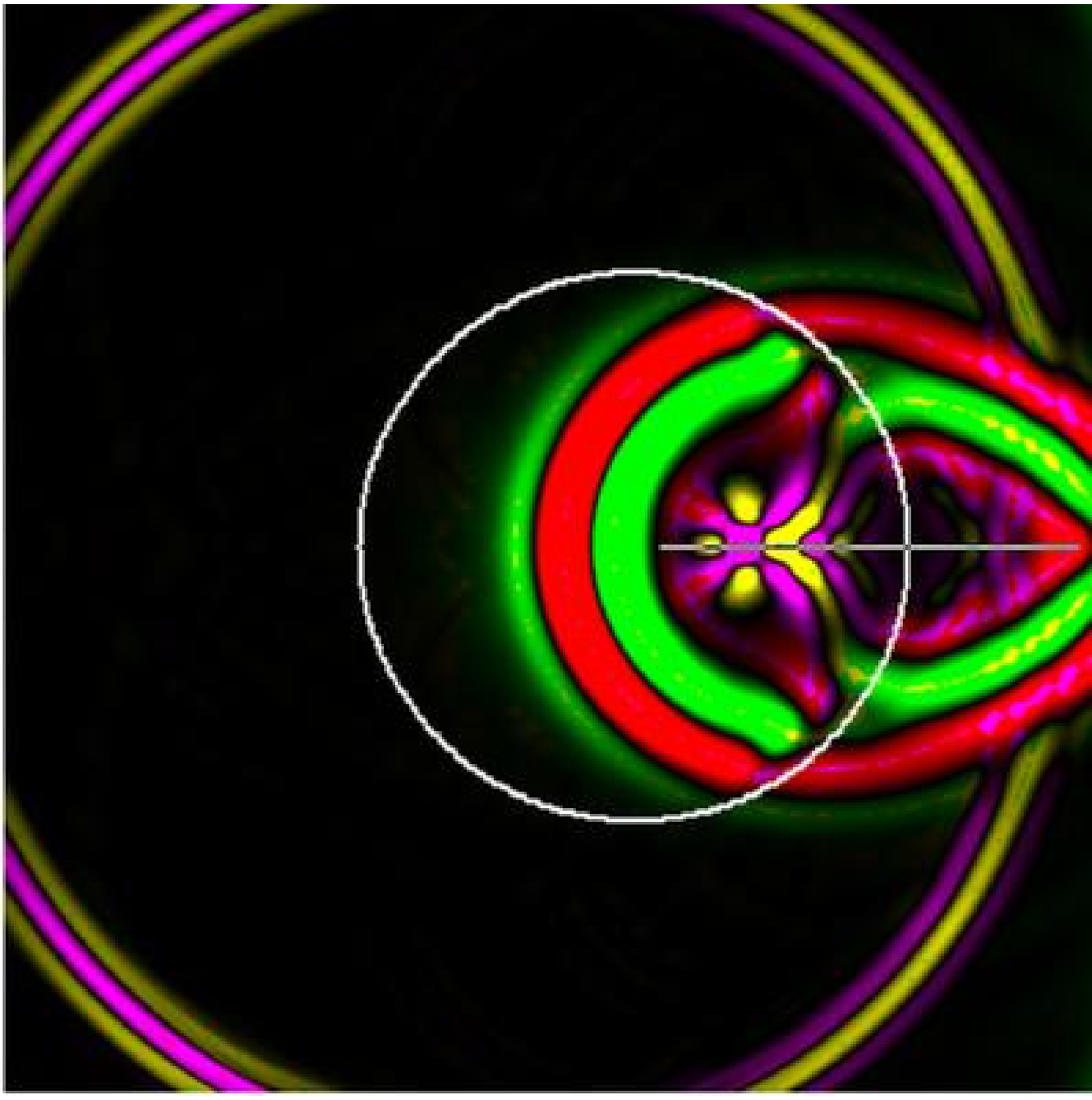}& 
\includegraphics[width=5.2cm,height=5.2cm]{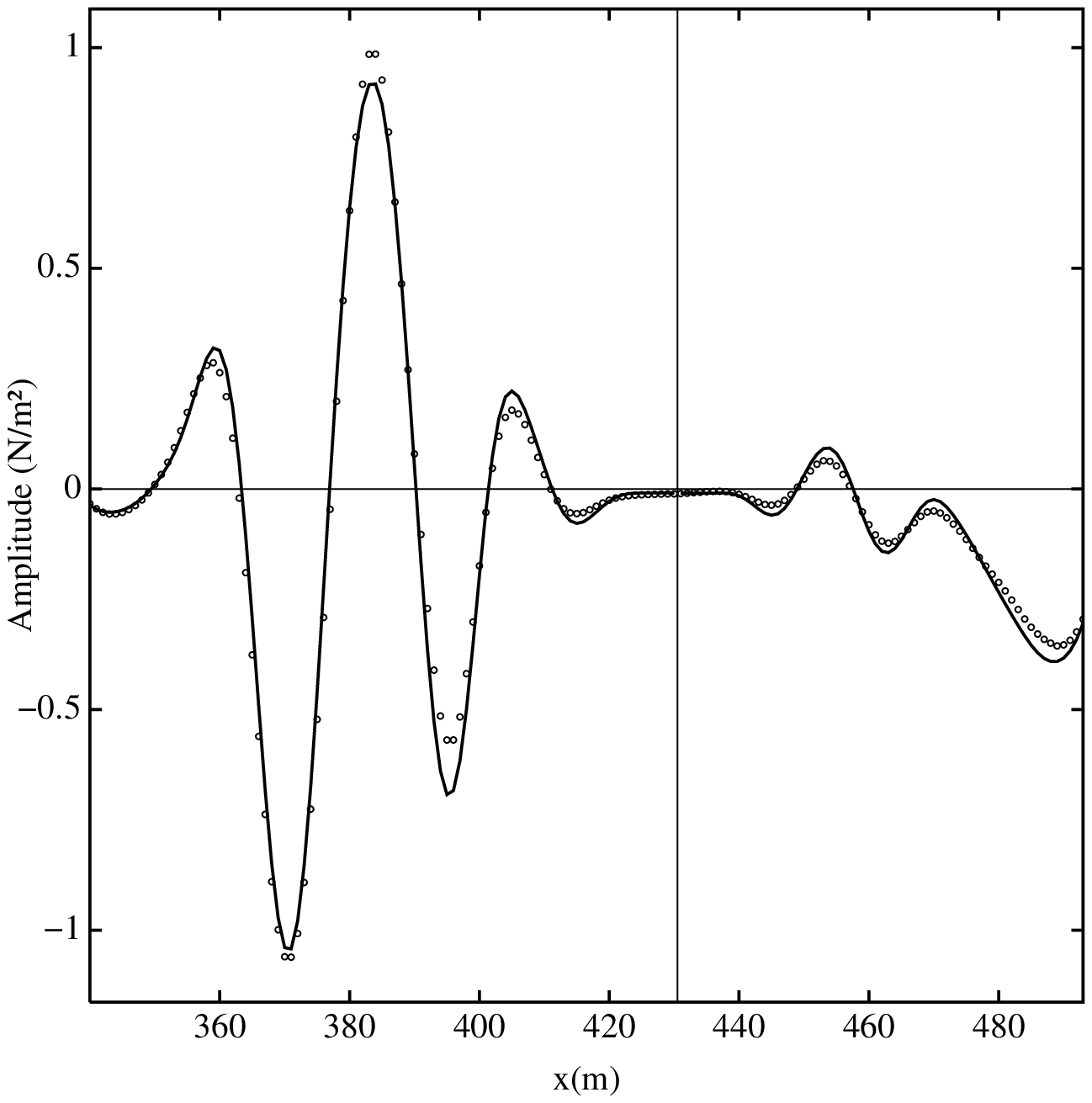}\\
&\\
$(i=5)$ & $(i=5)$\\
&\\
\includegraphics[width=5.2cm,height=5.2cm]{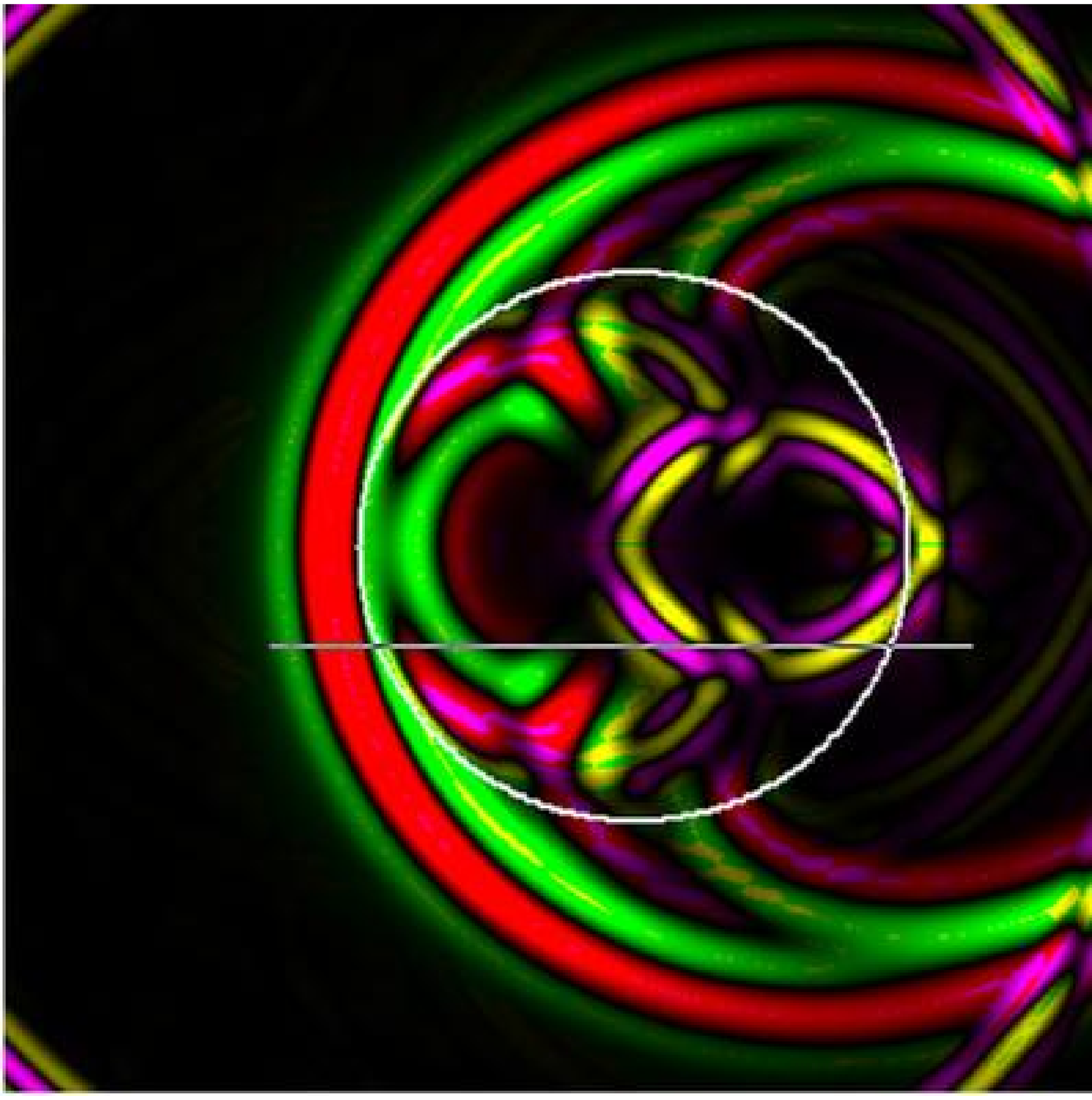}& 
\includegraphics[width=5.2cm,height=5.2cm]{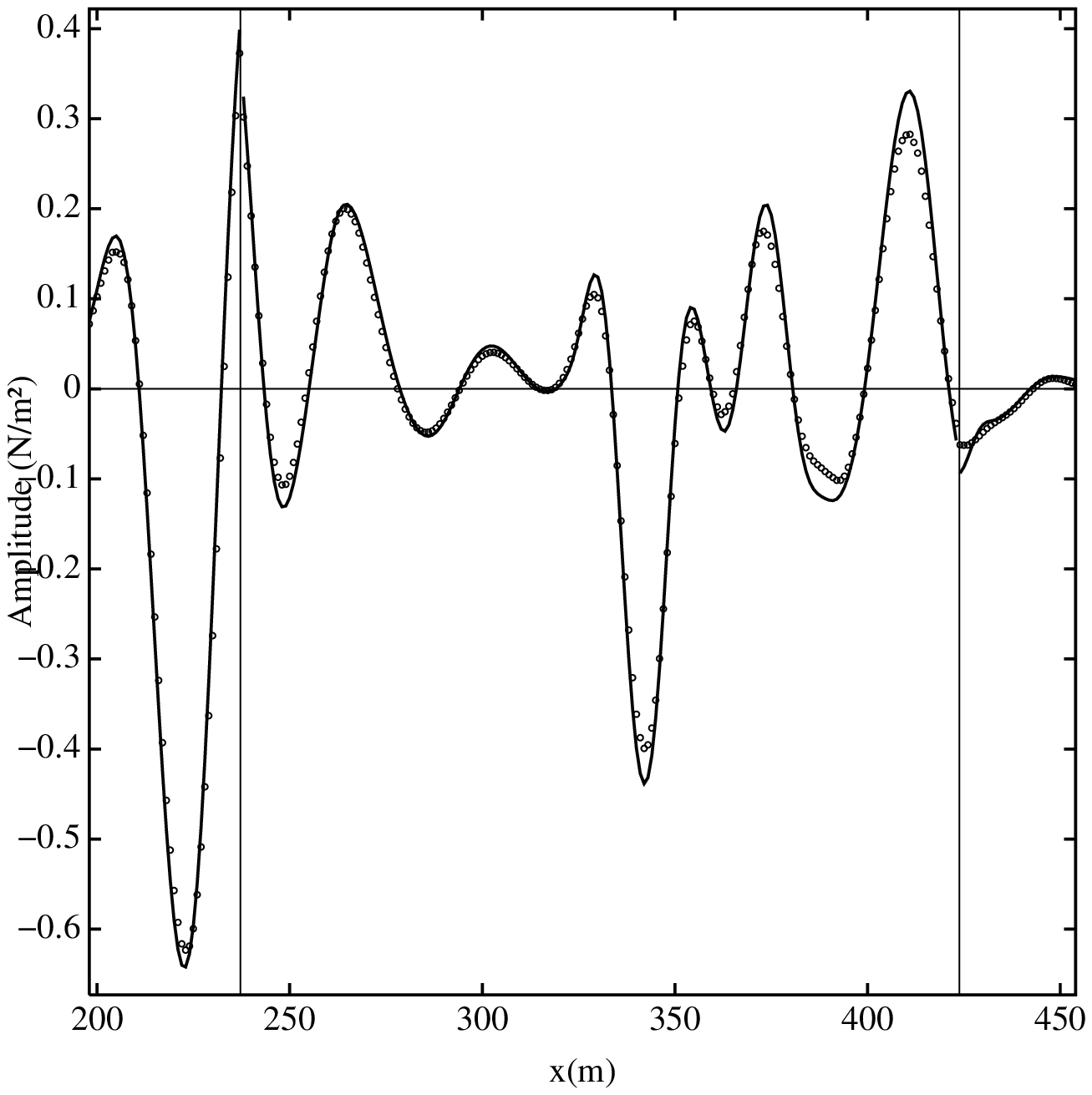}
\end{tabular}
\caption{Test 2-b: cylindar between identical media at $t=t_0+100\,i\,\Delta\,t$.}
\label{Test2_B}
\end{center}
\end{figure}

\subsection{Test 2: circular interface between identical media}\label{SEC_NUM2}

As a second example, we take the case of a $L_x \times L_y=600 \times 600$ m$^2$ domain and a circular interface (radius $a=100$ m, centred at $x_0=330.5$ m, $y_0=300$ m). The numerical experiments are performed on $N_x \times N_y=600 \times 600$ grid points. The physical parameters are the same as in the previous example. The spring-mass conditions are
$$
\left\{
\begin{array}{l}
K_N=K_T=10^{9}\,\mbox{ kg/s}^{2},\\
\\
M_N=M_T=1000\,\mbox{ kg/m}^{2}.
\end{array}
\right.
$$
Figures \ref{Test2_A} and \ref{Test2_B} show the solution at the initial instant $t_0=0.08$ s, and at $t_i=t_0+100\,i\,\Delta\,t$ ($i=1,...,5$) on a restricted domain $[100,500] \times [100, 500]$ m$^2$ centred on the middle of the computational domain. No special treatment is applied to simulate the wave propagation in infinite medium (such as absorbing boundary conditions or perfectly matched layers), but the integration times are sufficiently short to prevent spurious waves reflected by the edges of the computational region from appearing in the restricted domain (the same comment holds for the further figures). The slices are carried out from
\begin{itemize}
\item $i=0$: ($x=101$ m, $y=301$ m) to ($x=289$ m, $y=301$ m), 
\item $i=1$: ($x=144$ m, $y=300$ m) to ($x=243$ m, $y=300$ m), 
\item $i=2$: ($x=108$ m, $y=433$ m) to ($x=353$ m, $y=433$ m), 
\item $i=3$: ($x=339$ m, $y=339$ m) to ($x=413$ m, $y=339$ m), 
\item $i=4$: ($x=340$ m, $y=300$ m) to ($x=493$ m, $y=300$ m), 
\item $i=5$: ($x=198$ m, $y=264$ m) to ($x=454$ m, $y=264$ m). 
\end{itemize}
The analytical solutions are computed on $N_{Bessel}=90$ points (see (\ref{QX=Y}) in Appendix C), 256 Fourier points, with 1-Hz sampling frequency. 

The incident P-wave is converted into transmitted and reflected P- and SV-waves ($i=1,2$). A series of refraction-conversion events is then observed ($i=3,4,5$). The agreement between the numerical and analytical values is excellent.

\subsection{Test 3: circular interface between different media}\label{SEC_NUM3}

\begin{figure}[htbp]
\begin{center}
\begin{tabular}{cc}
$(i=0)$ & $(i=0)$\\
&\\
\includegraphics[width=5.2cm,height=5.2cm]{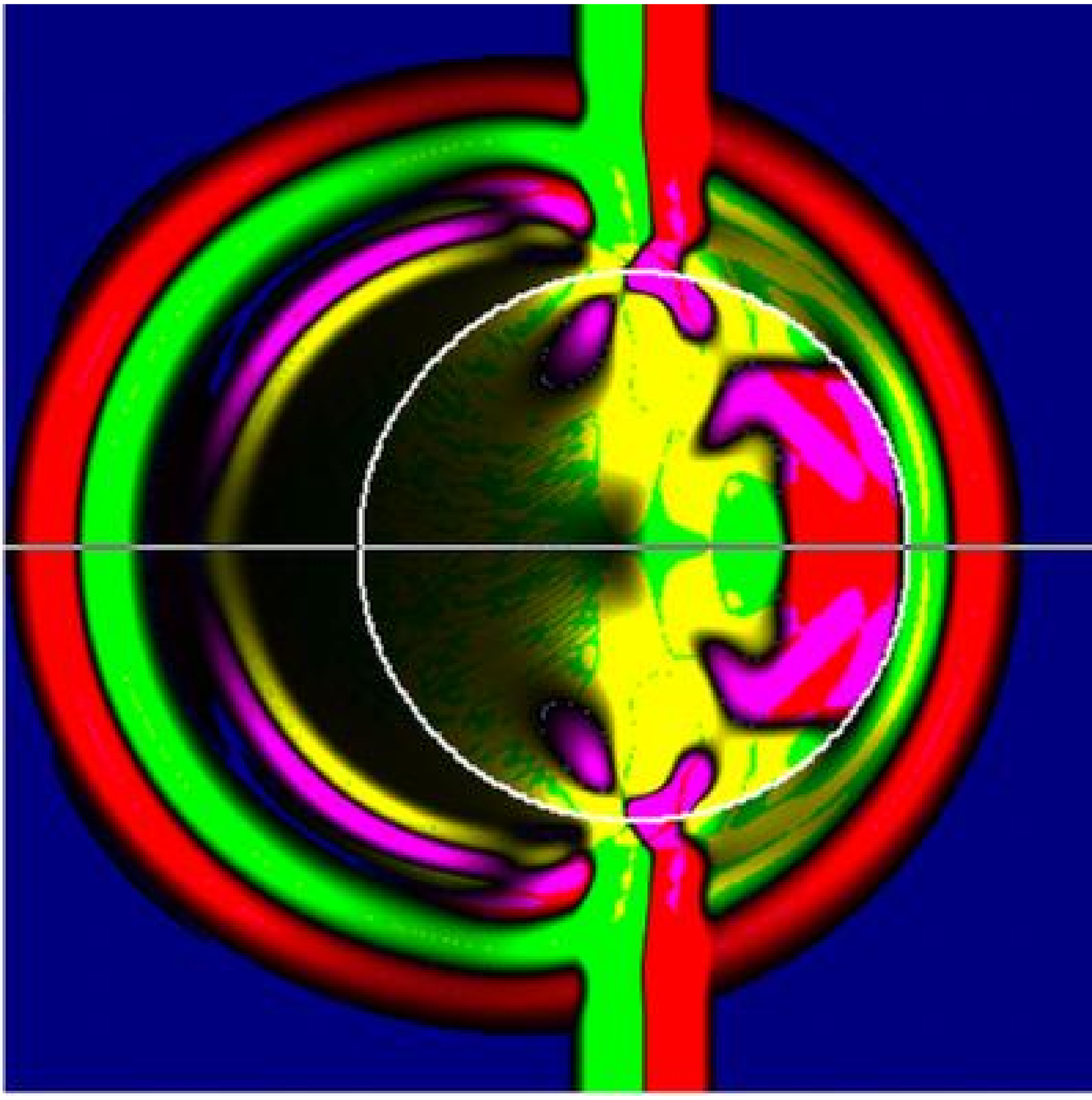}& 
\includegraphics[width=5.2cm,height=5.2cm]{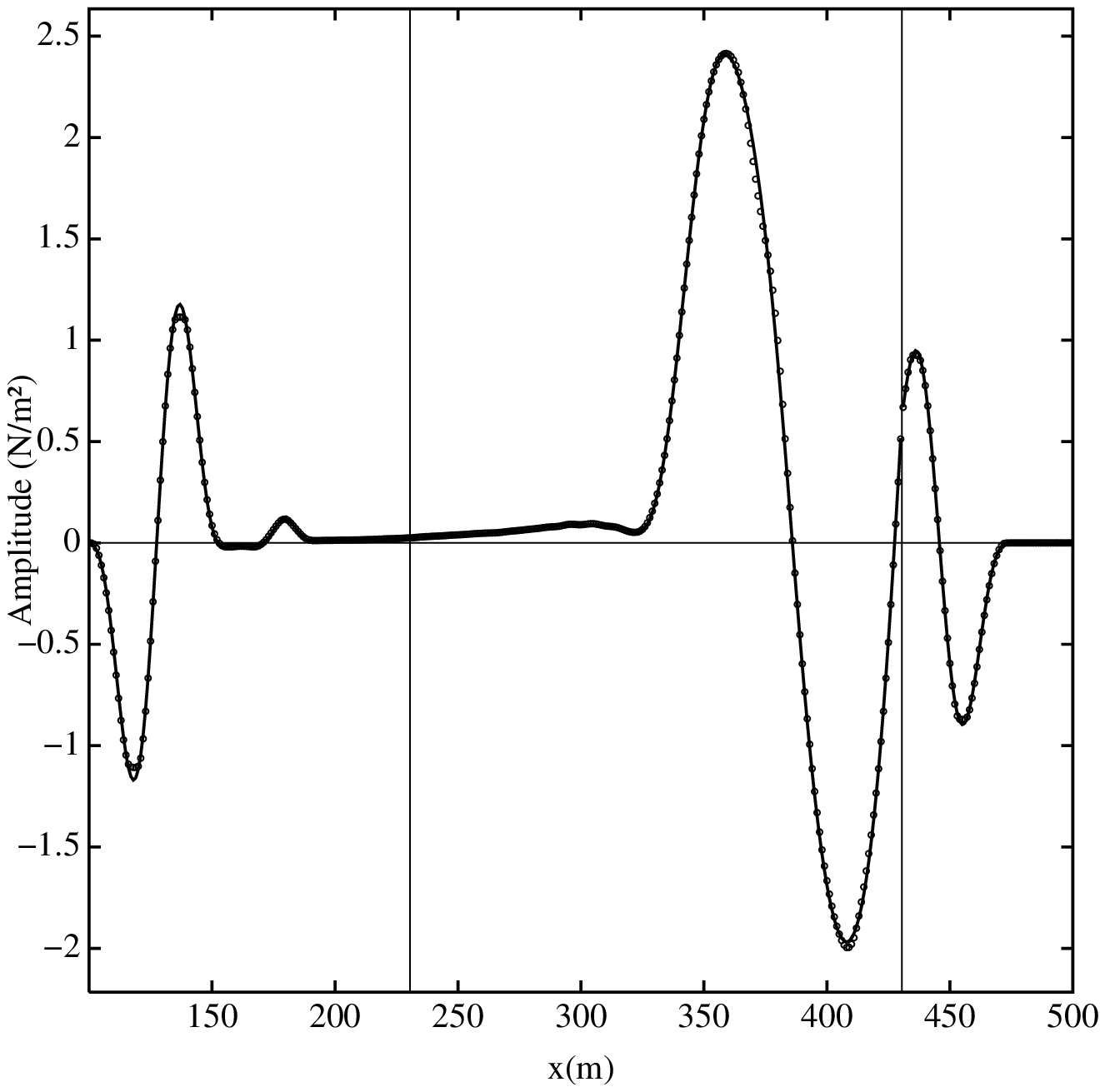}\\
&\\
$(i=1)$ & $(i=1)$\\
&\\
\includegraphics[width=5.2cm,height=5.2cm]{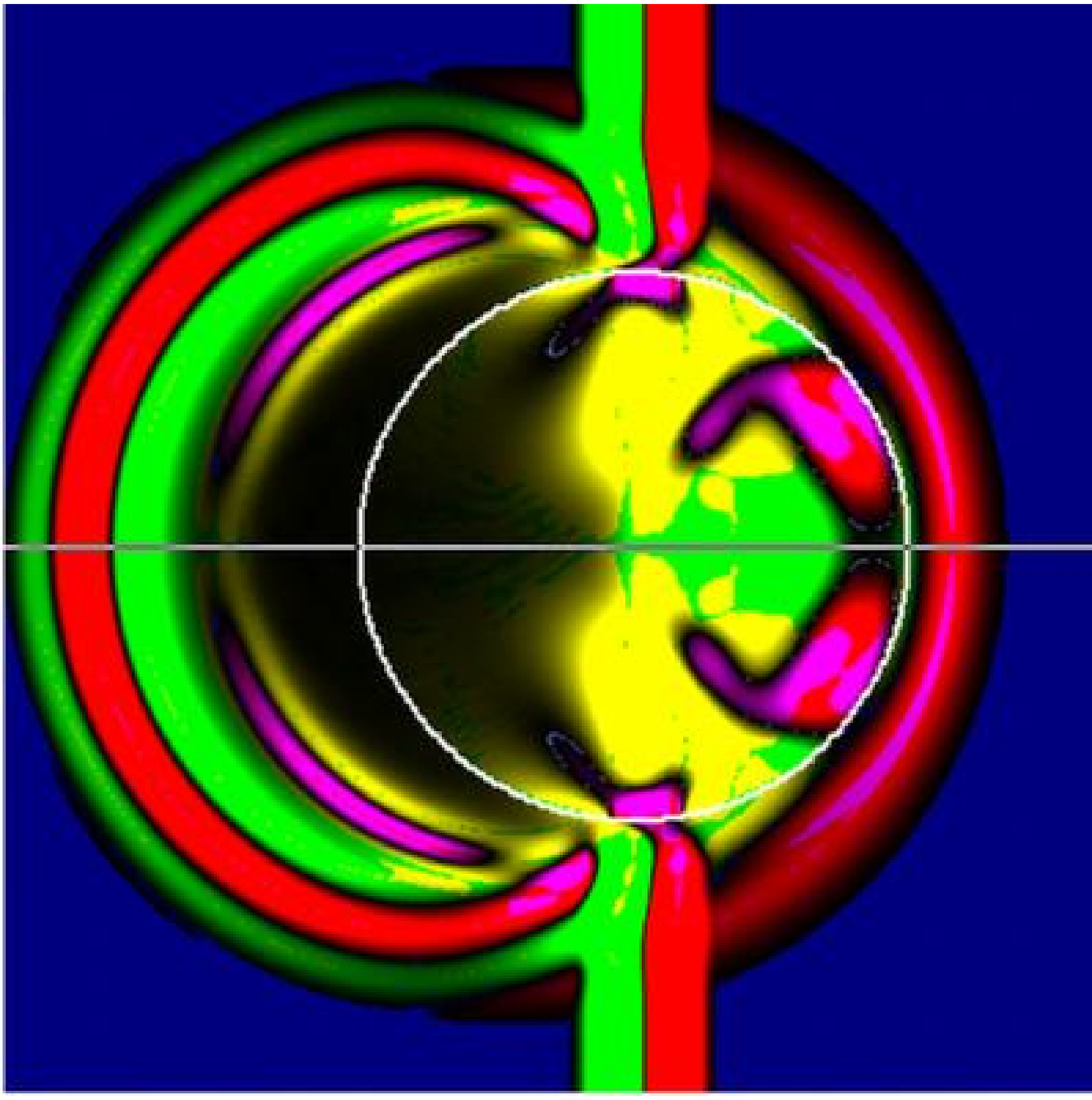}& 
\includegraphics[width=5.2cm,height=5.2cm]{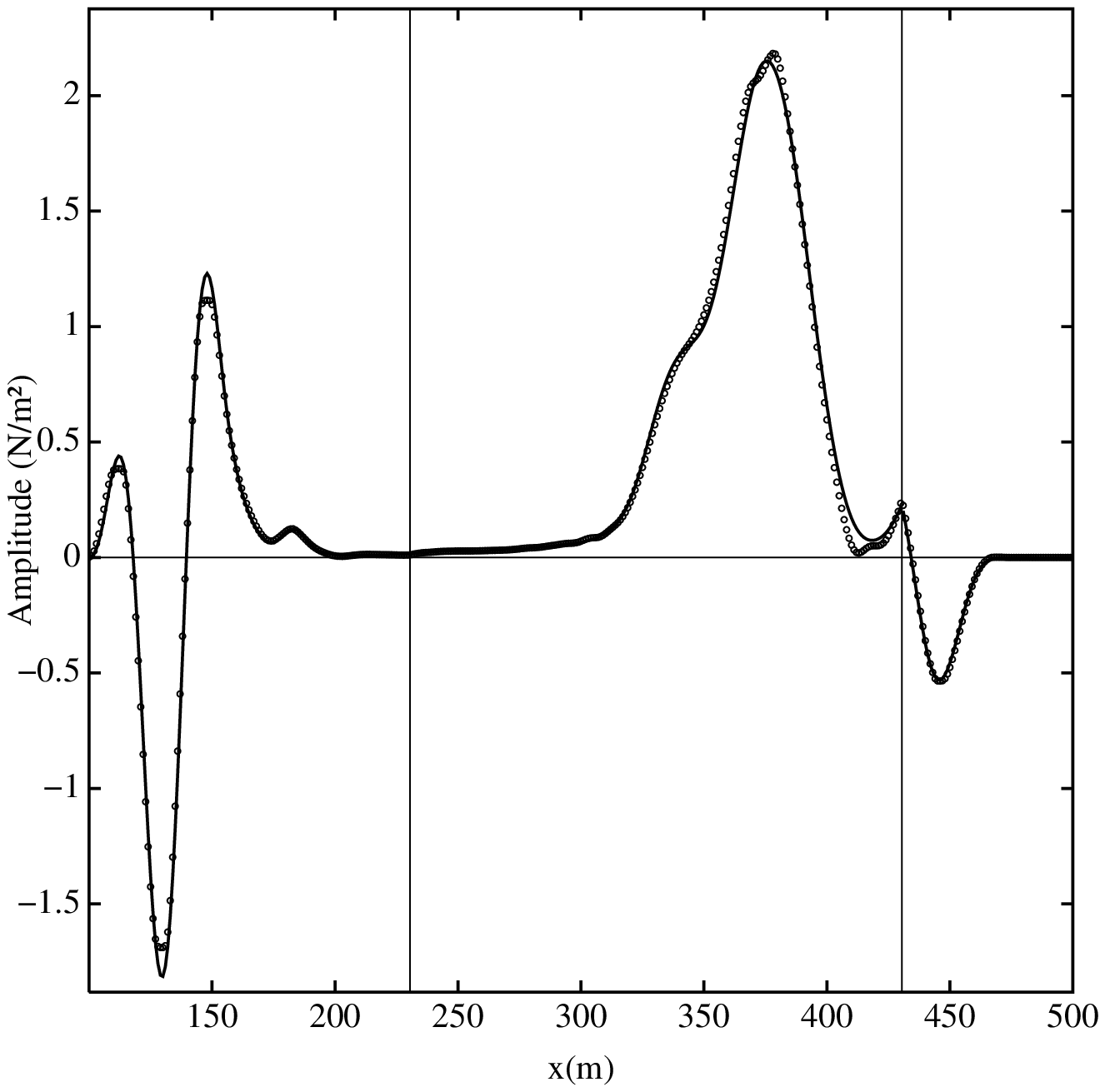}\\
&\\
$(i=2)$ & $(i=2)$\\
&\\
\includegraphics[width=5.2cm,height=5.2cm]{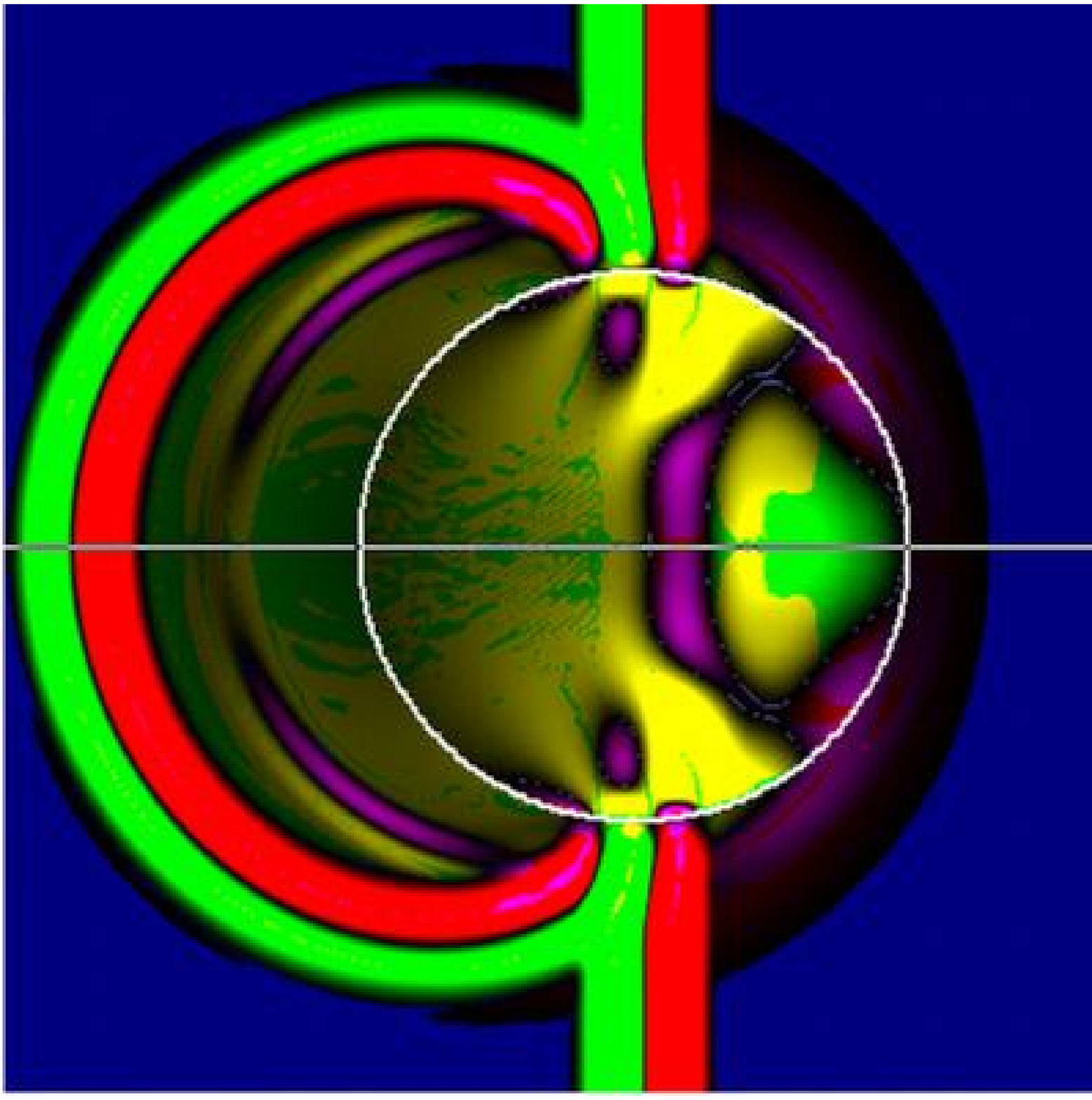}& 
\includegraphics[width=5.2cm,height=5.2cm]{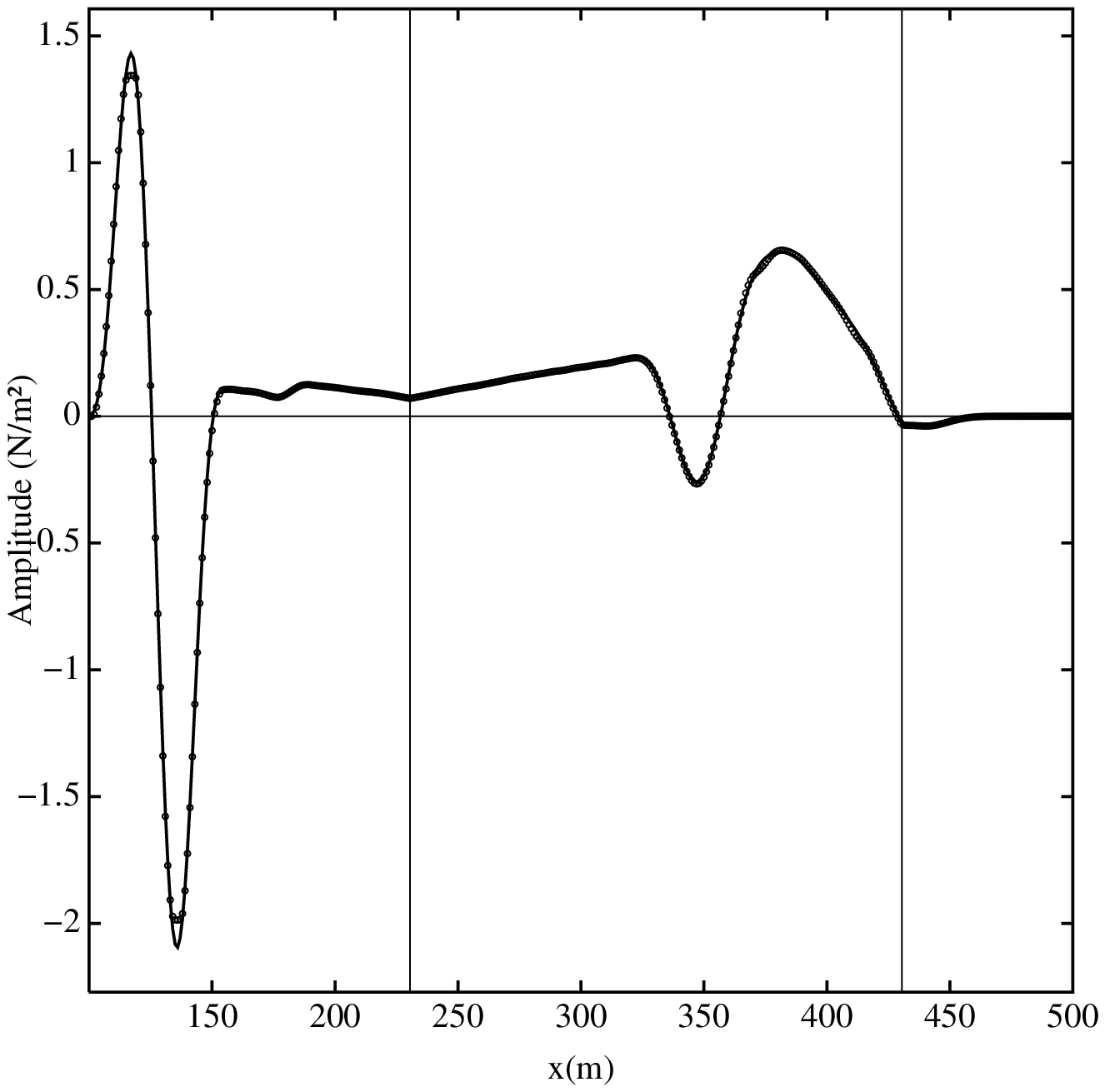}
\end{tabular}
\caption{Test 3: cylindar between different media at $t=t_0+360\,\Delta\,t$, with various values of normal stifness.}
\label{Test3}
\end{center}
\end{figure}

As a third example, we consider the same circular interface as in the previous test, but take it to have different physical parameters on both each sides of $\Gamma$
$$
(\rho,\,c_p,\,c_s)=
\left\{
\begin{array}{l}
\rho_0=2600\mbox{ kg/m}^3,\,c_{p0}=6400\mbox{ m/s},\,c_{s0}=3200\mbox{ m/s},\\
\\
\rho_1=1200\mbox{ kg/m}^3,\,c_{p1}=2800\mbox{ m/s},\,c_{s1}=1400\mbox{ m/s}.
\end{array}
\right.
$$
These parameters are those of aluminium (inside $\Gamma$) and Plexiglass (outside $\Gamma$). Apart from these physical parameters, the values used for the computations are the same as in section \ref{SEC_NUM2}. The initial values are the same as in figure \ref{Test2_A}, $i=0$.

Figure \ref{Test3} gives the results of a parametric study in terms of the normal stiffness $K_N$: perfect contact ($i=0$) which amounts to $K_N \rightarrow + \infty$, $K_N=10^9\,\mbox{ kg/s}^{2}$ ($i=1$), and $K_N=10^8\,\mbox{ kg/s}^{2}$ ($i=2$). The inertial effects and the tangential stiffness are not taken into account here
$$
\left\{
\begin{array}{l}
K_T\rightarrow +\infty,\\
\\
M_N=M_T=0\,\mbox{ kg/m}^{2}.
\end{array}
\right.
$$
The computations are shown in this figure after 350 time steps. The slices are performed from $x=100$ m to $x=500$ m, at $y=300$ m. Compared with the case of perfect contact ($i=0$), one can see in ($i=1,2$) that the signs of the reflected waves are reversed. The waves that have entirely crossed the interface also differ greatly between the three cases. The case ($i=2$) is much more difficult to handle than the cases ($i=0,1$) from the computational point of view. To obtain the good agreement shown in figure \ref{Test3} ($i=2$), it is necessary to use $k=4$ (for the numerical solution), and 32768 Fourier points with a sampling frequency of 0.018125 Hz (for the analytical solution).

Smaller values of $K_N$ have also been investigated (figures not shown here). Below $K_N=10^7\,\mbox{ kg/s}^{2}$, numerical instabilities have been observed when $k=2$ (see subsection \ref{SEC_OPEN}), whereas the computations remain stable when $k \geq 3$. Instabilities increase below $K_N=10^5\,\mbox{ kg/s}^{2}$ if $k=3$. Stable computations and excellent agreement with the analytical solutions are obtained down to $K_N=10^4\,\mbox{ kg/s}^{2}$ if $k=4$.  

\subsection{Test 4: cubic spline between identical media}\label{SEC_NUM4}

Up to now, we have dealt only with canonical objects with constant curvature. However, the use of the interface method is not restricted to such simple geometries. In this section, we study a more complex configuration corresponding to a cubic spline. The other parameters are the same as in section \ref{SEC_NUM2}, except $t_0=0.07$ s.

\begin{figure}[htbp]
\begin{center}
\begin{tabular}{c}
$(i=0)$\\
\\
\includegraphics[scale=0.4]{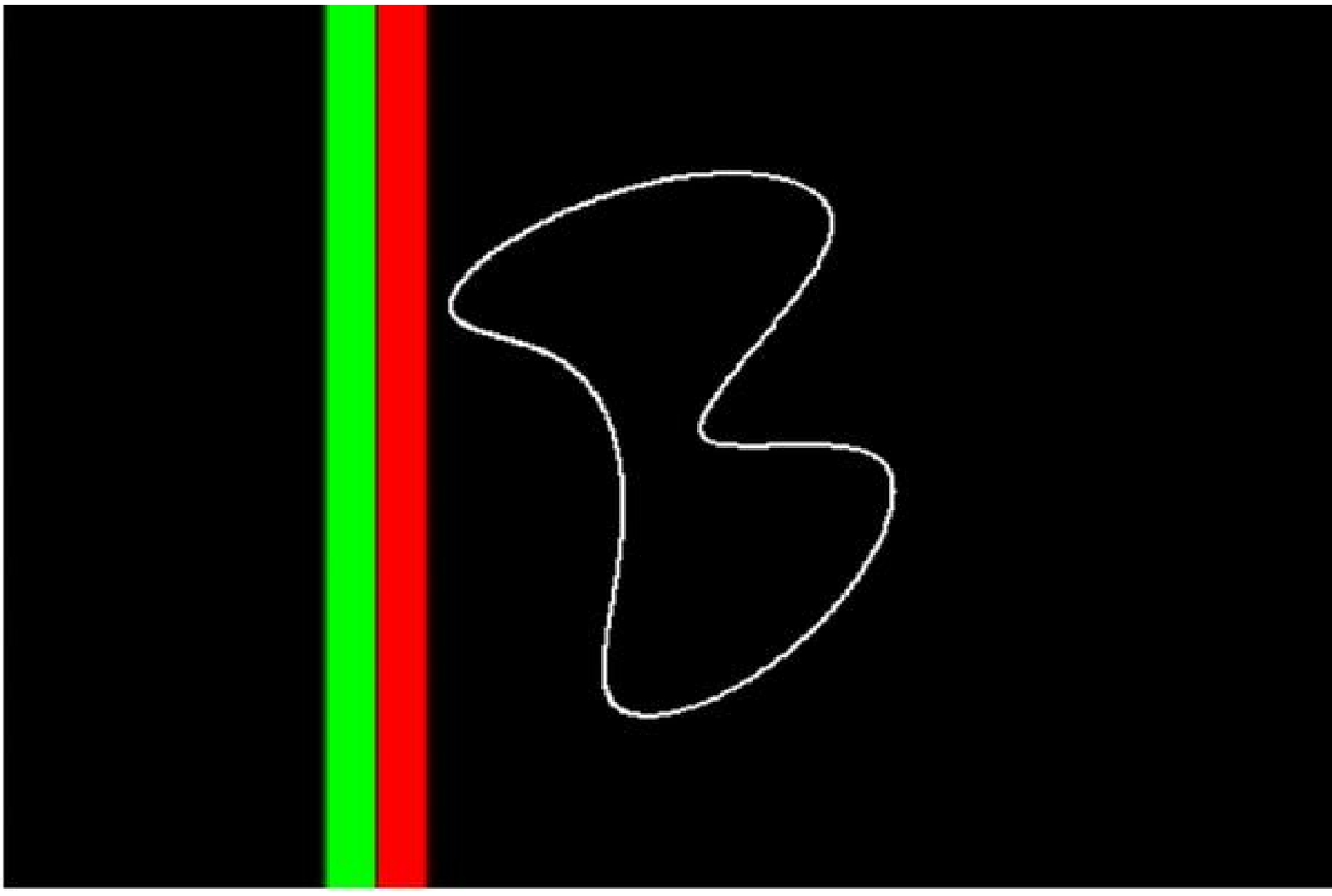}\\
\\
$(i=1)$\\
\\
\includegraphics[scale=0.4]{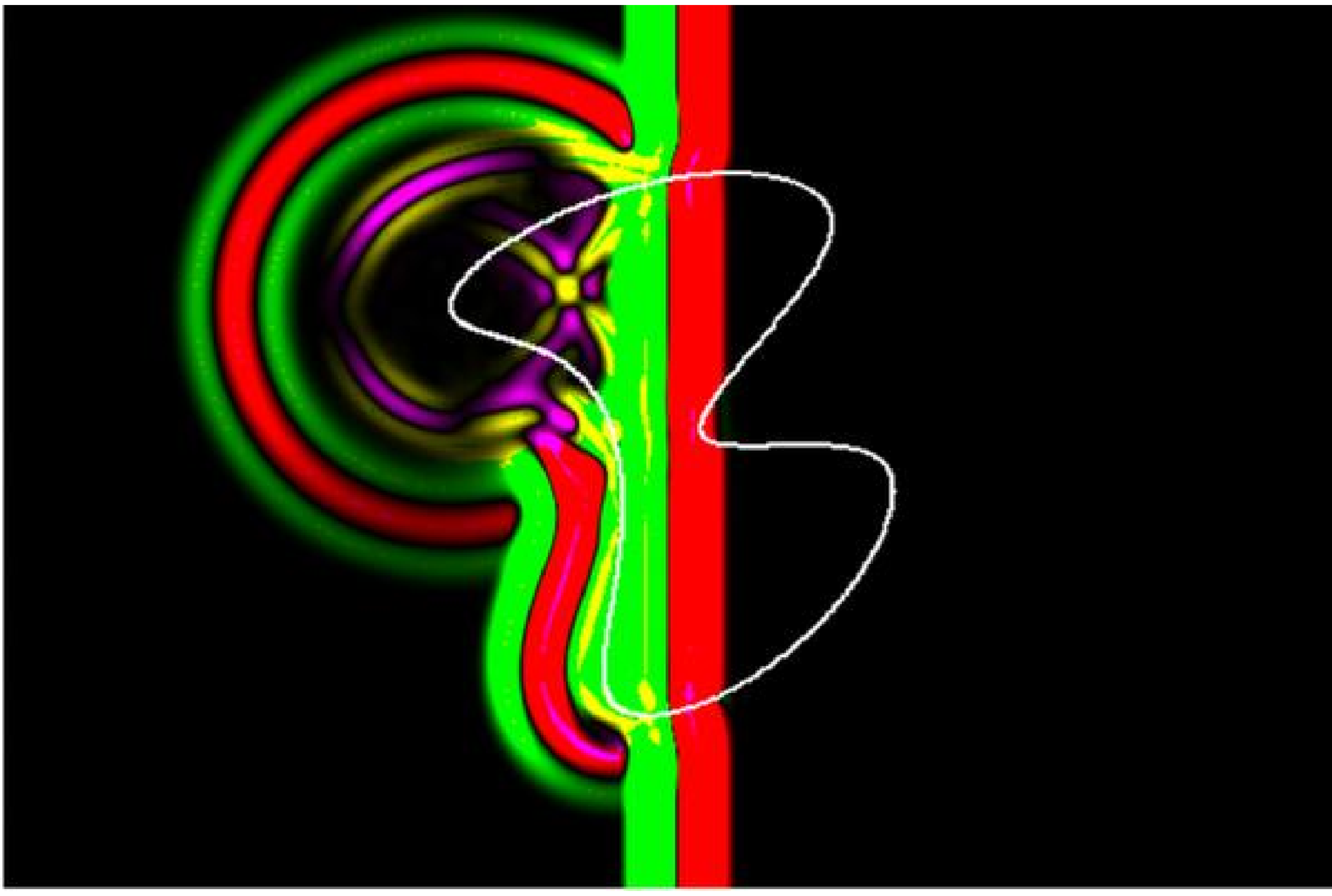}\\
\\
$(i=2)$\\
\\ 
\includegraphics[scale=0.4]{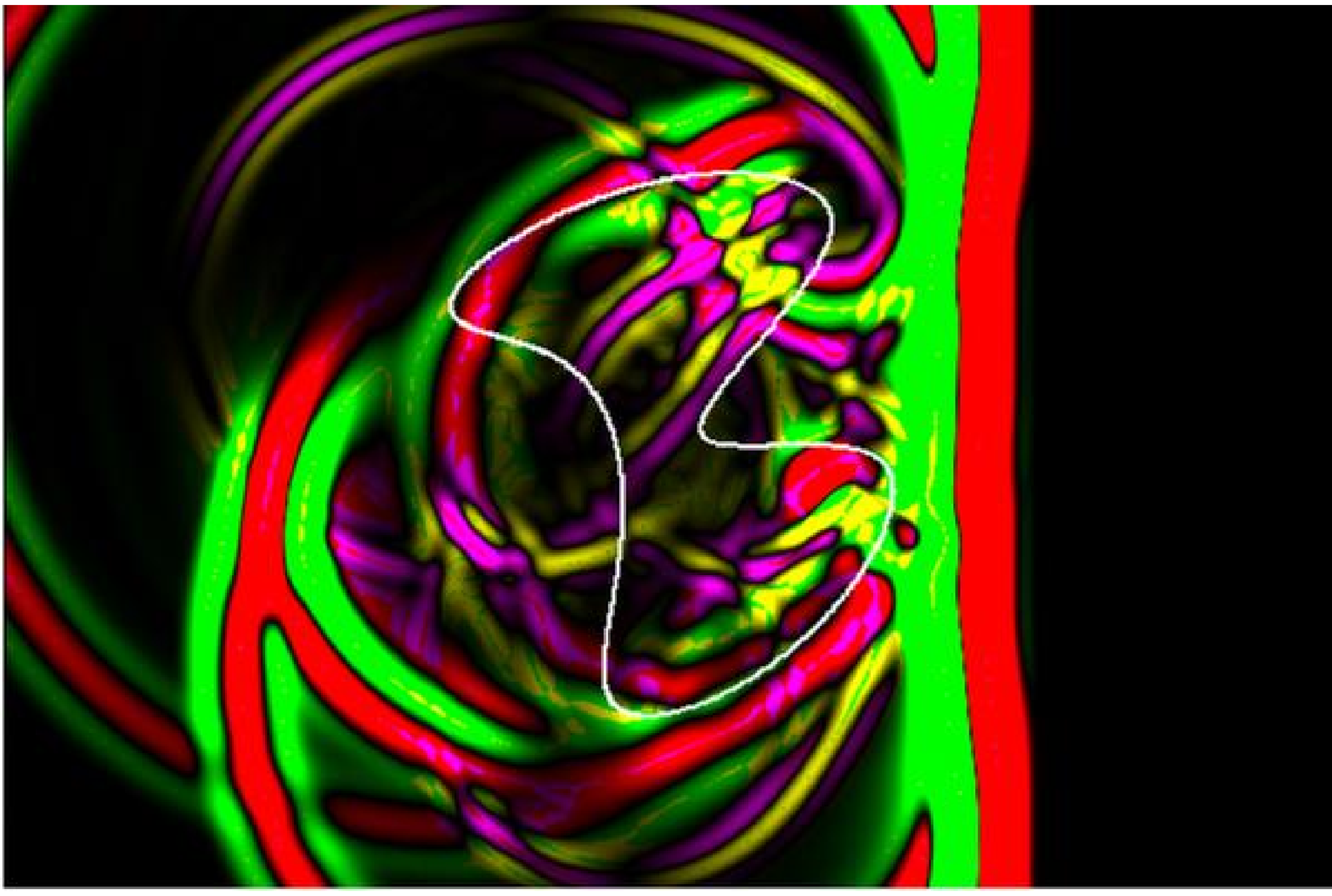}
\end{tabular}
\caption{Test 4: cubic spline between identical media, at $t=t_0+150\,i\,\Delta\,t$.}
\label{Test4}
\end{center}
\end{figure}

Figure \ref{Test4} shows snapshots of the solutions at times $t_i=t_0+150\,i\,\Delta \,t$ ($i=0,\,1,\,2$). We obviously have no analytical solutions to compare with the numerical values. However, the computations remain stable, and the wave phenomena seem to be realistic.

\subsection{Test 5: plane interface with variable jump conditions}\label{SEC_NUM5}

In the last experiment, we study the case of a plane interface with variable spring-mass conditions. The interface $\Gamma$ is placed horizontally at $y=130$ m on a $L_x \times L_y=600 \times 400$ m$^2$ domain. The physical parameters have the same values as in section \ref{SEC_NUM1}, and they are identical on both sides of $\Gamma$. The variable values of normal and tangential stiffness are
\begin{equation}
K_N=K_T=\left\{
\begin{array}{l}
10^{12}\,\mbox{ kg/s}^{2}, \,\mbox{ on } [x=0 \mbox{ m } ,\,x=270\mbox{ m} ],\\
\\
g(\tau) \,\mbox{ on } [x=270\mbox{ m } ,\,x=330\mbox{ m} ],\\
\\
10^{7}\,\mbox{ kg/s}^{2} , \,\mbox{ on } [x=330\mbox{ m } ,\,x=600\mbox{ m} ],
\end{array}
\right.
\label{Variable_KN}
\end{equation}
where $g(\tau)$ is a cubic polynomial that ensures a $C^1$ continuity ($\tau$ is the abscissa along $\Gamma$). The stiffness values (\ref{Variable_KN}) are those of almost perfectly-bonded media on the left part of the interface, and almost perfectly-disconnected media on the right part of the interface. The inertial effects are not taken into account: $M_N=M_T=0\,\mbox{ kg/m}^{2}$.

The computations are initialized by a transient explosive P-wave centred at ($x_s=300$ m, $y_s=200$ m), with an elastic potential \cite{MORSE}
\begin{equation}
\begin{array}{l}
\displaystyle
\Phi_{IP}(x,\,y,\,t)=\left({\tilde \Phi}_{IP}*f\right)(x,\,y,\,t),\\
\\
\displaystyle
{\tilde \Phi}_{IP}(r,\,t)=\frac{\textstyle 1}{\textstyle 2\,\pi}\,\frac{\textstyle H\left(t-\frac{\textstyle r}{\textstyle c_{p1}}\right)}{\textstyle \sqrt{t^2-\left(\frac{\textstyle r}{\textstyle c_{p1}}\right)^2}},\qquad r=\sqrt{\left(x-x_s\right)^2+\left(y-y_s\right)^2},
\end{array}
\label{Init_cyl}
\end{equation}
where $H$ is the Heaviside function, $*$ denotes the time convolution, and $f$ is the Ricker wavelet
\begin{equation}
f(t)=\frac{\textstyle 1}{\textstyle 2\,\pi^2\,f_c^2}\,e^{-\textstyle \left(\pi\,f_c\,(t-t_c)\right)^2}.
\label{Ricker}
\end{equation}
The velocities and stresses of the incident wave can be deduced from (\ref{Init_cyl}) thanks to classical elastodynamics relations \cite{ACHENBACH}. The numerical experiments are performed on $N_x \times N_y=600 \times 400$ grid points, with $f_c=70$ Hz, $t_0=0.03$ s, $t_c=0.01429$ s, CFL=0.9 and $k=3$.

\begin{figure}[htbp]
\begin{center}
\begin{tabular}{c}
$(i=0)$\\
\\
\includegraphics[width=7.5cm,height=5cm]{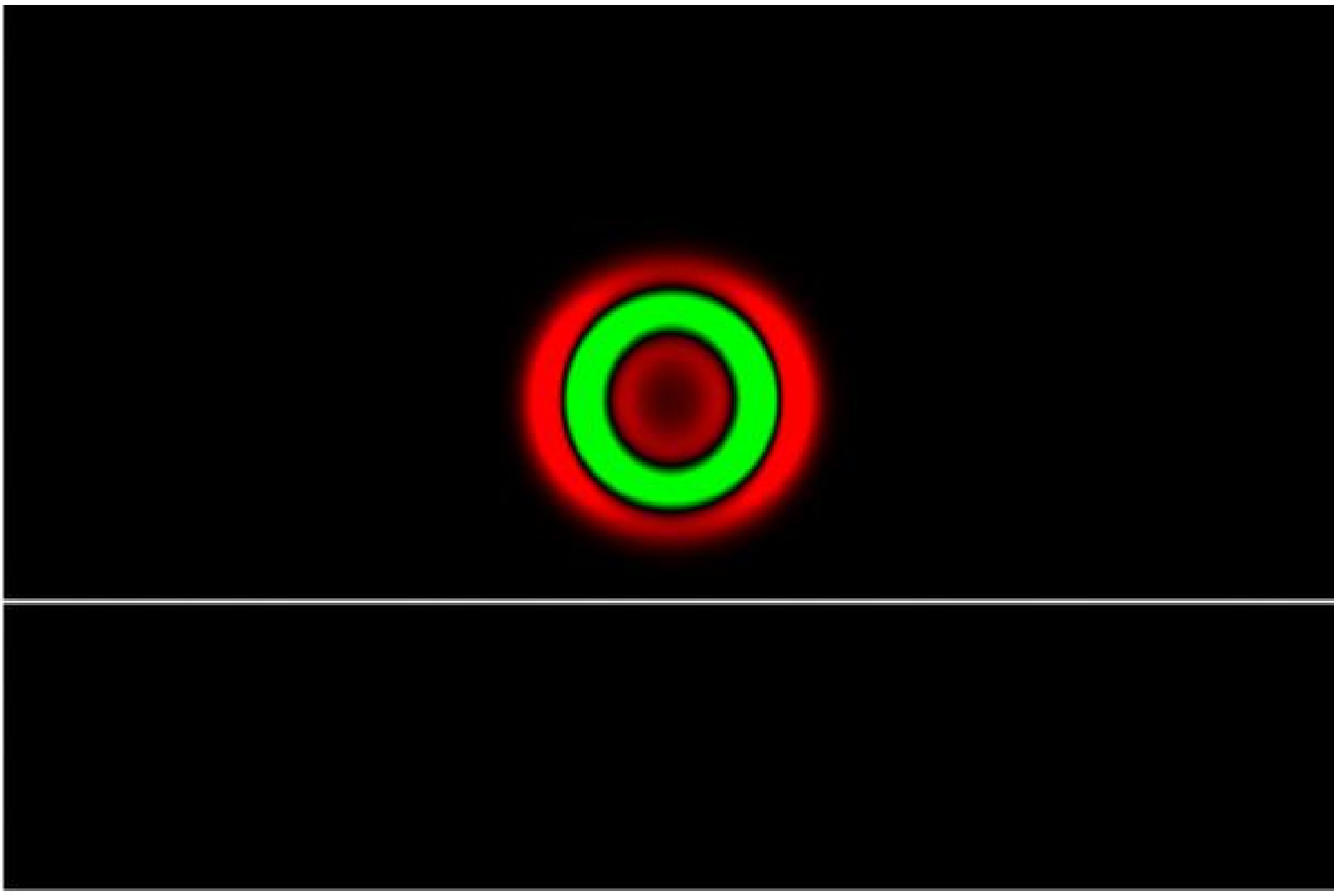}\\
\\
$(i=1)$\\
\\
\includegraphics[width=7.5cm,height=5cm]{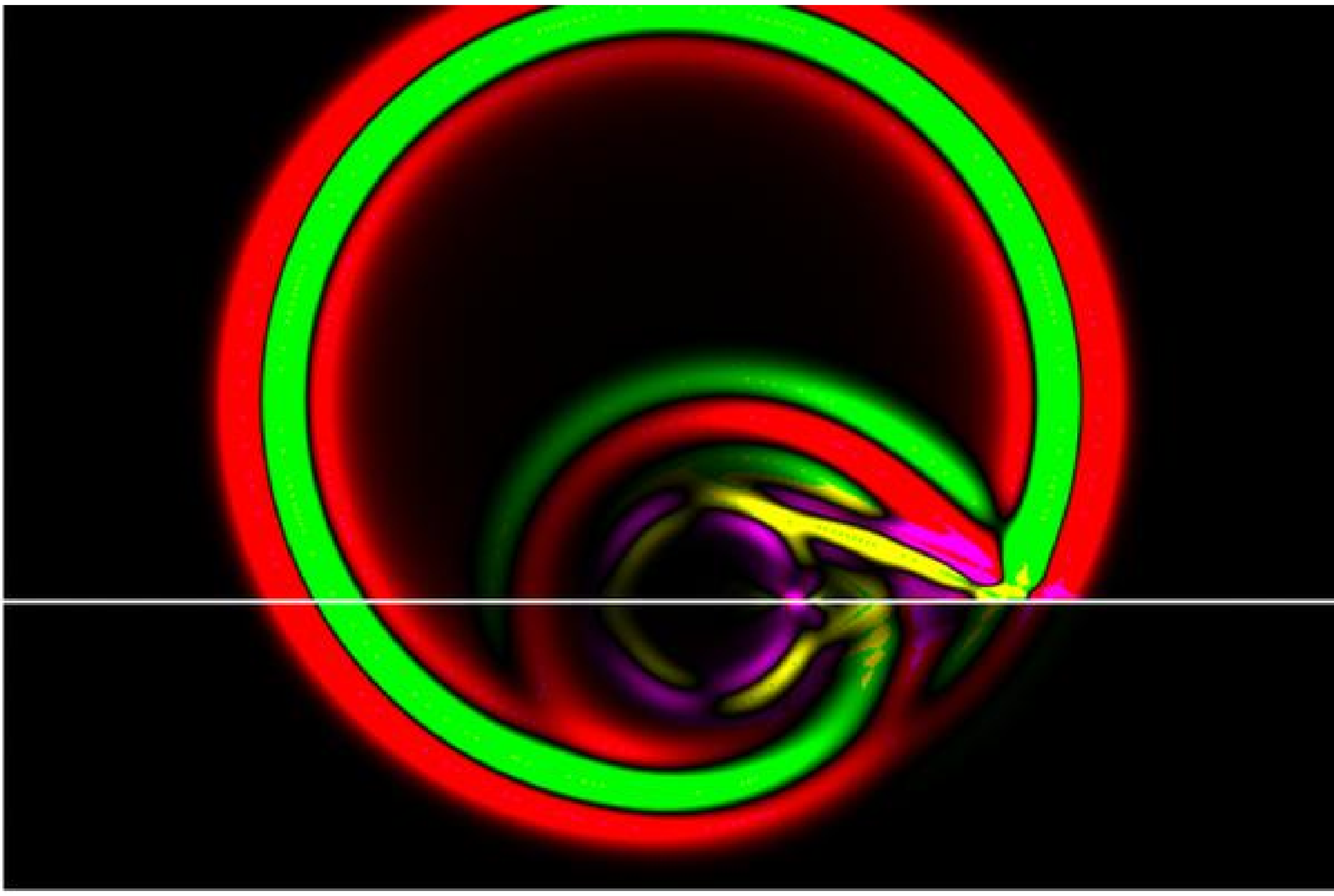}\\
\\
$(i=2)$\\
\\
\includegraphics[width=7.5cm,height=5cm]{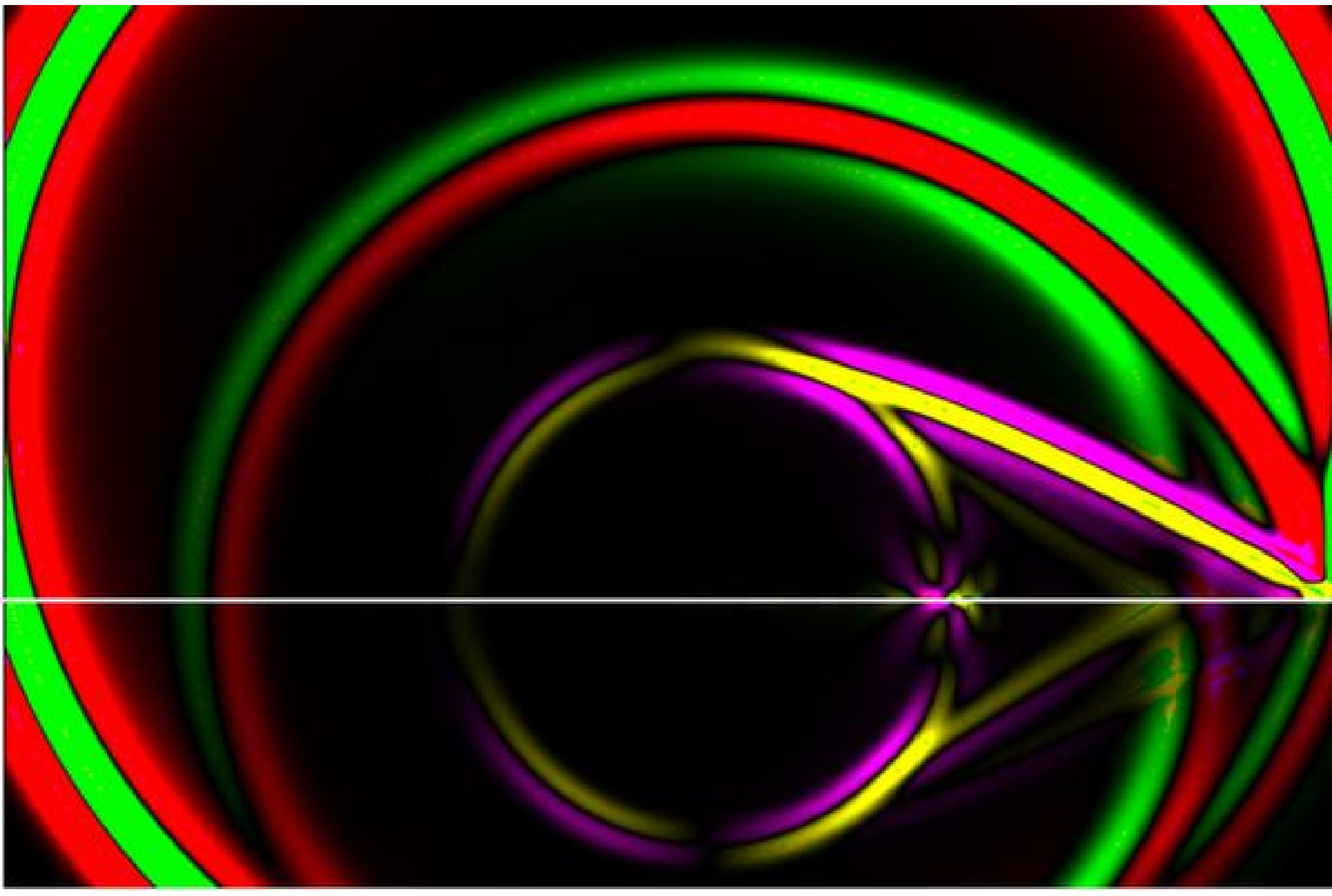}
\end{tabular}
\caption{Test 5: plane interface with variable contact at $t=t_0+150\,i\,\Delta\,t$.}
\label{Test5}
\end{center}
\end{figure}

Figure \ref{Test5} shows snapshots of the numerical solution at $t=t_0+150\,i\,\Delta\,t$. The palettes are changed in each snapshot to maintain a constant range of colours despite the geometrical spreading of the waves. On the left part of the second snapshot ($i=1$), the incident wave is totally transmitted; on the right part, the wave is almost totally reflected. The rapid change in the stiffness values occuring near $x=300$ m acts as a point source for cylindrical P- and SV-waves. On the last snapshot, one can also observe a rightward-moving surface wave, denoted by a yellow-magenta zone. Since both sides are almost stress-free surfaces in that place, this surface wave is a Rayleigh wave \cite{ACHENBACH}.

\section{Conclusion}

This paper deals with the numerical simulation of 2D wave propagation in elastic solids separated by imperfect contacts. These contacts are modeled by interfaces with linear jump conditions: the {\it spring-mass conditions} frequently used in geophysics \cite{PYRAK90} and nondestructive evaluation of materials \cite{ROUSSEAU03}. The spring-mass conditions are incorporated into finite-difference time-domain schemes on a uniform Cartesian grid, via an {\it interface method}. This is an extension of previous studies on perfect contacts in 1-D \cite{BIBLE1} and 2-D \cite{BIBLE2} contexts, and on imperfect contacts in 1-D context \cite{ALIMENTAIRE1}. The interface method also provides a fine description of the subcell geometry of the interfaces, at a negligible extra computational cost. Comparisons with original analytical solutions show the great accuracy of this approach.

Future paths of study might consist of taking into account numerically more realistic models for imperfect contact situations. A first improvement consists of describing the energy dissipation, as done, for example, in linear viscous bonding models \cite{FEHLER82}. A second improvement consists of extending the present approach for including nonlinear contact laws, which model realistic fractures and which prevent from interpenetration between the materials \cite{ZHAO01}.


\newpage

\appendix 

\section{On the asymmetry of 1-D spring-mass conditions}
The aim of this section is to show that, in some limit cases, the asymmetry of the jump conditions (\ref{SM_JC}) has a negligible influence.

Consider the following simplified case: that of a compressional plane wave propagating through $\Omega_0$ normally to the plane interface $\Gamma$ located at $x=\alpha$. Due to the symmetries involved in the problem, we have a 1-D configuration, with parameters $K=K_N$, $M=M_N$, $c_0=c_{p0}$, and $c_1=c_{p1}$ (a similar study could be performed with a shear plane wave and parameters $K=K_T$, $M=M_T$, $c_0=c_{s0}$, and $c_1=c_{s1}$). The normalized harmonic components of the elastic displacement $u$ and those of the elastic stress $\sigma$ in medium $\Omega_0$ (with $k_0=\omega/c_0$) are:
$$
\begin{array}{lll}
{\hat u}(x,\omega)&=&-i\,k_0\,e^{-ik_0 x}+i\,k_0\,R^{\pm}\,e^{ik_0 x},\\
&&\\
{\hat \sigma}(x,\omega)&=&-\rho_0\, \omega^2 \,e^{-ik_0 x}-\rho_0\, \omega^2\,R^{\pm}\,e^{ik_0 x},
\label{A1}
\end{array}
$$
and in medium $\Omega_1$ (with $k_1=\omega/c_1$), they are:
$$
\begin{array}{lll}
 {\hat u} (x,\omega)&=&-i\,k_1\,T^{\pm}\,e^{-ik_1 x},\\
&&\\
{\hat \sigma} (x,\omega)&=&-\rho_1\, \omega^2\, T^{\pm}\,e^{-ik_1 x},
\label{A2}
\end{array}
$$
where $\omega$ is the angular frequency, and $R^{\pm}$ and $T^{\pm}$ are the reflection and transmission coefficients it is required to determine (the superscripts $\pm$ are explained below). Under jump conditions "shifted to the left" (as in the present paper),
$$
\left[{\hat u}(\alpha,\,\omega)\right]=\frac{\textstyle 1}{\textstyle K}\,{\hat \sigma}(\alpha^-,\,\omega),\qquad
\left[{\hat \sigma}(\alpha,\,\omega)\right]=-M\,\omega^2\,{\hat u}(\alpha^-,\,\omega),
$$
a simple algebraic procedure gives
\begin{equation}
\begin{array}{l}
\displaystyle
R^-=\frac{\textstyle \rho_1\,c_1-\rho_0\,c_0-i\,\omega\left(\displaystyle \frac{\textstyle \rho_0\,c_0\,\rho_1\,c_1}{\textstyle K}-M\right)}{\textstyle \rho_0\,c_0+\rho_1\,c_1+i\,\omega\left(\displaystyle \frac{\textstyle \rho_0\,c_0\,\rho_1\,c_1}{\textstyle K}+M\right)}\,e^{-i\,\omega\frac{2\,\alpha}{c_0}},\\
\\
\displaystyle
T^-=\frac{\textstyle 2\,\rho_0\,c_1\left(\displaystyle 1+\frac{\textstyle M\,\omega^2}{\textstyle K}\right)}{\textstyle \rho_0\,c_0+\rho_1\,c_1+i\,\omega\left(\displaystyle \frac{\textstyle \rho_0\,c_0\,\rho_1\,c_1}{\textstyle K}+M\right)}\,e^{i\,\omega\left(\frac{1}{c_1}-\frac{1}{c_0}\right)\,\alpha}.
\end{array}
\label{RmTm}
\end{equation}
Under jump conditions "shifted to the right",
$$
\left[{\hat u}(\alpha,\,\omega)\right]=\frac{\textstyle 1}{\textstyle K}\,{\hat \sigma}(\alpha^+,\,\omega),\qquad
\left[{\hat \sigma}(\alpha,\,\omega)\right]=-M\,\omega^2\,{\hat u}(\alpha^+,\,\omega),
$$
we obtain in the same way
\begin{equation}
\begin{array}{l}
\displaystyle
R^+=\frac{\textstyle \rho_1\,c_1-\rho_0\,c_0-i\,\omega\left(\displaystyle \frac{\textstyle \rho_0\,c_0\,\rho_1\,c_1}{\textstyle K}-M\right)}{\textstyle \rho_0\,c_0+\rho_1\,c_1+i\,\omega\left(\displaystyle \frac{\textstyle \rho_0\,c_0\,\rho_1\,c_1}{\textstyle K}+M\right)}\,e^{-i\,\omega\frac{2\,\alpha}{c_0}},\\
\\
\displaystyle
T^+=\frac{\textstyle 2\,\rho_0\,c_1}{\textstyle \rho_0\,c_0+\rho_1\,c_1+i\,\omega\left(\displaystyle \frac{\textstyle \rho_0\,c_0\,\rho_1\,c_1}{\textstyle K}+M\right)}\,e^{i\,\omega\left(\frac{1}{c_1}-\frac{1}{c_0}\right)\,\alpha}.
\end{array}
\label{RpTp}
\end{equation}
Comparisons between (\ref{RmTm}) and (\ref{RpTp}) show that {\it the reflected waves are the same} under both types of jump conditions, and that
$$
\left|\frac{\textstyle T^--T^+}{\textstyle T^-}\right|=\frac{\textstyle M\,\omega^2}{\textstyle K}.
$$
With realistic contacts \cite{ALIMENTAIRE1,ROKHLIN1}, the stiffness and mass satisfy $K\sim \rho\,c^2\,/\,h$ and $M\sim \rho\,h$, where $\rho$ and $c$ are the parameters of the intermediate medium (or {\it interphase}) and $h$ is its thickness. Since $\omega=2\,\pi\,c/\lambda$ (where $\lambda$ is the wavelength) and $h\ll \lambda$ (a basic assumption underlying the spring-mass model), we obtain
$$
\left|\frac{\textstyle T^--T^+}{\textstyle T^-}\right| \sim 4\,\pi^2\left(\frac{\textstyle h}{\textstyle \lambda}\right)^2 \ll1.
$$
Hence, {\it the coefficients of transmission are very close together} under both types of jump conditions.

\section{Zero-th order matrices of spring-mass conditions}
The matrices $\boldsymbol{C}_l^0$ in (\ref{SM0}) are ($l=0,1$)
$$
\boldsymbol{C}_l^0=
\left(
\begin{array}{ccccc}
-y^{'} & x^{'} & 0 & 0 & 0\\[8pt]
x^{'} & y^{'} & 0 & 0 & 0\\[8pt]
0 & 0 & y^{'2} & -2\,x^{'}y^{'}& x^{'2}\\[8pt]
0 & 0 & -x^{'}y^{'} & x^{'2}-y^{'2} & x^{'}y^{'}\\[8pt]
\end{array}
\right).
$$
Setting
$$
\begin{array}{l}
\displaystyle
\beta_1=\frac{\textstyle \rho_0}{\textstyle K_N}\,\frac{\textstyle 1}{\textstyle \sqrt{x^{'2}+y^{'2}}},\quad
\beta_2=\frac{\textstyle \rho_0}{\textstyle K_T}\,\frac{\textstyle c_{s0}^2}{\textstyle \sqrt{x^{'2}+y^{'2}}},\\
\\
\displaystyle
\beta_3=\frac{\textstyle M_N}{\textstyle \rho_0}\,\sqrt{x^{'2}+y^{'2}},\quad
\beta_4=\frac{\textstyle M_T}{\textstyle \rho_0}\,\sqrt{x^{'2}+y^{'2}},
\end{array}
$$
the non-null components of $\boldsymbol{E}_0^0$ in (\ref{SM0}) are
$$
\begin{array}{ll}
\boldsymbol{E}_0^0[1,1]=\beta_1\left(y^{'2}c_{p0}^2+x^{'2}\left (c_{p0}^2-2\,c_{s0}^2\right)\right),\qquad &\boldsymbol{E}_0^0[3,3]=-\beta_3\,y^{'},\\
\boldsymbol{E}_0^0[1,2]=\beta_1\left(-2\,x^{'}y^{'}c_{s0}^2\right),&\boldsymbol{E}_0^0[3,4]=\beta_3\,x^{'},\\
\boldsymbol{E}_0^0[1,6]=\beta_1\left(-2\,x^{'}y^{'}c_{s0}^2\right),&\boldsymbol{E}_0^0[3,9]=-\beta_3\,y^{'},\\
\boldsymbol{E}_0^0[1,7]=\beta_1\left(x^{'2}c_{p0}^2+y^{'2}\left (c_{p0}^2-2\,c_{s0}^2\right)\right),& \boldsymbol{E}_0^0[3,10]=\beta_3\,x^{'},\\
\boldsymbol{E}_0^0[2,1]=\beta_2\left(-2\,x^{'}y^{'}\right),&\boldsymbol{E}_0^0[4,3]=\beta_4\,x^{'},\\
\boldsymbol{E}_0^0[2,2]=\beta_2\left(x^{'2}-y^{'2}\right),&\boldsymbol{E}_0^0[4,4]=\beta_4\,y^{'},\\
\boldsymbol{E}_0^0[2,6]=\beta_2\left(x^{'2}-y^{'2}\right),&\boldsymbol{E}_0^0[4,9]=\beta_4\,x^{'},\\
\boldsymbol{E}_0^0[2,7]=\beta_2\left(2\,x^{'}y^{'}\right),
&\boldsymbol{E}_0^0[4,10]=\beta_4\,y^{'}.
\end{array}
$$

\section{Exact solution for a plane wave on a circular interface with spring-mass conditions}\label{AnnexeExact}

This analytical solution is obtained in 6 steps:
\begin{enumerate}
\item Fourier-transforming the incident wave (\ref{Init});
\item writing the elastic potentials on the basis of circular functions;
\item expressing the elastic fields in terms of their potentials;
\item computing the reflection and transmission coefficients from (\ref{SM_JC});
\item returning to Cartesian coordinates;
\item inverting Fourier transform of the elastic fields (not described here).
\end{enumerate}
\textbf{Step 1.} The Fourier transform ${\cal F}$ of a function $s(t)$, and its inverse Fourier transform ${\cal F}^{-1}$, are denoted by
\begin{equation}
\begin{array}{l}
\displaystyle
{\hat s}(\omega)={\cal F}(s(t))=\frac{\textstyle 1}{\textstyle 2\,\pi}\int_{-\infty}^{+\infty}s(t)\,e^{-i\,\omega\,t}\,d\,t, \\
\\
\displaystyle
s(t)={\cal F}^{-1}({\hat s}(\omega))=\int_{-\infty}^{+\infty}{\hat s}(\omega)\,e^{i\,\omega\,t}\,d\,\omega,
\end{array}
\label{Fourier}
\end{equation}
where $\omega$ is the angular frequency. Applying a Fourier transform (\ref{Fourier}) to the following part of the initial wave (\ref{Init})
$$
\Phi_{IP}(x,y,t)=f\left(t-\frac{\textstyle x \cos \theta_1+y \sin \theta_1}{\textstyle c_{p1}}\right)
$$
gives the harmonic potential of the incident P-wave
\begin{equation}
{\hat \Phi}_{IP}(x,y,\omega)=e^{-i\,\frac{\omega}{c_{p1}}\left(x\,\cos \theta_1+y\,\sin\theta_1\right)}\,{\hat f}(\omega).
\label{Fourier_phi}
\end{equation}
The Fourier transform of the bounded sinusoid (\ref{JKPS}) is
\begin{equation}
{\hat f}(\omega)=A=\frac{\textstyle \omega_c}{\textstyle 2\,\pi}\left(\frac{\textstyle 1}{\textstyle \omega^2-\omega_c^2}+ \frac{\textstyle 1}{\textstyle \omega^2-4\,\omega_c^2}\right)\left(e^{-i\,\frac{2\,\pi}{\omega_c}\omega}-1 \right).
\label{Fourier_JKPS}
\end{equation}
\textbf{Step 2.} The circular interface is centred on $(x_0,y_0)$. A point $M(x,y)\in\Omega_1$ has polar coordinates $(x=x_0+r\,\cos\phi,y=y_0+r\,\sin\phi)$. We set ($l=0,\,1$)
$$
k_{pl}=\frac{\textstyle \omega}{\textstyle c_{pl}},\quad 
k_{sl}=\frac{\textstyle \omega}{\textstyle c_{sl}},\quad 
S=e^{-i\,k_{p1}\left(x_0\,\cos \theta_1+y_0\,\sin \theta_1\right)},\quad \theta=\phi-\theta_1.
$$
The first-kind Bessel functions $J_n$ satisfy the classical property \cite{MORSE}
\begin{equation}
e^{-i\,r\,\cos\theta} = \displaystyle \sum_{n=0}^{+\infty} \varepsilon_n\,i^n\,(-1)^n\,\cos n\,\theta\,J_n(r),
\label{Bessel}
\end{equation}
with $\varepsilon_n=1$ if $n=0$, 2 else. From (\ref{Fourier_phi}), (\ref{Fourier_JKPS}), and (\ref{Bessel}), we can therefore express the harmonic potential of the incident P-wave as
\begin{equation}
{\hat \Phi}_{IP}(x,y,\omega)=A\,S\,\sum_{n=0}^{+\infty}\varepsilon_n\,i^n(-1)^n\,\cos n\,\theta\,J_n(k_{p1}\,r).
\label{PhiInc}
\end{equation}
To satisfy the Sommerfeld condition, the harmonic elastic potential ${\hat \Phi}_{RP}$ of reflected P-waves and the harmonic pseudo-potential $\boldsymbol{{\hat \Psi}}_{RS}=(0,0,{\hat \Psi}_{RS})$ of reflected SV-waves are written on the basis of second-kind Hankel functions $H_n$. To prevent singularities from occuring at $r=0$, the harmonic potential ${\hat \Phi}_{TP}$ of transmitted P-waves and the harmonic pseudo-potential $\boldsymbol{{\hat \Psi}}_{TS}=(0,0,{\hat \Psi}_{TS})$ of transmitted SV-waves are written on the basis of first-kind Bessel functions. Hence, we write
\begin{equation}
\begin{array}{l}
\displaystyle
{\hat \Phi}_{RP}=\sum_{n=0}^{+\infty}R_n^p\,\cos n\,\theta\,H_n(k_{p1}\,r),\qquad
{\hat \Psi}_{RS}=\sum_{n=0}^{+\infty}R_n^s\,\sin n\,\theta\,H_n(k_{s1}\,r),\\
\\
\displaystyle
{\hat \Phi}_{TP}=\sum_{n=0}^{+\infty}T_n^p\,\cos n\,\theta\,J_n(k_{p0}\,r),\qquad
{\hat \Psi}_{TS}=\sum_{n=0}^{+\infty}T_n^s\,\sin n\,\theta\,J_n(k_{s0}\,r),
\end{array}
\label{PhiTra}
\end{equation}
where $R_n^p$, $R_n^s$, $T_n^p$ and $T_n^s$ are the unknown coefficients of reflection and transmission, which still remain to be determined. Note that numerical routines \cite{NRPAS} usually provide $Y_n$ and $Y^{'}_n$ (with $Y_n=J_n$ or $H_n$). Later in the text, we will need $Y^{''}_n$; to determine this last value, we use the differential equation satisfied by the Bessel and Hankel functions
$$
Y^{''}_n(x)+\frac{\textstyle 1}{\textstyle x}\,Y^{'}_n(x)+\left(1-\left(\frac{\textstyle n}{\textstyle x}\right)^2\right)\,Y_n(x)=0.
$$
\textbf{Step 3.} 
The elastic displacement $\boldsymbol{u}=\,^T(u_r,u_\theta)$ is deduced from $\Phi$ (for P-waves) or from $\boldsymbol{\Psi}=(0,0,\Psi)$ (for SV-waves) by
\begin{equation}
\boldsymbol{u}=\mbox{{\bf grad}}\, \Phi \,\mbox{  for P-waves},\qquad 
\boldsymbol{u}=\mbox{{\bf curl}}\,\boldsymbol{\Psi} \,\mbox{  for SV-waves}.
\end{equation}
In cylindrical coordinates, the {\bf grad} and {\bf curl} operators are
\begin{equation}
\mbox{{\bf grad}}\,\Phi=\left(
\frac{\textstyle \partial \,\Phi}{\textstyle \partial\,r},\,
\frac{\textstyle 1}{\textstyle r}\frac{\textstyle \partial \,\Phi}{\textstyle \partial\,\theta}
\right),
\qquad
\mbox{{\bf curl}}\,\boldsymbol{\Psi}(0,0,\Psi)=\left(
\frac{\textstyle 1}{\textstyle r}\frac{\textstyle \partial \,\Psi}{\textstyle \partial\,\theta},\,
-\frac{\textstyle \partial \,\Psi}{\textstyle \partial\,r}
\right).
\end{equation}
In cylindrical coordinates, the three independent components of $\boldsymbol{\sigma}$ are \cite{MORSE}
\begin{equation}
\begin{array}{l}
\displaystyle
\sigma_{rr}=\left(\lambda+2\,\mu\right)\, \frac{\textstyle \partial\,u_r}{\textstyle \partial\,r}+\lambda\left(\frac{\textstyle u_r}{\textstyle r}+\frac{\textstyle 1}{\textstyle r}\, \frac{\textstyle \partial\,u_\theta}{\textstyle \partial\,\theta}\right),\\
\\
\displaystyle
\sigma_{r\theta}=\mu\left(\frac{\textstyle \partial\,u_\theta}{\textstyle \partial\,r}-\frac{\textstyle u_\theta}{\textstyle r}+\frac{\textstyle 1}{\textstyle r}\, \frac{\textstyle \partial\,u_r}{\textstyle \partial\,\theta}\right),\\
\\
\displaystyle
\sigma_{\theta \theta}=\left(\lambda+2\,\mu\right)\, \left(\frac{\textstyle 1}{\textstyle r}\, \frac{\textstyle \partial\,u_\theta}{\textstyle \partial\,\theta}+\frac{\textstyle u_r}{\textstyle r}\right)+\lambda\,\frac{\textstyle \partial\,u_r}{\textstyle \partial\,r},
\end{array}
\label{ELASTO}
\end{equation}
where $\lambda$ and $\mu$ are the Lam\'e coefficients. From (\ref{PhiInc}), (\ref{PhiTra}), (\ref{ELASTO}), we easily deduce the harmonic fields over the whole domain. The components of the harmonic incident P-wave are
\begin{equation}
\begin{array}{l}
\displaystyle
{\hat u}_r^{ip}=A\,S\,k_{p1}\sum_{n=0}^{+\infty}\varepsilon_n\,i^{n}(-1)^n\cos n\,\theta\,J^{'}_n(k_{p1}\,r),\\
\displaystyle
{\hat u}_\theta^{ip}=-\frac{\textstyle A\,S}{\textstyle r}\sum_{n=0}^{+\infty}\varepsilon_n\,i^{n}\,\frac{\textstyle n\,\sin n\,\theta}{\textstyle r}\,J_n(k_{p1}\,r),\\
\displaystyle
{\hat \sigma}_{rr}^{ip}=A\,S\,\sum_{n=0}^{+\infty}\varepsilon_n\,i^{n}(-1)^n\cos n\,\theta\left(\left(\lambda_1+2\,\mu_1\right)k_{p1}^2\,J^{''}_n(k_{p1}\,r)\right.\\
\qquad \left.+\lambda_1\frac{\textstyle k_{p1}}{\textstyle r}\,J^{'}_n(k_{p1}\,r)-\lambda_1\left(\frac{\textstyle n}{\textstyle r}\right)^2\,J_n(k_{p1}\,r)\right),\\
\displaystyle
{\hat \sigma}_{r\theta}^{ip}=A\,S\,2\,\mu_1\sum_{n=0}^{+\infty}\varepsilon_n\,i^{n}(-1)^n\sin n\,\theta\left(\frac{\textstyle 1}{\textstyle r^2}J_n(k_{p1}\,r)-\frac{\textstyle k_{p1}}{\textstyle r}\,J^{'}_n(k_{p1}\,r)\right),\\
\displaystyle
{\hat \sigma}_{\theta\theta}^{ip}=A\,S\,\sum_{n=0}^{+\infty}\varepsilon_n\,i^{n}(-1)^n\cos n\,\theta\left(\lambda_1\,k_{p1}^2\,J^{''}_n(k_{p1}\,r)\right.\\
\qquad \left.+\left(\lambda_1+2\,\mu_1\right)\frac{\textstyle k_{p1}}{\textstyle r}\,J^{'}_n(k_{p1}\,r)-\left(\lambda_1+2\,\mu_1\right)\left(\frac{\textstyle n}{\textstyle r}\right)^2\,J_n(k_{p1}\,r)\right).
\label{CHAMP_INC}
\end{array}
\end{equation}
The components of harmonic reflected P-waves are 
\begin{equation}
\begin{array}{l}
\displaystyle
{\hat u}_r^{rp}=k_{p1}\sum_{n=0}^{+\infty}R_n^{p}\,\cos n\,\theta\,H^{'}_n(k_{p1}r),\\
\displaystyle
{\hat u}_\theta^{rp}=-\frac{\textstyle 1}{\textstyle r} \sum_{n=0}^{+\infty}R_n^{p}\,n\,\sin n\,\theta\,H_n(k_{p1}r),\\ 
\displaystyle
{\hat \sigma}_{rr}^{rp}=\sum_{n=0}^{+\infty} R_n^{p} \cos n\,\theta\left(\left(\lambda_1+2\,\mu_1\right)k_{p1}^2\,H^{''}_n(k_{p1}r)
+\lambda_1\,\frac{\textstyle k_{p1}}{\textstyle r}\,H^{'}_n(k_{p1}r)-\lambda_1\left(\frac{\textstyle n}{\textstyle r}\right)^2\,H_n(k_{p1}r)\right), \\
\displaystyle
{\hat \sigma}_{r\theta}^{rp}=2\,\mu_1\sum_{n=0}^{+\infty} R_n^{p}\,n\,\sin n\,\theta\left(\frac{\textstyle 1}{\textstyle r^2}H_n(k_{p1}r)-\frac{\textstyle k_{p1}}{\textstyle r}H^{'}_n(k_{p1}r)\right), \\
\displaystyle
{\hat \sigma}_{\theta\theta}^{rp}=\sum_{n=0}^{+\infty} R_n^{p}\cos n\,\theta \left(\lambda_1\,k_{p1}^2\,H^{''}_n(k_{p1}r)+\left(\lambda_1+2\,\mu_1\right)\,\frac{\textstyle k_{p1}}{\textstyle r}\,H^{'}_n(k_{p1}r)\right. \\
\displaystyle  \qquad \left.
-\left(\lambda_1+2\,\mu_1\right)\left(\frac{\textstyle n}{\textstyle r}\right)^2\,H_n(k_{p1}r)\right).
\label{CHAMP_REFP}
\end{array}
\end{equation}
The components of harmonic reflected SV-waves are 
\begin{equation}
\begin{array}{l}
\displaystyle
{\hat u}_r^{rs}=\sum_{n=0}^{+\infty}R_n^{s}\,\frac{\textstyle n\,\cos n\,\theta}{\textstyle r}\,H_n(k_{s1}r),\\
\displaystyle
{\hat u}_\theta^{rs}=-k_{s1}\sum_{n=0}^{+\infty}R_n^{s}\,\sin n\,\theta\,H^{'}_n(k_{s1}r),\\
\displaystyle
{\hat \sigma}_{rr}^{rs}=-2\,\mu_1\sum_{n=0}^{+\infty} R_n^{s}\,\cos n\,\theta\left(\frac{\textstyle 1}{\textstyle r^2}\,H_n(k_{s1}r)-\frac{\textstyle k_{s1}}{\textstyle r}\,H^{'}_n(k_{s1}r)\right),\\
\displaystyle
{\hat \sigma}_{r\theta}^{rs}=-\mu_1\sum_{n=0}^{+\infty} R_n^{s}\,\sin n\,\theta\left(k_{s1}^2\,H^{''}_n(k_{s1}r)-\frac{\textstyle k_{s1}}{\textstyle r}\,H^{'}_n(k_{s1}r)+\left(\frac{\textstyle n}{\textstyle r}\right)^2H_n(k_{s1}r)\right),\\
\displaystyle
{\hat \sigma}_{\theta\theta}^{rs}=-2\,\mu_1\,\sum_{n=0}^{+\infty} R_n^{s}\,n\,\cos n\,\theta\left(\frac{\textstyle k_{s1}}{\textstyle r}\,H^{'}_n(k_{s1}r)-\frac{\textstyle 1}{\textstyle r^2}\,H_n(k_{s1}r)\right).
\end{array}
\label{CHAMP_REFS}
\end{equation}
The components of harmonic transmitted P-waves are 
\begin{equation}
\begin{array}{l}
\displaystyle
{\hat u}_r^{tp}=k_{p0}\sum_{n=0}^{+\infty}T_n^{p}\,\cos n\,\theta\,J^{'}_n(k_{p0}r),\\
\displaystyle
{\hat u}_\theta^{tp}=-\frac{\textstyle 1}{\textstyle r} \sum_{n=0}^{+\infty}T_n^{p}\,n\,\sin n\,\theta\,J_n(k_{p0}r),\\ 
\displaystyle
{\hat \sigma}_{rr}^{tp}=\sum_{n=0}^{+\infty} T_n^{p} \cos n\,\theta\left(\left(\lambda_0+2\,\mu_0\right)k_{p0}^2\,J^{''}_n(k_{p0}r)\right. \\
\displaystyle  \qquad \left.
+\lambda_0\,\frac{\textstyle k_{p0}}{\textstyle r}\,J^{'}_n(k_{p0}r)-\lambda_0\left(\frac{\textstyle n}{\textstyle r}\right)^2\,J_n(k_{p0}r)\right), \\
\displaystyle
{\hat \sigma}_{r\theta}^{tp}=2\,\mu_0\sum_{n=0}^{+\infty} T_n^{p}\,n\,\sin n\,\theta\left(\frac{\textstyle 1}{\textstyle r^2}J_n(k_{p0}r)-\frac{\textstyle k_{p0}}{\textstyle r}J^{'}_n(k_{p0}r)\right), \\
\displaystyle
{\hat \sigma}_{\theta\theta}^{tp}=\sum_{n=0}^{+\infty} T_n^{p}\cos n\,\theta \left(\lambda_0\,k_{p0}^2\,J^{''}_n(k_{p0}r)+\left(\lambda_0+2\,\mu_0\right)\,\frac{\textstyle k_{p0}}{\textstyle r}\,J^{'}_n(k_{p0}r)\right. \\
\displaystyle  \qquad \left.
-\left(\lambda_0+2\,\mu_0\right)\left(\frac{\textstyle n}{\textstyle r}\right)^2\,J_n(k_{p0}r)\right).
\label{CHAMP_TRAP}
\end{array}
\end{equation}
Lastly, the components of harmonic transmitted SV-waves are 
\begin{equation}
\begin{array}{l}
\displaystyle
{\hat u}_r^{ts}=\sum_{n=0}^{+\infty}T_n^{s}\,\frac{\textstyle n\,\cos n\,\theta}{\textstyle r}\,J_n(k_{s0}r),\\
\displaystyle
{\hat u}_\theta^{ts}=-k_{s0}\sum_{n=0}^{+\infty}T_n^{s}\,\sin n\,\theta\,J^{'}_n(k_{s0}r),\\
\displaystyle
{\hat \sigma}_{rr}^{ts}=-2\,\mu_0\sum_{n=0}^{+\infty} T_n^{s}\,\cos n\,\theta\left(\frac{\textstyle 1}{\textstyle r^2}\,J_n(k_{s0}r)-\frac{\textstyle k_{s0}}{\textstyle r}\,J^{'}_n(k_{s0}r)\right),\\
\displaystyle
{\hat \sigma}_{r\theta}^{ts}=-\mu_0\sum_{n=0}^{+\infty} T_n^{s}\,\sin n\,\theta\left(k_{s0}^2\,J^{''}_n(k_{s0}r)-\frac{\textstyle k_{s0}}{\textstyle r}\,J^{'}_n(k_{s0}r)+\left(\frac{\textstyle n}{\textstyle r}\right)^2J_n(k_{s0}r)\right),\\
\displaystyle
{\hat \sigma}_{\theta\theta}^{ts}=-2\,\mu_0\,\sum_{n=0}^{+\infty} T_n^{s}\,n\,\cos n\,\theta\left(\frac{\textstyle k_{s0}}{\textstyle r}\,J^{'}_n(k_{s0}r)-\frac{\textstyle 1}{\textstyle r^2}\,J_n(k_{s0}r)\right).
\end{array}
\label{CHAMP_TRAS}
\end{equation}
\textbf{Step 4.} We now compute the coefficients $R_n^p$, $R_n^s$, $T_n^p$ and $T_n^s$. For this purpose, we deduce from the spring-mass conditions (\ref{SM_JC}) that
\begin{equation}
\begin{array}{l}
\displaystyle
\left({\hat u}_r^{ip}+{\hat u}_r^{rp}+{\hat u}_r^{rs}\right)(a^+,\theta)=\left({\hat u}_r^{tp}+{\hat u}_r^{ts}+\frac{\textstyle 1}{\textstyle K_N}\left({\hat \sigma}_{rr}^{tp}+{\hat \sigma}_{rr}^{ts}\right)\right)(a^-,\theta),\\[10pt]
\displaystyle
\left({\hat u}_\theta^{ip}+{\hat u}_\theta^{rp}+{\hat u}_\theta^{rs}\right)(a^+,\theta)=\left({\hat u}_\theta^{tp}+{\hat u}_\theta^{ts}+\frac{\textstyle 1}{\textstyle K_T}\left({\hat \sigma}_{r\theta}^{tp}+{\hat \sigma}_{r\theta}^{ts}\right)\right)(a^-,\theta),\\[12pt]
\displaystyle
\left({\hat \sigma}_{rr}^{ip}+{\hat \sigma}_{rr}^{rp}+{\hat \sigma}_{rr}^{rs}\right)(a^+,\theta)=\left({\hat \sigma}_{rr}^{tp}+{\hat \sigma}_{rr}^{ts}-M_N\,\omega^2\left({\hat u}_r^{tp}+{\hat u}_r^{ts}\right)\right)(a^-,\theta),\\[10pt]
\displaystyle
\left({\hat \sigma}_{r\theta}^{ip}+{\hat \sigma}_{r\theta}^{rp}+{\hat \sigma}_{r\theta}^{rs}\right)(a^+,\theta)=\left({\hat \sigma}_{r\theta}^{tp}+{\hat \sigma}_{r\theta}^{ts}-M_T\,\omega^2\left({\hat u}_\theta^{tp}+{\hat u}_\theta^{ts}\right)\right)(a^-,\theta),
\end{array}
\label{JC_Harm}
\end{equation}
for all $\theta$. Applying (\ref{JC_Harm}) to the fields (\ref{CHAMP_INC})-(\ref{CHAMP_TRAS}) yields an infinite number of linear systems. For computational purpose, these systems are computed up to $N_{Bessel}$ terms, giving the systems ($n=0,1,...,N_{Bessel}$)
\begin{equation}
\boldsymbol{Q}_n\,\boldsymbol{X}_n=\boldsymbol{Y}_n,\qquad \mbox{with }
\boldsymbol{X}_n=
\,^T\left(
R_n^p,\,R_n^s,\,T_n^p,\,T_n^s
\right),
\label{QX=Y}
\end{equation}
and
\begin{equation}
\boldsymbol{Y}_n=
-A\,S\,\varepsilon_n\,i^n(-1)^n
\left(
\begin{array}{l}
J^{'}_n(k_{p1}\,a)\\[10pt]
\displaystyle
\frac{\textstyle n}{\textstyle a}\,J_n(k_{p1}\,a)\\
\displaystyle
\left(\lambda_1+2\,\mu_1\right)k_{p1}^2\,J^{''}_n(k_{p1}\,a)+\frac{\textstyle \lambda_1}{\textstyle a}\,k_{p1}\,J^{'}_n(k_{p1}\,a)\\
\qquad \displaystyle -\lambda_1\left(\frac{\textstyle n}{\textstyle a}\right)^2\,J_n(k_{p1}\,a)\\
\displaystyle
2\,\mu_1\,n\left(\frac{\textstyle 1}{\textstyle a^2}\,J_n(k_{p1}\,a)-\frac{\textstyle k_{p1}}{\textstyle a}\,J^{'}_n(k_{p1}\,a)\right)
\end{array}
\right).
\end{equation}
The coefficients of the matrices $\boldsymbol{Q}_n$ are
\begin{equation}
\begin{array}{l}
\displaystyle
\boldsymbol{Q}_n[1,1]=k_{p1}\,H^{'}_n(k_{p1}\,a),\quad  
\boldsymbol{Q}_n[1,2]=\frac{\textstyle n}{\textstyle a}\,H_n(k_{s1}\,a),\\ 
\displaystyle
\boldsymbol{Q}_n[1,3]=-\left(k_{p0}\,J^{'}_n(k_{p0}\,a)+\frac{\textstyle 1}{\textstyle K_N}\left(
\left(\lambda_0+2\,\mu_0\right)k_{p0}^2\,J^{''}_n(k_{p0}\,a)\right.\right.\\
\displaystyle \qquad \qquad
\left.\left.+\frac{\textstyle \lambda_0}{\textstyle a}\,k_{p0}\,J^{'}_n(k_{p0}\,a)-\lambda_0\left(\frac{\textstyle n}{\textstyle a}\right)^2\,J_n(k_{p0}\,a)\right)\right),\\
\displaystyle
\boldsymbol{Q}_n[1,4]=-\left(\frac{\textstyle n}{\textstyle a}\,J_n(k_{s0}\,a)-\frac{\textstyle 2\,\mu_0\,n}{\textstyle K_N}\left(\frac{\textstyle 1}{\textstyle a^2}\,J_n(k_{s0}\,a)-\frac{\textstyle k_{s0}}{\textstyle a}\,J^{'}_n(k_{s0}\,a)\right)\right),
\\
\displaystyle
\boldsymbol{Q}_n[2,1]=\frac{\textstyle n}{\textstyle a} H_n(k_{p1}\,a),\quad  
\boldsymbol{Q}_n[2,2]=k_{s1}\,H^{'}_n(k_{s1}\,a),\\ 
\displaystyle
\boldsymbol{Q}_n[2,3]=-\left(\frac{\textstyle n}{\textstyle a}\,J_n(k_{p0}\,a)-\frac{\textstyle 2\,\mu_0\,n}{\textstyle K_T}\left(\frac{\textstyle 1}{\textstyle a^2}\,J_n(k_{p0}\,a)-\frac{\textstyle k_{p0}}{\textstyle a}\,J^{'}_n(k_{p0}\,a)\right)\right),\\
\displaystyle
\boldsymbol{Q}_n[2,4]=-\left(k_{s0}\,J^{'}_n(k_{s0}\,a)+\frac{\textstyle \mu_0}{\textstyle K_T}\left(
k_{s0}^2\,J^{''}_n(k_{s0}\,a)\right.\right.\\
\displaystyle \qquad \qquad
\left.\left.-\frac{\textstyle k_{s0}}{\textstyle a}\,k_{p0}\,J^{'}_n(k_{s0}\,a)+\left(\frac{\textstyle n}{\textstyle a}\right)^2\,J_n(k_{s0}\,a)\right)\right),
\\
\displaystyle
\boldsymbol{Q}_n[3,1]=\left(\lambda_1+2\,\mu_1\right)k_{p1}^2\,H^{''}_n(k_{p1}\,a)+\frac{\textstyle \lambda_1\,k_{p1}}{\textstyle a}\,H^{'}_n(k_{p1}\,a)-\lambda_1\left(\frac{\textstyle n}{\textstyle a}\right)^2\,H_n(k_{p1}\,a),\\
\displaystyle
\boldsymbol{Q}_n[3,2]=-2\,\mu_1\,n\left(\frac{\textstyle 1}{\textstyle a^2}\,H_n(k_{s1}\,a)-\frac{\textstyle k_{s1}}{\textstyle a}\,H^{'}_n(k_{s1}\,a)\right),\\
\displaystyle
\boldsymbol{Q}_n[3,3]=-\left(\left(\lambda_0+2\,\mu_0\right)k_{p0}^2\,J^{''}_n(k_{p0}\,a)+\left(\frac{\textstyle \lambda_1}{\textstyle a}-M_N\,\omega^2\right)k_{p0}\,J^{'}_n(k_{p0}\,a)\right.\\
\displaystyle \qquad \qquad 
\left.-\lambda_0\left(\frac{\textstyle n}{\textstyle a}\right)^2\,J_n(k_{p0}\,a)\right),\\
\displaystyle
\boldsymbol{Q}_n[3,4]=\left(\frac{\textstyle 2\,\mu_0\,n}{\textstyle a^2}+M_N\,\omega^2\,\frac{\textstyle n}{\textstyle a}\right)\,J_n(k_{s0}\,a)-\frac{\textstyle 2\,\mu_0\,n\, k_{s0}}{\textstyle a}\,J^{'}_n(k_{s0}\,a),
\\
\displaystyle
\boldsymbol{Q}_n[4,1]=2\,\mu_1\,n\left(\frac{\textstyle 1}{\textstyle a^2}\,H_n(k_{p1}\,a)-\frac{\textstyle k_{p1}}{\textstyle a}\,H^{'}_n(k_{s1}\,a)\right),\\
\displaystyle
\boldsymbol{Q}_n[4,2]=-\mu_1\left(k_{s1}^2\,H^{''}_n(k_{s1}\,a)-\frac{\textstyle k_{s1}}{\textstyle a}\,H^{'}_n(k_{s1}\,a)+\left(\frac{\textstyle n}{\textstyle a}\right)^2\,H_n(k_{s1}\,a)\right),\\
\displaystyle
\boldsymbol{Q}_n[4,3]=-\left(\left(\frac{\textstyle 2\,\mu_0\,n}{\textstyle a^2}+M_T\,\omega^2\,\frac{\textstyle n}{\textstyle a}\right)\,J_n(k_{p0}\,a)-\frac{\textstyle 2\,\mu_0\,n\, k_{p0}}{\textstyle a}\,J^{'}_n(k_{p0}\,a)\right),\\
\displaystyle
\boldsymbol{Q}_n[4,4]=\mu_0\,k_{s0}^2\,J^{''}_n(k_{s0}\,a)-\left(\frac{\textstyle \mu_0}{\textstyle a}+M_T\,\omega^2\right)k_{s0}\,J^{'}_n(k_{s0}\,a)+\mu_0\left(\frac{\textstyle n}{\textstyle a}\right)^2\,J_n(k_{s0}\,a).
\end{array}
\end{equation}
Since the values of Hankel functions can be huge (typically $10^{60}$), one must be careful when inverting (\ref{QX=Y}). Normalisation of $\boldsymbol{Q}_n$ and the use of extended arithmetic are required for this purpose.\\
\textbf{Step 5.} To return to Cartesian coordinates, we use the rotation formulas \cite{GERMAIN}
\begin{equation}
\left(
\begin{array}{c}
v_1\\[8pt]
v_2\\[8pt]
\sigma_{11}\\[8pt]
\sigma_{12}\\[8pt]
\sigma_{22}
\end{array}
\right)
=
\left(
\begin{array}{ccccc}
\cos \phi & -\sin \phi & 0 & 0 & 0\\[8pt]
\sin \phi & \cos \phi & 0 & 0 & 0\\[8pt]
0 & 0 & \cos^2 \phi & -2\,\sin \phi\,\cos \phi & \sin^2 \phi\\[8pt]
0 & 0 & \sin \phi \cos \phi & \cos^2 \phi - \sin^2 \phi & -\sin \phi \,\cos \phi\\[8pt]
0 & 0 & \sin^2 \phi & 2\,\sin \phi \,\cos \phi\ & \cos^2 \phi
\end{array}
\right)
\left(
\begin{array}{c}
v_r\\[8pt]
v_\theta\\[8pt]
\sigma_{rr}\\[8pt]
\sigma_{r\theta}\\[8pt]
\sigma_{\theta\theta}
\end{array}
\right).
\label{REPERE}
\end{equation}


\begin{thebibliography}{100} 

\bibitem{ACHENBACH}
{\sc J. D. Achenbach}, {\it Wave propagation in elastic solids}, North-Holland Publishing, Amsterdam, 1973.

\bibitem{BAIK1}
{\sc J.~M. Baik and R.~B. Thompson}, {\em Ultrasonic scattering from imperfect interfaces: a quasi-static model},
Journal of Nondestructive Evaluation, 4-3 (1985), pp.~177--196.

\bibitem{BAREN01}
{\sc G.~B.van Baren, W.~A. Mulder and G.C. Herman}, {\em Finite-difference modeling of scalar-wave propagation in cracked media},
Geophysics, 66-1 (2001), pp.~267--276.    

\bibitem{COATES95}
{\sc R.~T. Coates and M. Schoenberg}, {\em Finite-difference modeling of faults and fractures},
 Geophysics, 69-5 (1998), pp.~1514--1526.    

\bibitem{FEHLER82}
{\sc M. Fehler}, {\em Interaction of seismic waves with a viscous liquid layer},
Bull. Seism. Sos. Am., 72-1 (1982), pp.~55--72.

\bibitem{GERMAIN}
{\sc P. Germain, P. Muller}, {\it Introduction \`a la m\'ecanique des milieux continus},
Masson, 1995.  

\bibitem{GU2}
{\sc B. Gu, K. Nihei, and L. Myer}, {\em Numerical simulation of elastic wave propagation in fractured rock with the boundary integral equation method},
J. Geophys. Res., 101-B7 (1996), pp.~933--943.

\bibitem{HANEY03}
{\sc M. Haney and R. Sneider}, {\em Numerical modeling of waves incident on slip discontinuities}, Annual Report of the Consortium Project on Seismic Inverse Methods for Complex Structures (2003), pp.~1--10.

\bibitem{LEV90}
{\sc R.~J. LeVeque}, {\it Numerical Methods for Conservation Laws},
Birkhauser, 1990.

\bibitem{WPALG2}
{\sc R.~J. LeVeque}, {\em Wave propagation algorithms for multi-dimensional hyperbolic systems},
J. Comput. Phys., 131 (1997), pp.~327--353.

\bibitem{ALIMENTAIRE1}
{\sc B. Lombard, J. Piraux}, {\em How to incorporate the spring-mass conditions in finite-difference schemes}, SIAM J. Scient. Comput., 24-4 (2003), pp.~1379--1407.

\bibitem{BIBLE2} 
{\sc B. Lombard and J. Piraux}, {\em Numerical treatment of two-dimensional interfaces for acoustic and elastic waves}, 
J. Comput. Phys., 195-1 (2004), pp.~90--116. 

\bibitem{LOVE}
{\sc A.~E.~H. Love}, {\it A treatise on the mathematical theory of elasticity},
New-York Dover Publications (1944). 

\bibitem{MORSE}
{\sc P.~M. Morse, H. Feshbach}, {\it Methods of Theoretical Physics},
McGraw-Hill Book Company (1953). 
		  		  
\bibitem{BIBLE1}
{\sc J. Piraux and B. Lombard}, {\em A new interface method for hyperbolic problems with discontinuous coefficients. 1D acoustic example},
J. Comput. Phys.,168-1 (2001), pp.~227--248. 

\bibitem{NRPAS}
{\sc W.~H. Press, S.~A. Teukolskyn, W.~T. Vetterling, B.~P. Flannery}, {\it Numerical Recipes in Fortran: The Art of Scientific Computing},
Cambridge University Press (1992). 

\bibitem{PYRAK90}
{\sc L. Pyrak-Nolte, L. Myer, and N. Cook}, {\em Transmission of seismic waves across single natural fractures},
J. Geophys. Res., 95 (1990), pp.~8617--8638.

\bibitem{ROKHLIN1}
{\sc S.~I. Rokhlin and  Y.~J. Wang}, {\em Analysis of boundary conditions for elastic wave interaction with an interface between two solids}, J. Acoust. Soc. Am., 89-2 (1991), pp.~503--515.      

\bibitem{SAENGER00}
{\sc E. Saenger, N. Gold, A. Shapiro}, {\em Modeling the propagation of elastic waves using a modified finite-difference grid}, 
Wave Motion, 31 (2000), pp. 77-92.

\bibitem{SCHOENBERG80}
{\sc M. Schoenberg}, {\em Elastic wave behavior across linear slip interfaces}, 
J. Acoust. Soc. Am., 68-5 (1980), pp. 1516--1521.

\bibitem{TATTERSAL73}
{\sc H.G. Tattersall}, {\em The ultrasonic pulse-echo technique as applied to adhesion testing},
J. App. Phys., 6 (1973), pp.~819--832.
   
\bibitem{ROUSSEAU03}
{\sc V. Vlasie and M. Rousseau}, {\em Acoustical validation of the rheological models for a structural bond}, Wave Motion, 37 (2003), pp.~333--349.  

\bibitem{ZHAO01}
{\sc J. Zhao and J.~G. Cai}, {\em Transmission of elastic P-waves across single fractures with a nonlinear normal deformational behavior}, 
Rock Mech. Rock Engng., 34-1 (2001), pp. 3--22.
	  
\end{thebibliography}
\end{document}